\documentclass[aps, prx, reprint, superscriptaddress, floatfix]{revtex4-2}
\usepackage{bm}
\usepackage{graphicx}
\usepackage{xfrac}
\usepackage{float}
\usepackage{mathtools}
\usepackage{xcolor}

\begin{document}

\newcommand*{\SRO}{Sr$_{2}$RuO$_{4}$\space}
\newcommand*{\SROnospace}{Sr$_{2}$RuO$_{4}$}
\newcommand*{\NGO}{NdGaO$_{3}$\space}
\newcommand*{\NGOnospace}{NdGaO$_{3}$}
\newcommand*{\rT}{$\rho(T)$\space}
\newcommand*{\rO}{$\rho_{0}$\space}
\newcommand*{\uOc}{$\text{ }\mu\Omega\text{-cm}$}
\newcommand*{\mpicpfs}{Max Planck Institute for Chemical Physics of Solids, 01187 Dresden, Germany}
\newcommand*{\cornellphysics}{Department of Physics, Laboratory of Atomic and Solid State Physics, Cornell University, Ithaca, NY 14853, USA}
\newcommand*{\cornellmse}{Department of Materials Science and Engineering, Cornell University, Ithaca, NY 14853, USA}
\newcommand*{\cornellkavli}{Kavli Institute at Cornell for Nanoscale Science, Ithaca, NY 14853, USA}
\newcommand*{\lsisirius}{Laboratoire des Solides Irradi\'{e}s, CEA/DRF/IRAMIS, \'{E}cole Polytechnique, CNRS, Institut Polytechnique de Paris, 91128 Palaiseau, France}

\title{Controllable suppression of the unconventional superconductivity in bulk and thin-film \SRO via high-energy electron irradiation}

\author{Jacob P. Ruf}
\affiliation{\mpicpfs}
\affiliation{\cornellphysics}

\author{Hilary M. L. Noad}
\email[]{hilary.noad@cpfs.mpg.de}
\affiliation{\mpicpfs}

\author{Romain Grasset}
\affiliation{\lsisirius}

\author{Ludi Miao}
\affiliation{\cornellphysics}

\author{Elina Zhakina}
\affiliation{\mpicpfs}

\author{Philippa H. McGuinness}
\altaffiliation[Currently at ]{Institut f\"ur QuantenMaterialien und Technologien, Karlsruher Institut f\"ur Technologie, 76344 Eggenstein-Leopoldshafen}
\affiliation{\mpicpfs}

\author{Hari P. Nair}
\affiliation{\cornellmse}

\author{Nathaniel J. Schreiber}
\affiliation{\cornellmse}

\author{Naoki Kikugawa}
\affiliation{National Institute for Materials Science, Tsukuba 305-0003, Japan}

\author{Dmitry Sokolov}
\affiliation{\mpicpfs}

\author{Marcin Konczykowski}
\affiliation{\lsisirius}

\author{Darrell G. Schlom}
\affiliation{\cornellmse}
\affiliation{\cornellkavli}
\affiliation{Leibniz-Institut f\"{u}r Kristallzüchtung, 12489 Berlin, Germany}

\author{Kyle M. Shen}
\email[]{kmshen@cornell.edu}
\affiliation{\cornellphysics}
\affiliation{\cornellkavli}

\author{Andrew P. Mackenzie}
\email[]{andy.mackenzie@cpfs.mpg.de}
\affiliation{\mpicpfs}
\affiliation{SUPA, School of Physics and Astronomy, University of St Andrews, St Andrews KY16 9SS, United Kingdom}

\date{\today}

\begin{abstract}
In bulk \SROnospace, the strong sensitivity of the superconducting transition temperature $T_{\text{c}}$ to nonmagnetic impurities provides robust evidence for a superconducting order parameter that changes sign around the Fermi surface. In superconducting epitaxial thin-film \SROnospace, the relationship between $T_{\text{c}}$ and the residual resistivity $\rho_0$, which in bulk samples is taken to be a proxy for the low-temperature elastic scattering rate, is far less clear. Using high-energy electron irradiation to controllably introduce point disorder into bulk single-crystal and thin-film \SROnospace, we show that $T_{\text{c}}$ is suppressed in both systems at nearly identical rates. This suggests that part of $\rho_0$ in films comes from defects that do not contribute to superconducting pairbreaking, and establishes a quantitative link between the superconductivity of bulk and thin-film samples.
\end{abstract}

\maketitle

\section{Introduction \label{sec:new_intro}}

Conclusive identifications of the order parameter and of the interactions that induce superconductivity in \SRO have remained elusive despite receiving intense research interest for more than 25 years~\cite{mackenzie_superconductivity_2003, mackenzie_even_2017}. Among the first evidence for the unconventional nature of superconductivity in \SRO was the observation that the transition temperature $T_{\text{c}}$ could be completely suppressed with a minute concentration of nonmagnetic impurities~\cite{mackenzie_extremely_1998}, indicating that the phase of the superconducting order parameter changes sign in momentum space and, with sufficient scattering, averages to zero~\cite{mackenzie_superconductivity_2003}. Subsequent studies confirmed this extreme sensitivity to disorder, either by introducing variable defect densities via intentional chemical substitutions during growth~\cite{kikugawa_effect_2002, kikugawa_effects_2003, kikugawa_rigid-band_2004} or by taking advantage of native impurities and structural defects that exist in all nominally stoichiometric crystals~\cite{suderow_thermalcond_1998, mao_suppression_1999}.

While the basic fact of the unconventionality is well established, determining the specific symmetry of the order parameter in \SRO continues to be challenging, not least because of the intrinsically low energy scales of the problem, with $T_{\text{c}} \lesssim 1.5$ K. In principle, using epitaxial thin films of \SRO as a materials platform would offer a new angle of attack with numerous potential advantages: films can be patterned using standard optical lithography into devices and junctions for phase-sensitive magnetization and electrical transport measurements~\cite{uchida_characterization_2020, fang_quantum_2021}, and are naturally compatible with sophisticated scanned-probe microscopies~\cite{noad_variation_2016, watson_micron-scale_2018} and spectroscopic techniques~\cite{burganov_strain_2016, wang_separated_2021}. Indeed, in the cuprate high-temperature superconductors, scanning SQUID microscopy of thin-film samples provided conclusive evidence of the $d_{x^{2}-y^{2}}$ pairing symmetry~\cite{tsuei_symmetry_2000}.

The strong sensitivity of \SRO to disorder meant that it required many years of effort following the initial discovery of superconductivity in \SRO single crystals~\cite{maeno_superconductivity_1994} for the materials science community to realize epitaxial thin films of \SRO that were electronically clean enough to exhibit superconductivity~\cite{krockenberger_growth_2010}.  Now that superconducting thin-film samples can be reproducibly synthesized~\cite{uchida_molecular_2017, nair_demystifying_2018, garcia_pair_2020, kim_superconducting_2021}, there is an urgent need to establish a quantitative link between the physical properties of single crystals and the best thin films. The essential challenge in doing so is that the correlation between $T_{\text{c}}$ and the residual resistivity $\rho_{0}$---taken in bulk samples to be a proxy for the low-temperature elastic scattering rate---of as-grown films is not nearly as robust as that observed in single crystals. 

One hypothesis to explain the difference between the two forms of \SRO is that epitaxial thin films contain extended in-plane defects, such as small-angle grain boundaries, that increase the electrical resistivity but have a much smaller effect on the intrinsic intra-grain quasiparticle scattering rate and hence on $T_{\text{c}}$.  To test this hypothesis would require a means of introducing point disorder to such films while, hopefully, generating as small a change as possible to the resistivity of the extended defects.

Here, we use high-energy electron irradiation to controllably introduce point disorder into bulk single-crystal and thin-film \SROnospace.  We establish that the disorder from irradiation suppresses $T_{\text{c}}$ in bulk \SRO in the same way as chemical substitution or native point defects, but with far greater control over the disorder axis. We then demonstrate that thin films respond to irradiation-induced disorder in essentially the same way as bulk samples, with $T_{\text{c}}$ being suppressed in crystals and films at nearly identical rates.

\section{Methods \label{sec:methods}}

\subsection{FIB-sculpted single crystals of \SROnospace \label{subsec:methods_bulk}}

To prepare samples of bulk, single-crystal \SRO that were compatible with high-energy electron irradiations and that enabled reliable measurements of the low-temperature resistivity, we used a modified version of the epoxy-free method of mounting and preparing microstructures described in Ref.~\cite{sunko_controlled_2020} .  Procedures for the growth of the \SRO crystals are described in Ref.~\cite{bobowski_improved_2019}. 

From an oriented slice of the parent crystal, we cleaved a piece of \SRO having ${\langle}100{\rangle}$-oriented edges to expose a flat, $(001)$-oriented surface which later became the top surface of the finished microstructure. To make electrical contact, we sputtered 200~{nm} of gold onto the cleaved surface and annealed the crystal for 5 to 6 minutes at $500^{\circ}$C in air. We then used a focused ion beam (FIB) to extract lamellae (typical dimensions about $200~{\mu}\text{m} \times 160~{\mu}\text{m} \times 6~{\mu}\text{m}$), working from the uncleaved side inwards, and transferred these lamellae \textit{ex situ} onto mica substrates with the Au-coated surface facing upwards. After depositing 10~{nm} of Ti and another 150~{nm} of Au onto each lamella, we deposited Pt bridges in a xenon plasma-FIB to join the top of each lamella to the mica substrate. We then deposited a final 10\text~{nm} Ti/150~{nm} Au to ensure good electrical contact between the lamella and the gold film on the mica substrate. 

After mounting and contacting the lamellas, we used a gallium FIB to pattern each into a meandering microstructure suitable for four-point resistance measurements, as shown in Fig.~\ref{fig:rt_films_crystals_one_dose}(a). From the dimensions of each microstructured \SRO bar \cite{SupplementalMaterial}, we calculated the appropriate geometrical factors to convert the measured resistances into resistivities.  Such FIB-sculpted bars of \SRO have a well-defined shape and the positions of the voltage-sensing contacts on the crystal are independent of the wires and epoxy used for connecting the sample to external electronics equipment in subsequent resistance measurements.  The latter point is important because the wires attached to the mica substrate must be removed and remade every time the sample is transferred between the cryostat used for low-temperature electrical transport experiments and that used for electron irradiation experiments. Having a stable, reproducible geometry for the resistance measurement is essential for detecting small, irradiation-induced changes in the \textit{absolute} resistivity.

\subsection{Epitaxial thin films of \SROnospace \label{subsec:methods_films}}

\begin{figure}
\includegraphics{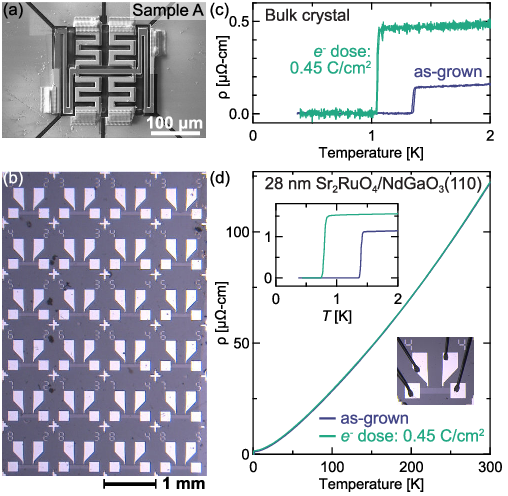}
\caption{\label{fig:rt_films_crystals_one_dose} Effects of high-energy electron irradiation on the electrical resistivity and superconducting $T_{\text{c}}$ of \SROnospace. (a) Scanning electron microscope image of a single crystal of bulk \SRO that was microstructured with a focused ion beam. (b) Optical microscope image of many resistivity bridges lithographically patterned on an epitaxial thin-film sample of \SROnospace. (c) Resistivity versus temperature at $T\leq2$~K for the crystal shown in (a), before and after the sample received a $0.45\text{ C}/\text{cm}^2$ dose of 2.5 MeV electron irradiation. The irradiation causes an increase in $\rho_{0}$ from $0.13$\uOc\space to $0.45$\uOc\space and a decrease in  $T_{\text{c}}$ from $1.35$ K to $0.97$ K. (d) Resistivity versus temperature data measured for a representative thin-film bridge, before and after a $0.45\text{ C}/\text{cm}^2$ dose of 2.5 MeV electron irradiation. Inset shows a zoomed-in view of $\rho(T)$ for $T < 2\text{ K}$: irradiation induces an increase in $\rho_{0}$ from $1.14$\uOc\space to $1.56$\uOc\space and a decrease in $T_{\text{c}}$ from $1.40$ K to $0.79$ K.}
\end{figure}

For our experiments on thin-film \SROnospace, we synthesized a (001)-oriented film of \SRO on a (110)-oriented \NGO substrate (Crystec, GmbH) by molecular beam epitaxy following the substrate preparation and adsorption-controlled film growth procedures detailed in Refs.~\cite{nair_demystifying_2018, nair_synthesis_2018}.  After preliminary structural characterization of the \SROnospace/\NGOnospace(110) wafer via x-ray diffraction \cite{SupplementalMaterial}, we used standard photolithography, sputter deposition, and ion milling techniques to pattern the \SRO film into many standalone devices for electrical resistivity measurements, as described in Ref.~\cite{fang_quantum_2021}.  As shown in Fig.~\ref{fig:rt_films_crystals_one_dose}(b), the region of the wafer patterned in this way is about $6\text{ mm} \times 4\text{ mm}$ in lateral extent, and each device occupies a total areal footprint of $1\text{ mm} \times 1 \text{ mm}$, thus allowing many bridges to be fabricated from the same film.  

Each resistivity bridge consists of two Pt~(25~{n}m)/Ti~(5~{n}m) contact pads at its ends for current injection and removal, and two Pt/Ti pads in the middle for attaching voltage-sensing wires, separated by a center-to-center distance of $L = 200~\mu\text{m}$.  Mismatch between the lattices of \SRO and the (110) surface of \NGO imposes in-plane, uniaxial and biaxial strains on the \SRO film; at 295~K, the two Ru-O-Ru bonding directions along $x$ and $y$ are compressed by $-0.39\%$ and $-0.16\%$ relative to unstrained, bulk single crystals of \SRO $(a = 3.8694~\text{\AA})$~\cite{walz_refinement_1993, schmidbauer_high-precision_2012}, corresponding to in-plane $A_{1g}$ and $B_{1g}$ strains of $-0.28\%$ and $0.23\%$. The longest dimension of every resistivity bridge is aligned with the slightly longer of the two in-plane Ru-O-Ru bond directions. Since the \NGO substrate is a wide-band-gap electrical insulator, the cross-sectional area through which the current is constricted to flow through is the \SRO film thickness, $t = 28.1\pm1.3$~{nm}, multiplied by the width of the resistivity bridges as determined by the lithography, $w = 80~\mu$m.  All of the physical dimensions of the thin-film resistivity bridges are significantly larger than the characteristic mean free paths of the charge carriers in \SRO along the corresponding crystallographic directions, so we expect Ohmic conduction to dominate the charge transport response at all temperatures.  

We diced the patterned wafer into individual pieces containing one or two devices and mechanically polished each of these smaller pieces down from the backside until the \NGO substrate was reduced to a thickness of 100 to 200 $\mu$m.  Each piece could then be handled and irradiated separately from the others; the substrate-thinning step allowed the irradiation to penetrate the full thickness of the film and substrate without causing excessive attenuation of the beam.  Throughout the main text and Supplemental Material \cite{SupplementalMaterial}, we refer to individual \SRO thin-film resistivity bridges separated out in this way from the same original wafer as distinct as-grown samples.  Comprehensive electrical characterization of these as-grown samples is described in the Supplemental Material \cite{SupplementalMaterial}. 

\subsection{Electrical transport measurements\label{subsec:methods_transport}}

We measured four-point resistances of the \SRO epitaxial thin films and FIB-sculpted single crystals using standard ac techniques at excitation frequencies less than 200~Hz, employing a multichannel lock-in amplifier (Synktek) and dual-ended current sources with active common mode rejection~\cite{BarberPhD17}. Unless otherwise noted, we used bias currents of $I_{\text{rms}} \leq 10~\mu\text{A}$ for resistance measurements. 

We controlled the sample temperature and external magnetic field using a Physical Properties Measurement System (Quantum Design) equipped with a helium-3 insert and a $14~\text{Tesla}$ superconducting magnet. To ensure uniform thermalization of the samples upon traversing the superconducting transitions, we incremented the temperature gradually at low temperatures: typical cooling/warming rates for thermal cycles between $0.4$~K and 2~K were about 1~K/hour. Detailed explanations of the definitions and analysis methods that we used to extract $\rho_{0}$ and $T_{\text{c}}$ from resistivity versus temperature data can be found in \cite{SupplementalMaterial}.

\subsection{High-energy electron irradiation\label{subsec:methods_irradiation}}

We performed high-energy electron irradiations ($E_{\text{incident}} = 2.5 \text{ MeV}$) at the SIRIUS Pelletron linear accelerator in Palaiseau, France, following the procedures described in~\cite{sunko_controlled_2020}.  The FIB-sculpted single-crystal and epitaxial thin-film samples all had the $c$-axis of the \SRO crystal structure aligned with the out-of-plane direction, along which the electron beam propagated. The samples were immersed in a bath of liquid hydrogen at a temperature of 22~K during irradiation in order to produce a random distribution of vacancy-interstitial Frenkel pairs~\cite{iseler_production_1966}. Maintaining low temperatures during electron irradiation promotes the formation of immobile, \textit{pointlike} defects, without appreciable defect clustering, cascades, columnar defects, or other extended spatial correlations~\cite{konczykowski_electron_1991}.  Measurements of the upper critical field of films before and after irradiation (Appendix~\ref{sec:appendix_magnetores}) confirm that the irradiation-induced defects are pointlike.

Using NIST tables \cite{nist_tables} to calculate the electron stopping power of \SRO in the continuous-slowing-down approximation, we estimate the stopping length of the $2.5$ MeV electrons to be 2.9 mm in \SROnospace, meaning that the irradiation-induced defects should be homogeneously distributed throughout the entire thickness of all samples studied here. According to the standard electron-scattering formalism, we expect that 2.5 MeV electrons transfer sufficient energy to the atoms in the crystal structure to create vacancy-interstitial Frenkel pairs on all four unique sublattices in \SRO (i.e., strontium, ruthenium, equatorial oxygen, and apical oxygen), albeit with higher production rates for the cations than for either oxygen site based on likely values for the interaction cross-sections (Appendix \ref{sec:appendix_insitu}). 

The electron irradiation facility in Palaiseau allows for \textit{in situ} electrical resistance measurements to be conducted during irradiation, with the sample held at constant $T = 22\text{ K}$.  Data of this kind are shown in Appendix~\ref{sec:appendix_insitu} for representative epitaxial thin-film and FIB-sculpted single-crystal samples of \SROnospace, and allow for perhaps the most straightforward comparison with theories that attempt to model cross sections for Frenkel pair production, based on tabulated cross sections for electron scattering from different nuclei and the displacement energies of the nuclei away from the unique chemical bonding sites in question. 

All low-temperature ($T < 22 \text{ K}$) electrical transport data on irradiated samples shown in this work, however, were acquired after \textit{ex situ} transfers of the samples from the cryostat at the irradiation facility to different cryostats at the Max Planck Institute for Chemical Physics of Solids in Dresden, Germany, which required warming up the samples to room temperature.  This initial ``room-temperature annealing'' step causes recombination of some of the irradiation-induced Frenkel pairs, as well as migration of some of the interstitials created during irradiation to various other sinks in the materials, such as surfaces, grain boundaries, and/or dislocations. Nevertheless, quasi-equilibrium populations of irradiation-induced vacancies and interstitials remain after the initial warmup to room temperature that are robust against further thermal cycles between 0 and 300~K.  It is the cumulative effect of the latter point-like scattering centers on the charge transport and superconductivity in \SRO that we attempt to quantify when investigating the dose dependencies of $\rho_{0}$ and $T_{\text{c}}$. 

\section{Results \label{sec:new_results}}

In Fig.~\ref{fig:rt_films_crystals_one_dose}(c,d), we show that a moderate $e^{-}$ dose  of 2.5 MeV electron irradiation ($D = 0.45 \text{ C}/\text{cm}^2 = 2.8 \times 10^{18} \, e^{-}/\text{cm}^2$) subtly perturbs the electrical resistivity $\rho(T)$ of \SROnospace. At low temperatures, where other, presumably inelastic, charge-carrier scattering mechanisms become frozen out by phase-space constraints, the effects of irradiation on $\rho$ become proportionally more relevant and noticeable. The irradiation-induced modification of the elastic transport scattering rate, as measured by the change in residual resistivity $\Delta\rho_{0}$, reduces the superconducting $T_{\text{c}}$  while largely preserving the sharpness of the superconducting transitions, indicating that the irradiation-induced defects are distributed homogeneously throughout the samples over the characteristic coherence length scales $\{\xi_{ab},\xi_{c}\}$ of the Cooper pairs.  The initial rates of $\rho_{0}$ increase and of $T_{\text{c}}$ suppression with $e^{-}$ dose are similar for the FIB-sculpted single crystal of \SRO and for the 28-nm \SROnospace/\NGOnospace(110) epitaxial thin-film sample.

Magnetoresistance measurements of the upper critical fields for superconductivity and of Shubnikov-de Haas oscillations for electron-irradiated \SRO thin films (Appendix~\ref{sec:appendix_magnetores}) provide additional evidence for the spatial uniformity and point-like nature of the irradiation-induced defects, and show that the primary effect of this additional disorder is to boost the rate at which the momentum of the charge-current-carrying excitations is relaxed, while preserving the itinerant charge-carrier densities (i.e., band occupancies) as well as other salient features of the normal-state electronic structure near the Fermi level, such as the quasiparticle velocities.

\begin{figure}
\includegraphics{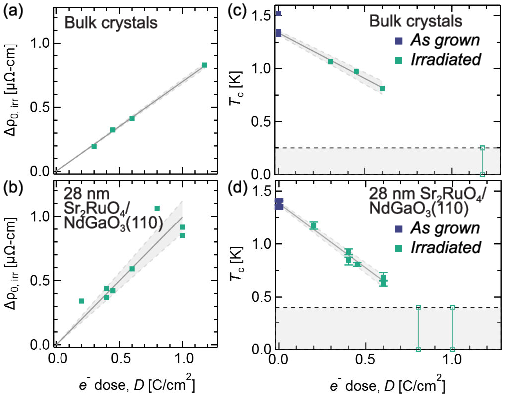}
\caption{\label{fig:dose_dependencies_rho_Tc} Irradiation dose dependencies of $\rho_{0}$ and $T_{\text{c}}$ in \SRO epitaxial thin films and single crystals. (a,b) Relative change in residual resistivity  $\Delta \rho_{0\text{, irr}}(D) \equiv \rho_{0\text{, after irr}}(D) - \rho_{0\text{, as grown}}$ induced per unit $e^{-}$ dose $D$ of 2.5 MeV electron irradiation in (a) microstructured bulk single crystals and (b) thin films after a single round of irradiation. Each data point represents a physically distinct sample.  Linear fits of all available data points are shown as solid gray lines; the shaded region bounded by dashed lines indicates $\pm 2\sigma$ uncertainties in the fitted slope of the best-fit line for each data set. (c,d) Resistively-measured superconducting $T_{\text{c}}$s versus accumulated $e^{-}$ dose $D$ for the irradiated samples displayed in panels (a,b) (teal markers), as well as the as-grown $T_{\text{c}}$s for the same samples (dark blue markers clustered at $D = 0$).  Here $T_{\text{c}}$ is defined as the temperature at which $\rho(T)$ crosses the 50\% threshold of $\rho_{0}$, and vertical error bars on each thin-film $T_{\text{c}}$ square in (d) indicate the temperatures at which $\rho(T)$ crosses the 10\% and 80\% thresholds of $\rho_{0}$; error bars defined analogously for the single-crystal $T_{\text{c}}$ in (c) are smaller than the heights of the corresponding midpoint $T_{\text{c}}$ squares.  In panels (c,d), irradiated samples having $T_{\text{c}}$s below the base temperatures of the measurement cryostats (horizontal black dashed lines) are drawn as open squares, and were not included in the linear fits of $T_{\text{c}}(D)$ drawn as solid gray lines [dashed gray lines have the same meaning as in panels (a,b)].   The vertical and horizontal scales of the single-crystal graphs in (a,c) are identical to the corresponding thin-film graphs in (b,d) to facilitate visual comparisons between the data sets.}
\end{figure}

Having established the qualitative effects of high-energy electron irradiation on some of the basic properties of the superconducting and correlated-metal states in \SROnospace, we now turn our attention to a detailed investigation of $e^{-}$-dose-dependent changes to $\rho_{0}$ and $T_{\text{c}}$.  We leave direct microscopic visualization of the irradiation-induced disorder in \SRO to future studies, because detecting dilute concentrations of point-like defects,  on the order of one part per thousand, is a notoriously challenging task, well beyond the current capabilities of state-of-the-art microstructural probes such as scanning transmission electron microscopy \cite{kim_electronic_2019, goodge_disentangling_2022}.  

Given that substitutional disorder on the Ru site has been shown to increase $\rho_{0}$ at approximately 10$\times$ the rate of substitional disorder on the Sr site \cite{kikugawa_effect_2002, kikugawa_effects_2003, kikugawa_rigid-band_2004}, we infer that Ru vacancies (produced at a rate of $1.2$ to $2.6 \times 10^{-3}$ displacements per Ru atom per $1 \text{ C}/\text{cm}^2$ of irradiation) account for most of the observed modifications to $\rho_{0}$. We provide a detailed discussion of the conversions between $D$ and Frenkel defect densities, and an analysis of the likely contributions of these defects to $\rho_{0}$, in Appendix \ref{sec:appendix_insitu}. 

In Fig.~\ref{fig:dose_dependencies_rho_Tc}(a), we plot the relative irradiation-induced changes in $\rho_{0}$ measured across four distinct FIB-sculpted single-crystal samples of \SRO as a function of the accumulated dose $D$ of 2.5 MeV electron irradiation, and in Fig.~\ref{fig:dose_dependencies_rho_Tc}(c) we plot the resistively measured $T_{\text{c}}$s versus $D$ for all of these samples, before and after irradiation. In Fig.~\ref{fig:dose_dependencies_rho_Tc}(b,d), we plot the same quantities as in Fig.~\ref{fig:dose_dependencies_rho_Tc}(a,c), except now for nine distinct 28-nm \SROnospace/\NGOnospace(110) thin-film samples. Raw $\rho(T)$ curves for all thin-film samples that underlie the data points in Fig.~\ref{fig:dose_dependencies_rho_Tc}(b,d) are presented in the Supplemental Material \cite{SupplementalMaterial}. For both types of \SRO samples, there are clear systematic effects of $D$ on both $\Delta\rho_{0\text{, irr}}$ and $T_{\text{c}}$, that further corroborate the trends suggested by the data and samples highlighted in Fig.~\ref{fig:rt_films_crystals_one_dose}: precisely controlled increases in $D$ cause nearly monotonic increases in $\rho_{0}$ that are accompanied by precisely controlled, smooth decreases in $T_{\text{c}}$.  

For the FIB-sculpted \SRO single-crystal sample that received the highest irradiation dose in Fig.~\ref{fig:dose_dependencies_rho_Tc}(a,c) ($D = 1.18 \text{ C}/\text{cm}^2$), such a dose was sufficient to reduce $T_{\text{c}}$ from its as-grown value of $1.52\text{ K}$ to a post-irradiation value that was less than the base temperature of the cryostat (in this case, an adiabatic demagnetization insert for the PPMS), $T_{\text{c}} < 0.25\text{ K}$. Accordingly, we schematically display this sample in Fig.~\ref{fig:dose_dependencies_rho_Tc}(c) as open square markers consistent with a range of possible $T_{\text{c}}$s between $0$ and $0.25 \text{ K}$. This sample was not included in fits to $T_{\text{c}}(D)$ described below, but it provides an illustrative example that highlights the stringent requirements on sample purity for observing superconductivity in \SROnospace: an extremely small (in absolute terms) increase of the quasiparticle scattering rate by $\Delta\rho_{0} = 0.83 \, \mu\Omega\text{-cm}$ is sufficient to suppress $T_{\text{c}}$ from near its maximally-achievable value to almost zero. Likewise, for the \SRO thin-film samples that accumulated the highest irradiation doses in Fig.~\ref{fig:dose_dependencies_rho_Tc}(b,d) ($D = 0.80\text{ C}/\text{cm}^2$ and $1.00\text{ C}/\text{cm}^2$), we did not observe superconductivity in $\rho(T)$ data collected down to the base temperature of the helium-3 cryostat, $0.4\text{ K}$. We do not include these points in subsequent fits to $T_{\text{c}}(D)$, and we leave a more comprehensive exploration of the regime of very low $T_{\text{c}}s$ to future work. 

To provide the simplest-possible quantitative parameterizations of these dose dependencies, we assume that $\Delta\rho_{0 \text{, irr}}$ and $T_{\text{c}}$ should initially vary linearly with $D$ over the investigated ranges of $e^{-}$ doses in Fig.~\ref{fig:dose_dependencies_rho_Tc}, because all $D$ here correspond to dilute concentrations of irradiation-induced defects, of the order of $0.1\%$ (Appendix~\ref{sec:appendix_insitu}).  Fitting the data points in Fig.~\ref{fig:dose_dependencies_rho_Tc} to lines, and taking systematic error bars on the fit coefficients to be $\pm 2\sigma$ surrounding the best-fit values, we obtain the curves summarized in Table~\ref{table:D_fits}. Alternatively, the explicit dependencies of $\Delta\rho_{0}$ and ${\Delta}T_{\text{c}}$ on $D$ can be eliminated by taking the ratios of the slopes of the best-fit lines, which yields $\Delta T_{\text{c}}/\Delta\rho_{0} = (-1.24 \pm 0.08) \text{ K}/\mu\Omega\text{-cm}$ for bulk crystals and $(-1.23 \pm 0.19) \text{ K}/\mu\Omega\text{-cm}$ for the thin films.

\begin{table*}[]
\caption{\label{table:D_fits} Linear fits to the dose dependencies of $\rho_{0}$ and of $T_{\text{c}}$ plotted in Fig.~\ref{fig:dose_dependencies_rho_Tc}(a,b) and Fig.~\ref{fig:dose_dependencies_rho_Tc}(c,d), respectively. Error bars on the fit coefficients are $\pm 2\sigma$. Units: $\rho_{0}$ $[\mu\Omega\text{-cm}]$, $T_{\text{c}}$ $[\text{K}]$, $D$ $[\text{C}/\text{cm}^{2}]$.}
\begin{ruledtabular}
\begin{tabular}{lcccc}
Sample type & $\Delta\rho_{0}(D)$ & Range of fit in $D$ & ${\Delta}T_{\text{c}}(D)$ & Range of fit in $D$ \\
\colrule
Bulk single crystal & $(0.70 \pm 0.02)D$ & $[0.3, 1.18]$ & $(1.34 \pm 0.02)-(0.86 \pm 0.07)D$ & $[0, 0.6]$\\
28 nm \SROnospace/\NGOnospace(110) films & $(1.00 \pm 0.13)D$ & $[0.2, 1.0]$ & $(1.38 \pm 0.02)-(1.22 \pm 0.06)D$ & $[0, 0.6]$\\
\end{tabular}
\end{ruledtabular}
\end{table*}

In other words, although the \textit{individual} magnitudes of both of the rates $\Delta \rho_{0}/\Delta D$ and $|\Delta T_{\text{c}}/\Delta D|$ for the \SROnospace / \NGOnospace(110) epitaxial thin-film samples are greater, in a statistically significant sense, than the corresponding rates for the FIB-sculpted single crystal \SRO samples, these differences essentially compensate each other when taking the \textit{ratio} of these quantities, resulting in initial rates of $T_{\text{c}}$ suppression with increasing residual resistivity, $\Delta T_{\text{c}}/\Delta \rho_{0}$, for the two types of \SRO samples that are indistinguishable within experimental precision.

\section{Discussion \label{sec:new_discussion}} 

\begin{figure} 
\includegraphics{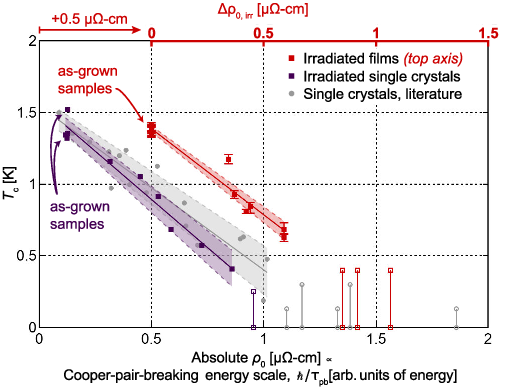}
\caption{\label{fig:Tc_vs_disorder_summary} Universal impact of elastic, momentum-relaxing scattering on superconductivity in \SROnospace. Gray circles represent previously published data for bulk single-crystal samples of \SRO containing native defects~\cite{mackenzie_extremely_1998}, as well as intentionally-substituted chemical impurities on the Ru and Sr sites\textemdash viz., Sr$_{2}$(Ru$_{1-x}$Ti$_{x}$)O$_{4}$~\cite{kikugawa_effects_2003}, Sr$_{2}$(Ru$_{1-x}$Ir$_{x}$)O$_{4}$~\cite{kikugawa_effects_2003}, and (Sr$_{2-y}$La$_{y}$)RuO$_{4}$~\cite{kikugawa_rigid-band_2004}. Dark purple squares represent results from this work for $e^{-}$-irradiated \SRO single crystals containing native defects plus irradiation-induced vacancy-interstitial Frenkel pairs on all atomic sublattices.  For all single-crystal data shown here, the measured residual resistivity in absolute units, $\rho_{0} \, [\mu\Omega\text{-cm}]$, is utilized as a proxy assumed to be directly proportional to the Cooper-pair-breaking energy scale, $\hbar/\tau_{\text{pb}}$, with no adjustments of the common zero point of these scales (bottom axis). On the other hand, results from this work for $e^{-}$-irradiated 28-nm \SROnospace/\NGOnospace(110) epitaxial thin-film samples are plotted as red squares against the irradiation-induced \textit{change} in residual resistivity, $\Delta\rho_{0\text{, irr}} \, [\mu\Omega\text{-cm}]$ (top axis).  As explained in the main text, a range of rigid shifts of the zero point of the top axis relative to the bottom axis are, in principle, compatible with the results of this work; here we chose an offset of $+0.5$\uOc\space for visual clarity, indicated by the red arrow in the upper left corner.  Color-coded straight lines are linear fits to each family of \SRO samples and the systematic uncertainties in the best-fit rates of $T_{\text{c}}$ suppression ($\pm 1\sigma$) are indicated by the color-coded shaded regions between dashed lines.} 
\end{figure}

In Fig.~\ref{fig:Tc_vs_disorder_summary}, we plot 22 ($\rho_{0}$, $T_{\text{c}}$) data points extracted from the literature as gray circles for bulk single crystals of \SRO containing natively occurring (as-grown) defects \cite{mackenzie_extremely_1998}, deliberate substitutions of Ti or Ir on the Ru site \cite{kikugawa_effects_2003}, as well as deliberate substitutions of La on the Sr site \cite{kikugawa_rigid-band_2004}.  Results for electron-irradiated FIB-sculpted single crystals of \SRO from the present work are overlaid as dark purple squares on these reference data points.  For the $e^{-}$-irradiated crystals, several repetitions of the cycle of \{measure $\rho(T)$ $\rightarrow$ irradiate $\rightarrow$ remeasure $\rho(T)$\} allowed us to use the same physical samples to acquire multiple ($\rho_{0}$, $T_{\text{c}}$) data points---hence the larger number of data points shown in Fig.~\ref{fig:Tc_vs_disorder_summary} (11 in total) than in Fig.~\ref{fig:dose_dependencies_rho_Tc}(b,d)---which, taken together, span nearly the entire range of disorder over which superconductivity is observed in unstrained bulk single crystals of \SROnospace. The overall trend of the previously reported $T_{\text{c}}(\rho_{0})$ dependence is reproduced well, with the more precisely clustered data from this work suggesting a slightly steeper slope of the best-fit line than previously, as seen by comparing the purple and gray solid lines in Fig.~\ref{fig:Tc_vs_disorder_summary}. 

The $T_{\text{c}}$ data points for all \SRO single-crystal samples in Fig.~\ref{fig:Tc_vs_disorder_summary} are plotted versus the absolute residual resistivity $\rho_{0}$, without any shift in the zero point of this scale. Implicit in this choice is the assumption that all sources of quasiparticle scattering that contribute to $\rho_{0}$, both intrinsically present from the original crystal growth and induced by irradiation, also contribute proportionally to the Cooper-pair-breaking energy scale $\hbar/\tau_{\text{pb}}$, and thus to the suppression of $T_{\text{c}}$. This assumption is justified by the limited scatter in the $T_{\text{c}}(\rho_{0})$ behavior observed across different studies, and by the inferred extrapolation of the experimental data to a common range of zero-disorder $T_{c0} = 1.5$ to $1.55$ K, which are furthermore consistent with the highest values of $T_{\text{c}}$ found in the literature for actual \SRO crystals under ambient conditions.

In contrast, the relatively large scatter in the initial absolute values of $\rho_{0}$ across nine distinct as-grown \SRO thin films~($1.1$ to $1.5 $ $\mu\Omega{\text{-cm}}$ \cite{SupplementalMaterial}) is incommensurate with the narrow distribution of as-grown $T_{\text{c}}$s for these same samples ($1.35$ to $1.41$ K). An assumption of direct proportionality between $\rho_{0}$ and $\hbar/\tau_{\text{pb}}$ is therefore not appropriate for the thin-film samples. Nevertheless, while there is no discernible correlation between \textit{absolute} values of $\rho_{0}$ and $T_{\text{c}}$, the \textit{changes} in $\rho_{0}$ and $T_{\text{c}}$ after being subjected to different doses of 2.5 MeV electron irradiation, $\Delta{\rho_{0}}$ and $\Delta{T_{\text{c}}}$, respond in a highly correlated fashion, in a way that is similar to the dose-dependent trends that we observed on FIB-sculpted single crystals. The most intuitive way to reconcile these observations is to assume that many of the natively occurring extended defects in as-grown \SRO thin films \cite{garcia_pair_2020, zurbuchen_suppression_2001, zurbuchen_defect_2003, zurbuchen_morphology_2007, fang_quantum_2021, uchida_molecular_2017, nair_demystifying_2018, kim_superconducting_2021, goodge_disentangling_2022} contribute to $\rho_{0}$ but do not contribute appreciably to $T_{\text{c}}$ suppression, whereas any pointlike defects, including those created by electron irradiation, cause both $\rho_{0}$ and $\hbar/\tau_{\text{pb}}$ to rapidly increase in lockstep, just as in bulk crystals.

The natural interpretation of the thin-film data as reflecting a separation of independent sources of scattering that couple differently to $\rho_{0}$ and $T_{\text{c}}$ therefore suggests that our irradiation experiments should still accurately probe the Cooper-pair-breaking rate of $T_{\text{c}}$ suppression, $\Delta T_{\text{c}}/\Delta(\hbar/\tau_{\text{pb}}) \propto \Delta T_{\text{c}}/\Delta \rho_{0}$, in thin films, provided that the \emph{relative} change in $\rho_{0}$ is measured pairwise for the same thin-film sample before and after each sample accumulates a given dose $D$ of electron irradiation (Appendix~\ref{sec:appendix_est_Tcmax} and Fig.~\ref{fig:Tc_vs_absRho0_films}). Following this line of reasoning, we combine the data for the as-grown and electron-irradiated 28-nm \SROnospace/\NGOnospace(110) thin-film samples previously shown in Fig.~\ref{fig:dose_dependencies_rho_Tc}(b,d) and replot these 18 $T_{\text{c}}$ data points as red squares against the top axis of Fig.~\ref{fig:Tc_vs_disorder_summary}, $\Delta \rho_{0\text{, irr}}$, thereby eliminating the explicit dependence on the dose $D$ of 2.5 MeV electron irradiation. By employing the same units of $[\mu\Omega\text{-cm}]$ for the quasiparticle scattering rates on the top and bottom axes in Fig.~\ref{fig:Tc_vs_disorder_summary}, we do not allow for any arbitrary scale factors in comparing the \SRO thin-film and single-crystal results. 

\begin{table}[]
\caption{\label{table:dTc_drho0_fits} Error bars on the fit coefficients are $\pm 1\sigma$.}
\begin{ruledtabular}
\begin{tabular}{lc}
Sample type & $\Delta T_{\text{c}}/\Delta \rho_{0}$ (K$/\mu\Omega\text{-cm}$) \\
\colrule
Single crystal (\cite{mackenzie_extremely_1998, kikugawa_effects_2003, kikugawa_effects_2003, kikugawa_rigid-band_2004}) & $-1.15 \pm 0.11$\\
Irradiated single crystal (this work)& $-1.34 \pm 0.09$\\
Irradiated films (this work)& $-1.20 \pm 0.07$\\
\end{tabular}
\end{ruledtabular}
\end{table}

Applying this procedure reveals a striking similarity between the best-fit slopes $\Delta T_{\text{c}}/\Delta \rho_{0}$ of the single-crystal and thin-film data sets, as depicted by the color-coded solid lines in Fig.~\ref{fig:Tc_vs_disorder_summary} and summarized in Table~\ref{table:dTc_drho0_fits}.  This result strongly suggests that momentum-relaxing elastic scattering from pointlike disorder plays the same microscopic role in Cooper-pair breaking in thin-film and single-crystal \SROnospace, a conclusion which is further supported by a comparison of the $T_{\text{c}}$ versus $\rho_{0}$ behavior observed here with previous impurity-scattering studies of bulk \SRO single crystals \cite{mackenzie_extremely_1998, mao_suppression_1999, kikugawa_effect_2002, kikugawa_effects_2003, kikugawa_rigid-band_2004}.  Despite those studies encompassing numerous distinct realizations of ``disorder''---i.e., variable densities of chemically different pointlike scattering centers, some of which couple weakly and some of which couple strongly to the charged quasiparticle excitations near the Fermi level---there is remarkable consistency across all studies when these microscopic details of the scattering processes are subsumed into their combined impact on the residual resistivity, and $\rho_{0}$ is treated as the relevant independent variable that ultimately controls the dependent variable, $T_{\text{c}}$. 

Quantitatively determining the appropriate rigid shift between the zero points of the top and bottom axes in Fig.~\ref{fig:Tc_vs_disorder_summary} would require more information than we have obtained in these experiments, since we do not know the clean-limit $T_{c0}$ of these thin-film samples with certainty because the films are subject to biaxial ($A_{\text{1g}}$) and uniaxial ($B_{\text{1g}}$) in-plane strains imparted by lattice matching to the substrate.  Indeed, the simplest-possible empirical estimates, described in Appendix~\ref{sec:appendix_est_Tcmax}, place a maximum range on $T_{c0}$ for \SROnospace/\NGOnospace(110) of $1.4$ to $2.7~\text{K}$; the illustrative choice of the $0.5 \text{ }\mu\Omega\text{-cm}$ horizontal offset employed in Fig.~\ref{fig:Tc_vs_disorder_summary} corresponds to $T_{c0} = 2.0~\text{K}$, and is furthermore consistent with a comparison to expectations from Abrikosov-Gor'kov pairbreaking theory (Appendix~\ref{sec:appendix_AG}).

Regardless of this remaining quantitative uncertainty in determining the exact $T_{c0}$ of \SROnospace/\NGOnospace(110) on a true plot of $T_{\text{c}}(\hbar/\tau_{\text{pb}})$, the results displayed in Fig.~\ref{fig:Tc_vs_disorder_summary} offer compelling evidence that quantitatively very similar mechanisms of superconductivity are operative in epitaxial thin films and single crystals of \SROnospace. We finish with a brief discussion of what insights are offered by the robust connection between $T_{\text{c}}$ and $\rho_{0}$ revealed by this work regarding open questions surrounding the materials science and the physics of superconducting \SRO thin films. 

Although the in-plane $B_{1g}$ orthorhombic strain imparted by the \NGOnospace(110) substrates on \SRO likely boosts $T_{c0}$ of these thin films slightly above that of unstrained bulk tetragonal \SROnospace~\cite{steppke_strong_2017, nair_demystifying_2018}, our data indicate that any strain-induced enhancement of the clean-limit pairing scale in our thin films (and concomitant symmetry-imposed anisotropies in the superconducting gaps caused by the reduction of the crystal point-group symmetry from tetragonal~($D_{4h}$) to orthorhombic~($D_{2h}$)) are modest effects. They likely result from subtle changes in the superconducting order parameter that become irrelevant in the experimentally accessible range of quasiparticle scattering rates, because samples of either variety having as-grown $T_{\text{c}}$s of about $1.4 \text{ K}$ exhibit nearly identical rates of $T_{\text{c}}$ suppression with increasing $\rho_{0}$ when additional point-like defects are introduced via high-energy electron irradiation.  

Turning this observation around, the relatively ``high'' $T_{\text{c}}$s achieved for the as-grown films, which approach the clean-limit value for bulk \SRO of $1.50$ to $1.55 \text{ K}$, indicate minuscule concentrations of native in-plane defects that are known to be strongly pair-breaking (for example, of order $0.01\%$ Ru vacancies). This attests to the remarkable degree of vacancy control that has been achieved in the best \SRO thin films, which sets a benchmark for modern oxide molecular-beam epitaxy.  This result is also encouraging from a physics point of view, particularly for ongoing research efforts using \SRO thin films to gain new perspectives on the outstanding puzzles regarding the order parameter and mechanism of the enigmatic unconventional superconductivity in this material.  Finally, we note that the smooth progression from ``high'' to ``low'' $T_{\text{c}}$s in all cases is consistent with the presence of a sign-changing gap $\Delta(\mathbf{k})$ on all bands (likely with symmetry-enforced, non-accidental nodes) if intraband scattering dominates the pair breaking \cite{mackenzie_extremely_1998}, or with the absence of large differences between the average magnitudes of $|\Delta(\mathbf{k})|$ on distinct bands if interband scattering dominates the pair breaking (see also Appendix~\ref{sec:appendix_AG}, Figs.~\ref{fig:Tc_vs_rho0_AGcomp_crystals} \& \ref{fig:Tc_vs_rho0_AGcomp_films}).

In conclusion, the results displayed in Figs.~\ref{fig:rt_films_crystals_one_dose}-\ref{fig:Tc_vs_disorder_summary} show the insights that can be gained by precisely tuning the point defect concentrations in \SRO using high-energy electron irradiation.  These experiments can be performed equally well on epitaxial thin films and single crystals of \SROnospace, and have the extremely high precision demonstrated here.  Other methods commonly used to locate samples along a horizontal ``disorder'' axis, such as x-ray spectroscopy in electron microprobes \cite{mackenzie_extremely_1998, mackenzie_recent_1993}, require large sample volumes to accurately quantify dilute impurity concentrations, and would not have the sensitivity to detect such minute deviations from ideal stoichiometry.  The precision available in low-volume stoichiometry techniques based on transmission electron microscopy \cite{sunko_controlled_2020, goodge_disentangling_2022}, Raman scattering \cite{kim_electronic_2019}, or Rutherford backscattering spectrometry, is lower still, so the high-energy electron irradiation enables investigation of a part of parameter space that cannot currently be reached by any other technique.  Moreover, the ability in irradiation studies to directly compare measurements performed on the same samples, before and after irradiation, renders moot concerns about inevitable sample-specific peculiarities that could obscure the effects of the pairbreaking point disorder that we wish to study. 

With a controlled and reproducible knob now available for finely adjusting the quasiparticle scattering rate and $T_{\text{c}}$ in \SROnospace, it may be possible to definitively resolve some of the longstanding controversies regarding the superconductivity, such as the interpretation of heat capacity data for $T < T_{\text{c}}$ \cite{nishizaki_effect_1999, li_high-sensitivity_2021, li_elastocaloric_2022} in light of other thermodynamic probes that evidence a multi-component order parameter~\cite{ghosh_thermodynamic_2021, benhabib_ultrasound_2021, kivelson_proposal_2020}, as well as the interpretation of signals in muon spin relaxation~\cite{luke_time-reversal_1998,grinenko_split_2021,grinenko_unsplit_2021} and polar Kerr effect data~\cite{xia_high_2006}, which evidence some kind of time-reversal symmetry breaking onsetting near $T_{\text{c}}$. Looking forward, it will also be interesting to see what new experiments \SRO thin films enable in this field; for example, performing irradiations using a collimated $e^{-}$ beam inside an electron microscope~\cite{tolpygo_effect_1996, tolpygo_universal_1996} may provide a straightforward way to spatially pattern the disorder landscape so as to create superconductor/normal-metal all-\SRO tunnel junctions for use in phase-sensitive measurements.

% Data availability:
Data plotted in Figs. 1-9 of the main text and Figs. S1-S5 of the Supplemental Material will be available at~\footnote{Dataset for Ruf \textit{et al.} ``Controllable suppression of the unconventional superconductivity in \SRO via high-energy electron irradiation'' (2024); DOI to be inserted here}. Additional data available upon reasonable request.

\begin{acknowledgments}
We thank Cyrus Dreyer for useful discussions. J.P.R., H.M.L.N., E.Z., P.H.M., D.S., and A.P.M. thank the Max Planck Society for financial support.  H.M.L.N. was supported by the Alexander von Humboldt Foundation Research Fellowship for Postdoctoral Researchers.  N.K. is supported by JSPS KAKENHI (No. JP18K04715, No. JP21H01033, and No. JP22K19093). Research in Dresden benefits from the environment provided by the DFG Cluster of Excellence ct.qmat (EXC 2147, project ID 390858490). Electron irradiation experiments performed at the SIRIUS beamline were supported by the EMIR\&A French network (FR CNRS 3618). This work was supported by the National Science Foundation Platform for the Accelerated Realization, Analysis, and Discovery of Interface Materials (PARADIM) under Cooperative Agreement No. DMR-2039380. This research was funded in part by the Gordon and Betty Moore Foundation's EPiQS Initiative through Grant Nos. GBMF3850 and GBMF9073 to Cornell University. Sample preparation was facilitated in part by the Cornell NanoScale Facility, a member of the National Nanotechnology Coordinated Infrastructure (NNCI), which is supported by the National Science Foundation (Grant No. NNCI-2025233).

J.P.R. and H.M.L.N. contributed equally to this work.

\end{acknowledgments}

\begin{appendix}

\section{Conversions between accumulated dose and Frenkel defect density \label{sec:appendix_insitu}} 

\begin{figure}
\includegraphics{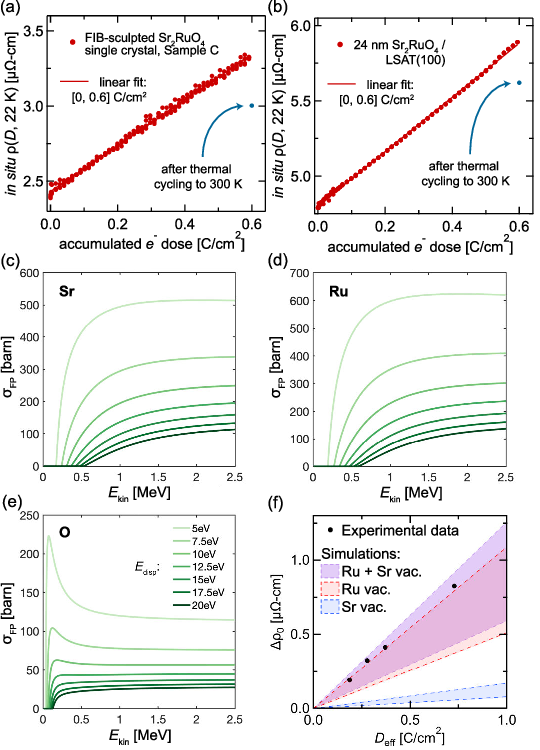}
\caption{\label{fig:frenkel_defect_cs} Dependence of $\rho(22\text{ K})$ on irradiation dose, cross sections for creating Frenkel defects, and the effects of these defects on the residual resistivity in \SROnospace.  (a,b) $\rho(D,\,22\text{ K})$ measured \textit{in situ} during 2.5 MeV $e^{-}$ irradiations, for a representative (a)~FIB-sculpted single-crystal and (b)~epitaxial thin-film sample of \SROnospace. (c-e) Calculated cross sections for creating vacancy-interstitial Frenkel pairs from (c)~strontium, (d)~ruthenium, and (e)~oxygen atoms via electron irradiation. Each panel considers a range of possible displacement energies $E_{\text{disp}}$ for creating each type of Frenkel pair (light to dark green color scale). (f) Data for FIB-sculpted \SRO single crystals showing the experimentally measured increments in residual resistivity after accumulating variable doses of $E_{\text{kin}} = 2.5 \text{ MeV}$ $e^{-}$ irradiation (black markers).  These data are to be compared with the expected changes in $\rho_{0}$ that are attributable to irradiation-induced Sr vacancies (blue dashed lines), Ru vacancies (red dashed lines), and the sum of Ru + Sr vacancies (purple dashed lines), given the Frenkel pair production rates displayed in (c,d). The region between each pair of dashed lines represents how the systematic uncertainties in $E_{\text{disp}}$ convert into ranges of possible $\Delta \rho_{0}$ versus $D_{\text{eff}}$ behavior; the physically relevant parameter ranges are likely $E_{\text{disp}}$ ranging from $7.5$ to 15~{eV} for all sublattices in \SROnospace.}
\end{figure}

\subsection{\textit{In situ} resistivity experiments}

In Fig.~\ref{fig:frenkel_defect_cs}(a,b), we show the resistivity $\rho(D)$ measured \textit{in situ} (during 2.5 MeV electron irradiations) at $T = 22\text{ K}$ as a function of accumulated $e^{-}$ dose, for representative (a)~FIB-sculpted single-crystal samples and (b)~epitaxial thin-film samples of \SROnospace, respectively. Increasing $D$ causes monotonic, nearly-linear increases in $\rho(D,\,22\text{ K})$.  The best-fit lines to data collected over the range $D$:~$[0, 0.60] \text{ C}/\text{cm}^2$ are:

\begin{equation*}
\rho(D,\,22\text{ K}) \,\, [\mu\Omega\text{-cm}] = \begin{cases*} 
4.8 + 1.8 \cdot D \, [\text{C}/\text{cm}^2] & \text{(films)} \\
2.4 + 1.5 \cdot D \, [\text{C}/\text{cm}^2] & \text{(crystals)} 
\end{cases*} 
\end{equation*}  

\noindent After both samples accumulated $e^{-}$ doses of $0.60 \text{ C}/\text{cm}^2$, we thermally cycled the samples to $300\text{ K}$ and back to $22\text{ K}$ without removing the samples from the irradiation cryostat, and remeasured the resistivities indicated by the blue markers in Fig.~\ref{fig:frenkel_defect_cs}(a,b).  This initial ``room-temperature annealing'' of the irradiated samples caused $\rho(22\text{ K})$ to decrease from $3.3$ to $ 3.0$\uOc \space ($5.89$ to $5.62$\uOc) for the FIB-sculpted single-crystal (thin-film) \SRO samples, respectively, or to about 91\% (95\%) of the starting values. Such reductions of $\rho$ are commonly observed upon initially warming up electron-irradiated crystals, including materials that have similar solid-state chemistries to \SRO (e.g. SrRuO$_{3}$~\cite{klein_negative_2001,haham_testing_2013} and high-$T_{\text{c}}$ cuprates~\cite{konczykowski_electron_1991,legris_effects_1993}), and are typically attributed to some fractions of the original concentrations of irradiation-induced defects becoming mobile at elevated temperatures.  Finally, we note that due to technical difficulties with conducting \textit{in situ} transport measurements during some irradiation beamtimes, the thin-film sample measured in Fig.~\ref{fig:frenkel_defect_cs}(b) is a resistivity bridge patterned on a 24-nm \SROnospace/LSAT(100) sample, rather than the 28-nm \SROnospace/\NGOnospace(110) samples characterized elsewhere in the manuscript. Although this likely affects certain quantitative details of the measurements, such as the rate of resistivity increase $\Delta\rho(22\text{ K})/\Delta D$, we believe that all qualitative features of the observed $\rho(D)$ behavior in Fig.~\ref{fig:frenkel_defect_cs}(b), including the annealing of some defects upon the first thermal cycle to 300~{K}, would also be observed for irradiated \SROnospace/\NGOnospace(110) samples measured \textit{in situ}. 

\subsection{Conversions between accumulated electron irradiation dose and Frenkel defect densities}

We used the methods outlined in Ref.~\cite{sunko_controlled_2020} to estimate the rates at which Frenkel defects (i.e., vacancy-interstitial pairs) are produced in \SRO by high-energy electron irradiation. Creating a defect requires a certain amount of energy, known as the displacement energy $E_{\text{disp}}$, which is specific to a given atomic site and type of defect. The probability that such a defect will be formed during irradiation is given by the relativistic cross section for the transfer of $E \ge E_{\text{disp}}$ from an electron having kinetic energy $E_{\text{kin}}$ to a nucleus of that atom. In our irradiation experiments, the probability that a high-energy electron ($E_{\text{kin}} \approx E_{\text{incident}} = 2.5 \text{ MeV}$) interacts with \textit{any} atom in the crystal is extremely low; we are operating in the ``long mean free path'' limit. Therefore, we calculate irradiation-induced defect concentrations for each atomic site independently, using the same global, measured irradiation dose in each case, without self-consistently accounting for energy losses and/or multiple scattering events that reduce $E_{\text{kin}}$ below $E_{\text{incident}}$ as the electron beam propagates through the samples. 

In Fig.~\ref{fig:frenkel_defect_cs}(c-e), we plot the calculated cross sections for creating Frenkel pairs $\sigma_{\text{FP}}$ on the strontium, ruthenium, and oxygen sites, respectively.  The dependence of $\sigma_{\text{FP}}$ on the electron kinetic energy $E_{\text{kin}}$ (horizontal axis) is modeled using values for the relativistic Mott scattering cross section tabulated in the literature~\cite{lijian_analytic_1995, boschini_expression_2013, sunko_controlled_2020}, and the simulations are performed considering a range of displacement energies $E_{\text{disp}}$ (individual traces, green colorscale) for the formation of each type of Frenkel pair.  Density-functional-based calculations of Frenkel defect energetics suggest that in \SROnospace, $E_{\text{disp, Sr}} \geq 7.6 \text{ eV}$ and $E_{\text{disp, Ru}} \geq 8.2 \text{ eV}$~(C. E. Dreyer, private communication).  Meanwhile, previous electron irradiation studies of high-$T_{\text{c}}$ copper oxides (specifically, YBa$_{2}$Cu$_{3}$O$_{7-\delta}$) that are structurally similar to \SRO have found that $E_{\text{disp, eqO}} = 10 \text{ eV}$~\cite{legris_effects_1993} or $8.4 \text{ eV}$~\cite{tolpygo_effect_1996} for creating Frenkel defects on the planar oxygen site.  Accordingly, we propose that displacement energies in the range of $E_{\text{disp}} = 7.5$ to 15~{eV} are the physically relevant parameter regimes for producing Frenkel defects on each of the strontium, ruthenium, and oxygen sublattices; for simplicity, we neglect any distinction between $E_{\text{disp, O}}$ for equatorial and apical oxygen sites.  Over these ranges of $E_{\text{disp}}$, the cross sections for Frenkel pair production at $E_{\text{kin}} = 2.5 \text{ MeV}$ can be read off from Fig.~\ref{fig:frenkel_defect_cs}(c-e) as $\sigma_{\text{FP, Sr}} =$ 339 to 159 barn, $\sigma_{\text{FP, Ru}} =$ 410 to 193 barn, and $\sigma_{\text{FP, O}} =$ 76 to 37 barn.  

Note that $1\text{ barn} = 10^{-24} \text{ cm}^{2}$; thus, to convert these cross sections to the number of displacements per atom~(dpa) for a specific dose $D$ of 2.5 MeV electron irradiation, we simply need to multiply by $D$ given in units of [number of electrons per unit area].  It follows that Frenkel defect densities ($n_{\text{FP}}$) are produced on the unique sublattices per unit dose $D$ at rates of

\begin{eqnarray*}
\frac{dn_{\text{FP, Sr}}}{dD} &=& \frac{(2.1 \text{ to } 1.0) \times 10^{-3} \text{ dpa}}{\text{C}/\text{cm}^2} \,\, (\times \, 2 \text{ Sr/unit cell}),\\
\frac{dn_{\text{FP, Ru}}}{dD} &=& \frac{(2.6 \text{ to }1.2) \times 10^{-3} \text{ dpa}}{\text{C}/\text{cm}^2} \,\, (\times \, 1 \text{ Ru/unit cell}),  \\
\frac{dn_{\text{FP, O}}}{dD} &=& \frac{(0.47 \text{ to }0.23) \times 10^{-3} \text{ dpa}}{\text{C}/\text{cm}^2} \,\, (\times \, 4 \text{ O/unit cell}).
\end{eqnarray*}

\noindent As mentioned in the main text, direct visualization of such dilute point-like defect concentrations, $n_{\text{FP}} \lesssim \mathcal{O}(0.1\%) =$ one part per thousand, is a formidable challenge, well beyond the current capabilities of state-of-the-art microstructural probes such as scanning transmission electron microscopy~\cite{goodge_disentangling_2022, sunko_controlled_2020}.  Nonetheless, by making a few more assumptions, it is possible to indirectly confirm the general consistency of these estimated defect densities by relating $n_{\text{FP}}$ to one of the observables measured in our current study\textemdash namely, the residual resistivity in the absence of superconductivity, $\rho_{0} \equiv \rho_{ns}(T \rightarrow 0\text{ K})$.  

In Fig.~\ref{fig:dose_dependencies_rho_Tc}(a) of the main text, we displayed the results of \textit{ex situ} measurements of the irradiation-induced increment in residual resistivity $\Delta \rho_{0}$ for four distinct FIB-sculpted single crystals of \SRO that accumulated variable doses $D$ of 2.5 MeV $e^{-}$ irradiation. As discussed above, some fraction of the defects created by irradiation are lost when a sample is first warmed to room temperature following irradiation. We model this loss as a simple multiplicative reduction of the dose received by each sample, calculating the reduction factor using the 22 K in-situ data [red points in Fig.~\ref{fig:frenkel_defect_cs}(a)] and the resistivity measured at 22 K after thermally cycling the sample to room temperature and back [blue point in Fig.~\ref{fig:frenkel_defect_cs}(a)]. In Fig.~\ref{fig:frenkel_defect_cs}(f), we re-plot the data points from Fig.~\ref{fig:dose_dependencies_rho_Tc}(a) of the main text versus the effective irradiation dose received, $D_{\text{eff}} = 0.62 \times D$, where $0.62$ is the calculated reduction factor. We used the reduction factor determined from sample C for all four samples because most of the other samples had incomplete \textit{in situ} resistivity datasets. Although this adjustment procedure implicitly appeals to Matthiessen's-Rule-type reasoning, which is known to not be strictly valid in perovskite-based ruthenates~\cite{klein_negative_2001, wang_separated_2021}, the errors associated with this approximation are small compared with other systematic uncertainties in these estimates of $n_{\text{FP}}$ and in how efficiently the equilibrium $n_{\text{FP}}$ for Sr, Ru, and O are transduced into corresponding increases in $\rho_{0}$. 

Previous studies of \SRO single crystals that were intentionally doped with chemical impurities on the Ru site, namely, Sr$_{2}$Ru$_{1-x}$(Ti, Ir)$_{x}$O$_{4}$, demonstrated that such defects act as nearly unitary-limit scatterers that increase the residual resistivity at a rate of $d\rho_{0}/dx \approx 425 $\uOc, where $x$ is the fractional dopant density per unit cell~\cite{kikugawa_effect_2002, kikugawa_effects_2003}.  Likewise, previous studies of \SRO single crystals that were intentionally doped with chemical impurities on the Sr site, namely, Sr$_{2-y}$La$_{y}$RuO$_{4}$, demonstrated that such defects act as nearly Born-limit scatterers that increase the residual resistivity at a much smaller rate of $d\rho_{0}/dy \approx 40 $\uOc, where $y$ is the fractional dopant density per unit cell~\cite{kikugawa_rigid-band_2004}.  Assuming that Sr and Ru vacancies scatter the charge-carrying quasiparticles in an identical fashion to chemically substituted impurities on the respective atomic sites, we can then multiply the above estimates of $dn_{\text{FP, Sr}}/dD$ ($dn_{\text{FP, Ru}}/dD$) by the empirical value of $d\rho_{0}/dy$ ($d\rho_{0}/dx$) to obtain:

\begin{eqnarray*}
\frac{d\rho_{0}}{dD} &=& \frac{(0.17 \text{ to } 0.079) \text{\uOc}}{\text{C}/\text{cm}^2} \,\, \text{(from Sr vacancies)}\\
\frac{d\rho_{0}}{dD} &=& \frac{(1.1 \text{ to } 0.51) \text{\uOc}}{\text{C}/\text{cm}^2} \,\, \text{(from Ru vacancies)}\\
\frac{d\rho_{0}}{dD} &=& \frac{(1.3 \text{ to } 0.59) \text{\uOc}}{\text{C}/\text{cm}^2} \,\, \text{(summed total)}
\end{eqnarray*}

These simulated rates of residual resistivity increase per unit of 2.5 MeV $e^{-}$ irradiation dose are drawn as dashed lines in Fig.~\ref{fig:frenkel_defect_cs}(f). The experimentally measured $\Delta\rho_{0}$ versus $D_{\text{eff}}$ data points for FIB-sculpted single crystals of \SRO agree reasonably well with these simulations, which suggests that the estimated cross sections for Frenkel defect production are realistic.  We note that in arriving at these simulated rates of $d\rho_{0}/dD$, we neglected any possible contributions of Ru and Sr interstitials to increases in $\rho_{0}$, and we ignored entirely all irradiation-induced defects on the O site.  Although these choices were made out of convenience---namely, because there are not any available data in the literature to indicate how much such defects might contribute to $\rho_{0}$ in \SROnospace---Fig.~\ref{fig:frenkel_defect_cs}(f) indicates that considering these additional types of irradiation-induced disorder is not strictly necessary to account for the experimentally observed rate $d\rho_{0}/dD$.  

It remains an open question to determine whether the apparent level of agreement between simulation and experiment in Fig.~\ref{fig:frenkel_defect_cs}(f) is somewhat artificial, in the sense that the true displacement energies for Sr and Ru are actually towards the higher end of our estimated range $(15\text{ eV})$ and that the contributions of interstitials and oxygen defects to $\Delta \rho_{0}$ are not negligible, or whether it is suggestive that the effects on $\Delta \rho_{0}$ of the explicitly neglected defects are already implicitly accounted for by the $D_{\text{eff}} \rightarrow D$ ``room-temperature annealing'' conversion of the experimental data.  Along these lines, we note that the extent to which Frenkel defects on the oxygen sites affect the resistivity and superconductivity in \SRO might be addressable in a more direct fashion in future irradiation studies that utilize lower incident electron kinetic energies, which would transfer sufficient energy to create defects on the O sublattices, but not to create defects on the Sr or Ru sites (cf.~the low $E_{\text{kin}} \lesssim 0.5 \text{ MeV}$ regions of Fig.~\ref{fig:frenkel_defect_cs}(c-e)); a similar approach has been employed previously for high-$T_{\text{c}}$ cuprates~\cite{tolpygo_effect_1996}.

\section{Magnetoresistance of thin films \label{sec:appendix_magnetores}}

\subsection{Superconducting upper critical field}

\begin{figure}
\includegraphics{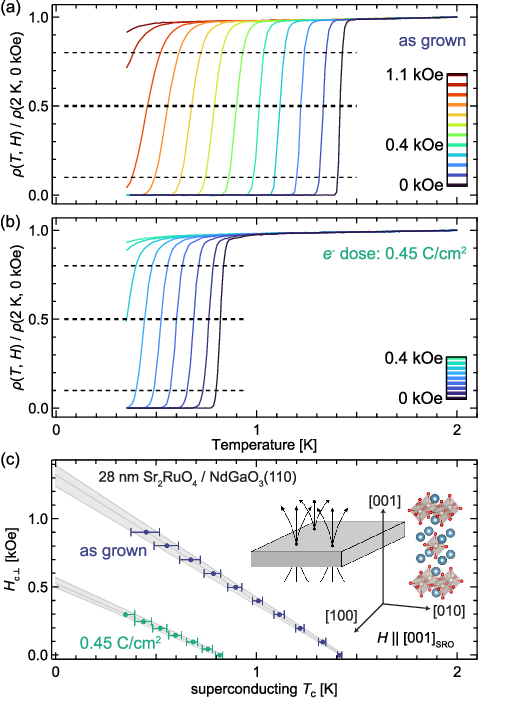}
\caption{\label{fig:hcperp_film_one_dose} Effect of electron irradiation on the superconducting upper critical fields of \SRO thin films. (a) Normalized resistivity versus temperature data for an as-grown \SRO thin-film resistivity bridge, acquired at external magnetic fields ranging from 0 to $1.1$~{kOe} in discrete steps of 0.1~{kOe}. (b) Same measurements as in panel (a), but after the sample had received a $0.45\text{ C}/\text{cm}^2$ dose of 2.5 MeV electron irradiation. Note that the external magnetic field steps in (b) are half of those displayed in (a), and that the maximum applied field is now merely $0.4\text{ kOe}$, which is sufficient to suppress $T_{\text{c}}$ to substantially below the base temperature of the helium-3 cryostat, $\approx 0.4\text{ K}$. Currents of $1 {\mu}\text{A}$, rather than the usual $10 {\mu}\text{A}$,  were used in the acquisition of the data shown in (a,b). (c) Upper critical magnetic fields for superconductivity ($H_{c\perp}$), before and after irradiation, extracted from the resistivity data in panels (a,b), along with linear fits (grey solid lines) used for estimating $H_{c\perp}(T\rightarrow0)$. The inset schematically depicts the measurement geometry and orientation of $H$ with respect to the \SRO crystal structure. Solid markers indicate the $T_{\text{c}}$s where $\rho(T, H)$ crosses the 50\% threshold of $\rho(2\text{ K},\, 0 \text{ kOe})$ [bold dashed line in (a,b)], and the horizontal error bars on these $T_{\text{c}}$s represent where $\rho(T, H)$ crosses the 10\% and 80\% thresholds of $\rho(2\text{ K},\, 0 \text{ kOe})$ [dashed lines in (a,b)].} 
\end{figure}

In Fig.~\ref{fig:hcperp_film_one_dose}, we show that a moderate $e^{-}$ dose ($D = 0.45 \text{ C}/\text{cm}^2$) of 2.5 MeV electron irradiation substantially reduces the upper critical magnetic fields required to suppress superconductivity in thin-film \SROnospace.  For these electrical transport measurements, the external magnetic field is directed along the crystallographic $c$-axis of \SROnospace, which is also the out-of-plane direction of epitaxial thin-film samples studied here [see inset to Fig.~\ref{fig:hcperp_film_one_dose}(c)]. Thus, the critical field probed is $H_{c\perp}$ in the terminology of Refs.~\cite{tinkham_effect_1963,harper_mixed_1968}, which is essentially equivalent to $H_{c2}$ (with $H\,||\,c$) for a bulk specimen of a type-II superconductor such as \SROnospace.  Hereafter we refer to $H_{c\perp}$ and $H_{c2||c}$ interchangeably.

The simultaneous decreases of both $T_{\text{c}}$ and $H_{c2||c}$ upon high-energy electron irradiation provides further evidence that the irradiation-induced defects are point-like scattering centers that uniformly suppress the characteristic intra-granular condensation energy scale of the superconductivity in \SROnospace. By contrast, if the electron irradiation created columnar and/or other types of extended defects, we would generally expect $H_{c2||c}$ to remain unchanged, or perhaps even to be enhanced, due to increased pinning of vortices at the extended defects, while properties that depend on the details of the inter-granular Josephson network of superconducting grains~\cite{konczykowski_electron_1991}, such as critical currents, would change. 

Without a clear understanding of the effective pairing interactions that cause superconductivity to emerge in \SROnospace, it is difficult to quantitatively interpret the observed magnitudes of $T_{\text{c}}$ and $H_{c2||c}$ reduction upon irradiation, since both of these quantities depend on some appropriately weighted averages of the superconducting order parameter over the quasiparticle states near the Fermi level~$E_{F}$.  Nevertheless, making the standard assumption that $H_{c2||c}$ is orbitally limited, we can convert the measured upper critical fields into phenomenological in-plane superconducting coherence lengths ($\xi_{ab}$) using the Ginzburg-Landau relation:
 
\begin{equation*}
\mu_{0}H_{c2||c} = \frac{\phi_{0}}{2\pi\xi_{ab}^2}  \;\;\; \Longrightarrow \;\;\; \xi_{ab} \, [\text{\AA}] = \sqrt{\frac{329106}{H_{c2||c} \, [\text{kOe}]}}
\end{equation*}

\noindent where $\phi_{0} = h/(2e)$ is the superconducting flux quantum.  

To account for the temperature dependence of the quantities in this expression, we follow previous studies of epitaxial thin films of \SROnospace, which have shown that the nearly linear scaling of $H_{c\perp}$ versus $T_{\text{c}}$\textemdash observed in Fig.~\ref{fig:hcperp_film_one_dose} for $T$ down to $0.4\text{ K}$, i.e., down to $T/T_{\text{c}} = 0.3$ for the as-grown sample\textemdash\space in fact persists over at least another decade in $T$, down to the lowest temperatures accessible in a dilution refrigerator, $\mathcal{O}[0.06\text{ K}]$~\cite{uchida_anomalous_2019}.  Accordingly, we fit all available $H_{c\perp}$ versus $T_{\text{c}}$ data points for the as-grown and irradiated 28-nm \SROnospace/\NGOnospace(110) samples to separate lines, and extrapolate the fit results to zero temperature, as shown by the gray lines in Fig.~\ref{fig:hcperp_film_one_dose}(c), to obtain $\xi_{ab}(T \rightarrow 0 \text{ K}) = 500 \pm 20\text{ \AA}$ for as-grown films, and $790 \pm 30\text{ \AA}$ after a dose of $0.45 \text{ C}/\text{cm}^2$.

The error bars quoted here on $\xi_{ab}$ refer to the spread of $H_{c\perp}(T \rightarrow 0 \text{ K})$ values obtained by employing $10\%$ or $80\%$ thresholds of $\rho(2\text{ K}, 0\text{ kOe})$ to define $T_{\text{c}}$, rather than a $50\%$ threshold, and propagating these systematic uncertainties through the linear fits and extrapolations of the measured $H_{c\perp}(T_{\text{c}})$ phase boundary [cf. gray-shaded regions in Fig.~\ref{fig:hcperp_film_one_dose}(c)]. While the sensitivity of $\rho(T, H)$ to percolating, but otherwise filamentary, superconducting pathways implies that electrical transport measurements inherently tend to overestimate upper critical fields compared with bulk-sensitive measurements of superconductivity~\cite{tinkham_effect_1963, hsu_superconducting_1992}, such considerations only affect the absolute values of $H_{c2}$ that are sensed by different probes; by contrast, the relative change observed here in \SRO upon electron irradiation is unambiguous: increasing the disorder scattering rate suppresses $H_{c2||c}$ (enhances $\xi_{ab}$).  This result is qualitatively consistent with previous studies of \SRO that deduced such a correlation by analyzing $H_{c2||c}$ data across a collection of different as-grown samples, in both bulk~\cite{mao_suppression_1999} and thin-film form~\cite{uchida_anomalous_2019}.   

\subsection{Effect of electron irradiation on Shubnikov-de Haas oscillations}

\begin{figure}
\includegraphics{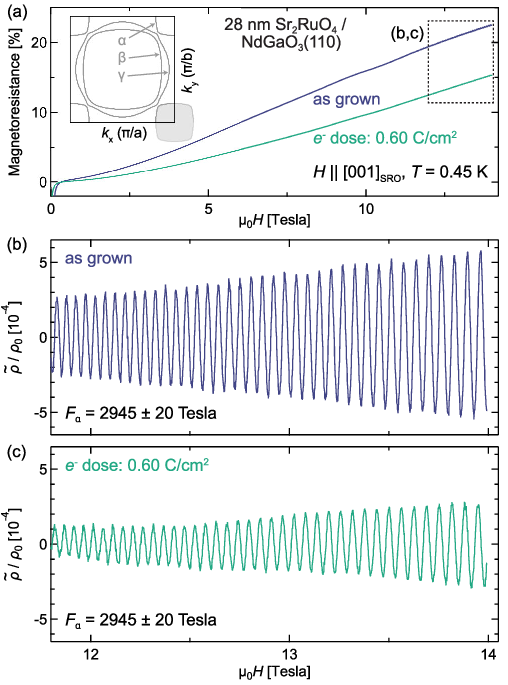}
\caption{\label{fig:SdH_film_one_dose} Effect of electron irradiation on the magnetoresistance of \SROnospace. (a) Normalized magnetoresistance (MR) plotted versus the externally applied magnetic field for an as-grown \SRO thin-film resistivity bridge~(blue), and the same sample after receiving a $0.60\text{ C}/\text{cm}^2$ dose of 2.5 MeV electron irradiation~(teal).  Here we define $\text{MR } [\%] \equiv 100 \times [\rho(H) - \rho(0)]/\rho(0)$, where $\rho(0)$ is the extrapolated resistivity as  $H \rightarrow 0$ in the absence of superconductivity. Currents of $100 {\mu}\text{A}$, rather than the usual $10 {\mu}\text{A}$,  were used in the acquisition of the data shown in this figure. (b,c) Zoomed-in views of the resistivity measured over the high-field range indicated by the dashed box in panel (a), after subtracting and normalizing to empirically determined quadratic polynomial backgrounds $\rho_{\text{bg}}(H)$ for both data sets: $\tilde{\rho}/\rho_{0} \equiv [\rho(H) - \rho_{\text{bg}}(H)]/\rho_{\text{bg}}(H)$. Superimposed on these slowly and smoothly varying backgrounds are rapidly oscillating Shubnikov-de Haas contributions to the magnetoresistance, that are periodic in inverse field and have dominant frequencies $F_{\alpha}$ characteristic of charge carriers completing phase-coherent cyclotron motion around the $\alpha$-sheet Fermi surface of \SRO [cf.~upper left inset to panel (a)].}
\end{figure}

To place the irradiation-induced modifications to $\rho_{0}$, $T_{\text{c}}$, and $H_{c2||c}$ in proper context, it is important to note that certain features of the normal-state near-$E_{F}$ electronic structure in \SROnospace, such as the sizes and shapes of the Fermi surfaces, are \textit{not} appreciably affected by the doses of high-energy electron irradiation administered in this work. To investigate the Fermiology in detail, in Fig.~\ref{fig:SdH_film_one_dose} we plot low-temperature magnetotransport data acquired for an as-grown 28-nm \SROnospace/\NGOnospace(110) sample, and for the same sample after accumulating a moderate $e^{-}$ dose ($D = 0.60 \text{ C}/\text{cm}^2$) of 2.5 MeV electron irradiation.  The external magnetic field is again directed out-of-plane, along the crystallographic $c$ axis of \SROnospace, and the temperature is held constant at a nominal value of $T = 0.45\text{ K}$.  We used rms excitation currents of $I = 100~\mu\text{A}$ for transport measurements performed during these field sweeps, to further enhance the signal-to-noise ratio.

The overall magnitude of the normalized magnetoresistance (MR) in the normal state (i.e., for $B = \mu_{0}H >> \mu_{0}H_{c\perp}$) is somewhat reduced upon increasing $D$ [Fig.~\ref{fig:SdH_film_one_dose}(a)].  This behavior is generically expected whenever more than one decay rate enters the charge-carrier momentum relaxation dynamics (see, e.g., Ref.~\cite{sunko_controlled_2020} and references contained therein), and indeed time-domain THz spectroscopy measurements on these \SRO epitaxial thin films have directly shown that the low-energy optical conductivity at low temperatures contains contributions from more than one Drude-like oscillator~\cite{wang_separated_2021}. Although detailed accounts of how electron irradiation couples to the multiple transport lifetimes that determine the dc MR (as well as the ac $\sigma(\omega)$ for small $\omega \neq 0$) are beyond the scope of the present work, there is another feature of the data that is simpler to interpret quantitatively: namely, rapidly oscillating Shubnikov-de Haas (SdH) contributions to the MR. 

To better isolate and visualize the SdH signal, we fit the raw MR data within the 12 to 14 Tesla field range [dashed box in Fig.~\ref{fig:SdH_film_one_dose}(a)] to the locally quadratic form $\rho_{{bg}}(H) = A_{0} + A_{1}H + A_{2}H^{2}$, and then compute the background-subtracted and background-normalized MR, $\tilde{\rho}/\rho_{0} \equiv [\rho(H) - \rho_{\text{bg}}(H)]/\rho_{\text{bg}}(H)$.  We plot the results thus obtained for $\tilde{\rho}/\rho_{0}$ for the as-grown and electron-irradiated \SRO samples in panels (b) and (c) of Fig.~\ref{fig:SdH_film_one_dose}, respectively. Both $\tilde{\rho}/\rho_{0}$ traces clearly exhibit SdH oscillations that are periodic in inverse field.  

The dominant oscillation frequency in the data collected at the temperatures and fields shown here is $F_{\alpha}$, which results from charge carriers completing phase-coherent cyclotron orbits around the $\alpha$-Fermi surface of \SROnospace.  Counting peaks and troughs in Fig.~\ref{fig:SdH_film_one_dose}(b,c), we measure $N = 37$ complete oscillation periods over the field range $\mu_{0}H = 11.880$ to $13.964 \text{ Tesla}$ for both the as-grown and the electron-irradiated samples, corresponding to oscillation frequencies of $F_{\alpha} = N/(1/11.880\text{ T} - 1/13.964\text{ T}) = 2945 \pm 20 \text{ Tesla}$ for both samples. In other words, moderate $e^{-}$ doses do not modify $F_{\alpha}$, within the experimental resolution of approximately 1\%.  We estimated the systematic error bars quoted here on $F_{\alpha}$ by our ability to detect $\pm \sfrac{1}{4}$ of an oscillation period of the $\alpha$-sheet-derived SdH signal over the (inverse) magnetic field range probed in Fig.~\ref{fig:SdH_film_one_dose}(b,c).  

In the inset to Fig.~\ref{fig:SdH_film_one_dose}(a), we schematically sketch the well-known $\{ \alpha, \beta, \gamma \}$ Fermi surfaces of \SROnospace, projected onto the two-dimensional first Brillouin zone, which is bounded by solid black lines. The area $A_{k}$ enclosed by each Fermi surface in $\mathbf{k}$-space follows directly from the corresponding primary quantum oscillation frequency $F_{i}$ $(i \in \{ \alpha, \beta, \gamma \})$, according to the Onsager relation $A_{k,i} = (2 \pi e/\hbar)F_{i}$.  Therefore, the invariance of $F_{\alpha}$ noted above implies that electron irradiation does not alter the $\mathbf{k}$-space area enclosed by the $\alpha$-sheet Fermi surface, i.e., the gray-shaded region drawn in the inset to Fig.~\ref{fig:SdH_film_one_dose}(a).  Invoking Luttinger's theorem, this observation also implies that moderate doses of electron irradiation do not alter the itinerant carrier density in \SROnospace. 

\section{Estimating clean-limit $T_{c0}$ for thin films \label{sec:appendix_est_Tcmax}}

\begin{figure}
\includegraphics{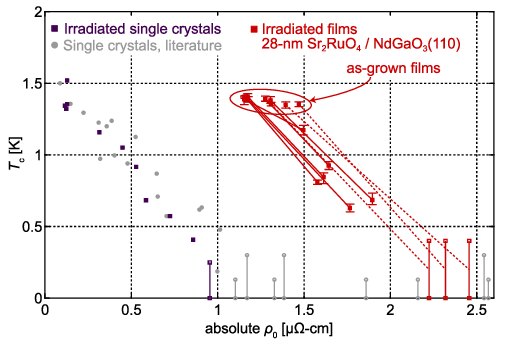}
\caption{\label{fig:Tc_vs_absRho0_films} $T_{\text{c}}$ versus absolute $\rho_{0}$ for \SRO single crystals and epitaxial thin films.  For \SRO single crystals, purple square markers from the electron-irradiation experiments detailed in this work and gray circular markers extracted from impurity-scattering studies in the literature~\cite{mackenzie_extremely_1998, kikugawa_effects_2003, kikugawa_rigid-band_2004} are reproduced from Fig.~\ref{fig:Tc_vs_disorder_summary} of the main text.  For \SRO thin films, the red square markers plotted here are obtained directly from  $\rho(T)$, without the shift in the horizontal axis used in Fig.~\ref{fig:Tc_vs_disorder_summary}.  Red lines connect pairs of $(\rho_{0}, T_{\text{c}})$ points measured before and after 2.5 MeV electron irradiation for the \emph{same} physical thin-film sample: solid lines for irradiated samples having measurable $T_{\text{c}}$, and dashed lines terminating at $T_{\text{c}} = 0.2\text{ K}$ for irradiated samples having $T_{\text{c}}$ below the base temperature of our cryostat, $0.4 \text{ K}$. The two-point slopes suggested by this pairwise grouping of the data are remarkably consistent, despite differing ``initial conditions'' (i.e., differing absolute $\rho_{0}$ values); moreover, the $\Delta T_{\text{c}}/\Delta \rho_{0}$ Cooper-pair-breaking effect induced by irradiation is at least $10$ times larger in magnitude than whatever subtle trend may be discernible in the $T_{\text{c}}(\rho_{0})$ behavior observed across distinct as-grown thin-film samples, which are all encircled by a red oval.}
\end{figure}

In the discussion surrounding Fig.~\ref{fig:Tc_vs_disorder_summary} of the main text, we noted that in \SRO thin films, $T_{\text{c}}$ responds sensitively to relative changes in the residual resistivity induced by electron irradiation, $\Delta \rho_{0, \text{ irr}}$, yet $T_{\text{c}}$ appears to depend much less regularly on $\rho_{0}$ when all residual resistivities for as-grown and irradiated thin-film samples are plotted in absolute units.  Figure~\ref{fig:Tc_vs_absRho0_films} illustrates both of these observations graphically.  All 18 $(\rho_{0}, T_{\text{c}})$ red square markers for \SRO thin films plotted in Fig.~\ref{fig:Tc_vs_absRho0_films} are extracted in absolute units from the corresponding $\rho(T)$ data traces plotted in Fig.~S4 of the Supplemental Material \cite{SupplementalMaterial}.  To serve as side-by-side references, the same $(\rho_{0}, T_{\text{c}})$ gray points and purple squares for \SRO bulk single crystals from Fig.~\ref{fig:Tc_vs_disorder_summary} of the main text are also included in Fig.~\ref{fig:Tc_vs_absRho0_films}.  

Viewed merely as a collection of independent points, the $T_{\text{c}}$ versus $\rho_{0}$ behavior of the thin-film data in Fig.~\ref{fig:Tc_vs_absRho0_films} appears rather scattered, and it is difficult to establish any clear correlation between these two variables.  By contrast, when one draws lines between each pair of $(\rho_{0}, T_{\text{c}})$ points that correspond to the same physical sample, measured before and after irradiation, visual inspection of the solid lines in Fig.~\ref{fig:Tc_vs_absRho0_films} reveals that electron irradiation has very similar, proportional effects on $\Delta T_{\text{c}}$ and $\Delta \rho_{0 \text{, irr}}$ across all distinct \SRO thin-film samples.  Furthermore, as emphasized in the main text, collecting these independent measurements of $\Delta T_{\text{c}}/\Delta \rho_{0}$ into a single number for the rate of $T_{\text{c}}$ suppression with increasing quasiparticle scattering rate results in a number that is quantitatively very similar between these electron-irradiated \SRO thin-film samples and single-crystal \SRO samples containing various types of defects (Table~\ref{table:dTc_drho0_fits}).

In Fig.~\ref{fig:Tc_vs_disorder_summary} of the main text, we proposed that the most natural interpretation and framing of the raw $T_{\text{c}}(\rho_{0})$ data for thin films can be summarized as follows: (i.)~the quantitative similarity of $T_{\text{c}}$ across all as-grown thin-film samples ($1.35$ to $1.41\text{ K}$) implies that their $\hbar/\tau_{\text{pb}}$ values are nearly identical; (ii.)~based on the detailed comparisons between \SRO thin films and single crystals presented in the main text, using electron irradiation is a valid way to produce pure (quasi-elastic and quasi-isotropic) Cooper-pair-breaking scattering, and thereby suppress $T_{\text{c}}$ at an initial \emph{rate} of $\approx1.2 \text{ K}/\mu\Omega\text{-cm}$; thus, (iii.)~the remaining outstanding question is to determine the relative rigid offset $\rho_{0, \text{ offset}}$ of these as-grown thin-film samples relative to the \emph{zero point} of an appropriately-defined $\hbar/\tau_{\text{pb}}$ Cooper-pair-breaking energy scale.  Equivalently, since our electron irradiation studies have now precisely established the magnitude of $T_{\text{c}}$ suppression for a given change in $\rho_{0}$ that also proportionally increments $\hbar/\tau_{\text{pb}}$, determining $\rho_{0, \text{ offset}}$ requires knowledge of the clean-limit $T_{c0} \equiv T_{\text{c}}(\hbar / \tau_{\text{pb}} \rightarrow 0)$ of \SROnospace/\NGOnospace(110).  

At present, the simplest-possible empirical estimates constrain $\rho_{0, \text{ offset}}$ to be within the following extreme limits: (i.)~$\rho_{0, \text{ offset}} = 0$, in which case \emph{none} of the experimentally-measured $\rho_{0}$ in as-grown thin-film samples actually contributes to $\hbar/\tau_{\text{pb}}$ (i.e., $T_{c0}$ of \SROnospace/\NGOnospace(110) is equal to the values measured in this work on as-grown samples, $\approx 1.4 \text{ K}$), and (ii.)~$\rho_{0, \text{ offset}} = 1.1$\uOc, which is equal to the experimentally-measured $\rho_{0}$ for as-grown $T_{\text{c}} \approx 1.4 \text{ K}$ thin-film samples exhibiting the lowest absolute $\rho_{0}$s in the present work \cite{SupplementalMaterial}. In this case, the hypothetical $T_{c0}$ of \SROnospace/\NGOnospace(110) would equal $1.4\text{ K} + (1.1 \, \mu\Omega\text{-cm} \times 1.2 \text{ K}/\mu\Omega\text{-cm}) \approx 2.7 \text{ K}$.  In Fig.~\ref{fig:Tc_vs_disorder_summary} of the main text, we arbitrarily chose a rigid shift of the top axis by $\rho_{0, \text{ offset}} = 0.5 $\uOc \space relative to the bottom axis, about halfway between these two extremes, corresponding to a \SROnospace/\NGOnospace(110) $T_{c0} \approx 2.0 \text{ K}$.  

Based on the method proposed here of interpreting the thin-film data, it would be possible to further narrow down the range of potential $T_{c0}$ values for \SROnospace/\NGOnospace(110) in the future either by synthesizing such \SRO thin-film samples with appreciably higher as-grown $T_{\text{c}} > 1.4\text{ K}$, or by maintaining ``high'' $T_{\text{c}} \approx 1.4\text{ K}$ in as-grown samples with appreciably lower as-grown $\rho_{0} < 1 \, \mu\Omega\text{-cm}$.  While some limited evidence for the former exists in the literature, e.g. Ref.~\cite{nair_demystifying_2018} reported a broad (about $0.4 \text{ K}$-wide) resistively-measured superconducting transition centered at a midpoint $T_{\text{c}} = 1.8 \text{ K}$ in a 55-nm-thick \SROnospace/\NGOnospace(110) sample, to the best of our knowledge, there are no examples in the literature of \SRO thin-film samples synthesized on any perovskite-based substrate that exhibit the latter property. The latter observation suggests that the levels of $\rho_{0}$ currently achieved are pushing up against a fundamental length scale for charge-carrier momentum-relaxing scattering from extended structural defects in heteroepitaxial thin film/substrate material platforms~\cite{fang_quantum_2021}.

\section{Comparison to Abrikosov-Gor'kov theory \label{sec:appendix_AG}}

\begin{figure}
\includegraphics{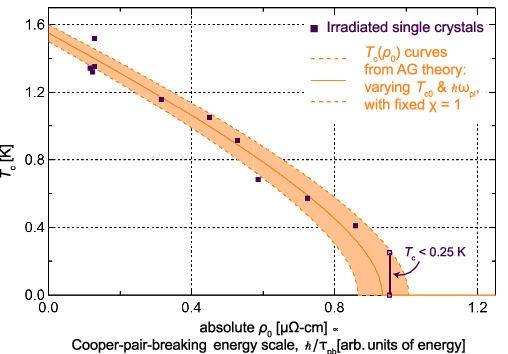}
\caption{\label{fig:Tc_vs_rho0_AGcomp_crystals} Abrikosov-Gor'kov pair-breaking theory simulations of $T_{\text{c}}(\rho_{0})$ for bulk \SROnospace.  Purple square markers are data collected in this work for as-grown and electron-irradiated FIB-sculpted \SRO single crystals, reproduced from Fig.~\ref{fig:Tc_vs_disorder_summary} of the main text.  The orange curves and shaded region are simulations obtained by solving the implicit AG-like equation for $T_{\text{c}}$ as a function of $\rho_{0}$, varying $\{T_{c0}, \text{ } \hbar\omega_{pl}\}$ while keeping $\chi = 1$ fixed; for quantitative values of the parameters employed here, see the text of Appendix~\ref{sec:appendix_AG}}
\end{figure}

\begin{figure}
\includegraphics{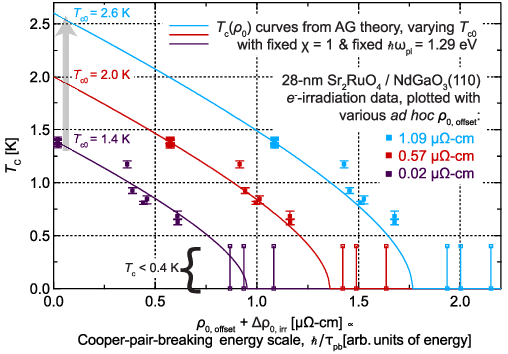}
\caption{\label{fig:Tc_vs_rho0_AGcomp_films} Abrikosov-Gor'kov pair-breaking theory simulations of $T_{\text{c}}(\rho_{0})$ for \SRO thin films. Square markers are data collected in this work for as-grown and electron-irradiated 28-nm \SROnospace/\NGOnospace(110) thin-film samples.  Differently-colored sets of markers are the same underlying $T_{\text{c}}$ versus $\Delta\rho_{0\text{, irr}}$ data set reproduced from Fig.~\ref{fig:Tc_vs_disorder_summary} of the main text, plotted with different \emph{ad hoc} rigid shifts of the horizontal axis, $\rho_{0\text{, offset}}$, as indicated in the inset legend.  Solid color-coded lines are simulations obtained by solving the implicit AG-like equation for $T_{\text{c}}$ as a function of the Cooper-pair-breaking energy scale $\hbar/\tau_{pb} \propto \rho_{0} \equiv \rho_{0\text{, offset}} + \Delta\rho_{0\text{, irr}}$, varying $T_{c0}$ while keeping both $\hbar\omega_{pl} = 1.29 \text{ eV}$ and $\chi = 1$ fixed.}
\end{figure} 

In Fig.~\ref{fig:Tc_vs_rho0_AGcomp_crystals}, we show that the results of our electron irradiation experiments on FIB-sculpted bulk single crystals of \SRO can be described well by theoretical $T_{\text{c}}(\rho_{0})$ curves calculated according to the standard Abrikosov-Gor'kov (AG) Cooper-pair-breaking theory of $T_{\text{c}}$ suppression~\cite{abrikosov_contribution_1961}, applied to the case of non-magnetic point-defect scattering in an effectively single-band unconventional superconductor~\cite{hohenberg_anisotropic_1964, abrikosov_influence_1993, radtke_predictions_1993, borkowski_distinguishing_1994, openov_critical_1998}.  

In such theories, $T_{\text{c}}$ is found by solving an implicit equation of the general form (see, e.g., Eqn.~16 of Ref.~\cite{abrikosov_influence_1993} or Eqn.~35 of Ref.~\cite{openov_critical_1998}):

\begin{equation}
\ln \left( \frac{T_{c0}}{T_{\text{c}}} \right) = \chi \left[\Psi\left( \frac{1}{2} + \frac{\hbar/\tau}{4 \pi k_{B} T_{\text{c}}} \right) - \Psi\left( \frac{1}{2} \right) \right],  \label{eq:GL}
\end{equation}

\noindent where $\Psi$ is the digamma function, $\hbar$ is Planck's constant, $1/\tau$ is the quasiparticle scattering rate, $k_{B}$ is Boltzmann's constant, $T_{c0}$ is the so-called ``clean-limit'' superconducting transition temperature in the disorder-free limit $\hbar/\tau \rightarrow 0$, and  

\begin{equation}
\chi = 1 - \frac{\left< \Delta(\mathbf{k}) \right>^{2}_{\text{FS}}}{\left< \Delta^{2}(\mathbf{k}) \right>_{\text{FS}}}
\end{equation}

\noindent is a constant that describes the anisotropy of the superconducting order parameter (SCOP), $\left< \text{averaged} \right>$ over the Fermi surface (FS).  For specific limiting examples of the latter, a purely sign-changing (e.g., ``$d$-wave'') SCOP that averages to zero over the Brillouin zone $(\left< \Delta(\mathbf{k}) \right>_{\text{FS}} = 0)$ has $\chi = 1$, whereas a purely uniform/isotropic ($\Delta(\mathbf{k}) = \text{constant}$, ``$s$-wave'') SCOP has $\chi = 0$.  

Although more complicated relationships are often observed in real materials, in theory the single-particle scattering rate that parameterizes the Cooper-pair-breaking energy scale in Eq.~\ref{eq:GL} is typically assumed to be directly proportional to the experimentally measured planar residual resistivity $\rho_{0}$, $1/\tau = \rho_{0}(\omega_{pl}^{2}/4 \pi)$, where $\hbar\omega_{pl}$ is a suitably averaged characteristic plasma frequency describing in-plane motion of the charge carriers~\cite{radtke_predictions_1993, openov_critical_1998}.

The quantitative parameters that describe the range of theoretical $T_{\text{c}}(\rho_{0})$ curves for bulk \SRO drawn in Fig.~\ref{fig:Tc_vs_rho0_AGcomp_crystals} are $\chi = 1$, $T_{c0} = 1.50$ K to $1.60$ K, and $\hbar\omega_{pl} = 1.40$~eV to $1.34$ eV.  Taken together, these parameter choices determine critical scattering rates $\rho_{0}^{\text{crit.}} = 0.86  \text{ } \mu\Omega\text{-cm}$ to $1.00 \text{ } \mu\Omega\text{-cm}$. For $\rho_0 > \rho_{0}^{\text{crit.}}$, only the trivial solution $T_{\text{c}} = 0$ exists to the above AG-like equation with $\chi = 1$.  

We consider the agreement observed here between theory and experiment to be essentially phenomenological in nature; there are too many outstanding unknowns about the superconductivity of \SRO for this type of effective description to be interpreted too literally.  In the regime of \textquoteleft small\textquoteright \space $\rho_0 << \rho_{0}^{\text{crit.}}$ where many of the experimental $(\rho_{0}, T_{\text{c}})$ data points we have collected exist, there are redundant dependencies of the initial rate of $T_{\text{c}}$ suppression~\cite{hohenberg_anisotropic_1964, abrikosov_influence_1993, borkowski_distinguishing_1994, openov_critical_1998}, both in how increases in $\rho_0$ transduce into increases in $\hbar/\tau$ (i.e., on the effective $\omega_{pl}$)~\cite{radtke_predictions_1993, openov_critical_1998}, and in how the fully $\mathbf{k}$-resolved SCOP of a multi-orbital system such as \SRO reduces to an effectively single-band SCOP anisotropy $\chi$~\cite{allen_new_1978, golubov_effect_1997}. Nevertheless, while the solutions presented here are by no means uniquely determined, the overall phenomenological $T_{\text{c}}(\rho_{0})$ curves shown in Fig.~\ref{fig:Tc_vs_rho0_AGcomp_crystals} may still prove useful as initial guesses for modeling strain-induced changes to $T_{c0}$ across different thin-film variants of \SROnospace, as discussed in more detail below.   

In principle, the systematic deviations from purely linear $T_{\text{c}}(\rho_{0})$ behavior expected in AG pairbreaking theory should allow one to infer the proximity of a given set of $(\rho_{0}, T_{\text{c}})$ data points to the critical $\rho_{0}^{\text{crit.}}$ at which superconductivity is suppressed completely.  This method of discerning $\rho_{0}^{\text{crit.}}$ in turn allows for an absolute determination of $T_{c0}$, because increases in $T_{c0}$ cause linear increases in $\rho_{0}^{\text{crit.}}$ in simple Cooper-pair-breaking theories that are effectively single-band, with one dominant superconducting pairing scale, $|\Delta_{\text{max.}}|$.

To apply this type of reasoning to our \SRO thin-film electron-irradiation data, we started with a phenomenological AG-like $T_{\text{c}}(\rho_{0})$ curve similar to those that describe well the $(\rho_{0}, T_{\text{c}})$ data points we collected for electron-irradiated \SRO FIB-sculpted single crystals in this work,  Fig.~\ref{fig:Tc_vs_rho0_AGcomp_crystals}.  To optimize the general agreement of these calculated $T_{\text{c}}(\rho_{0})$ curves with the experimental thin-film $(\rho_{0}, T_{\text{c}})$ data points, we adjusted the effective plasma frequency $\hbar\omega_{pl} = 1.37 \text{ eV}$ used for the bulk \SRO AG simulations in Fig.~\ref{fig:Tc_vs_rho0_AGcomp_crystals} to ${\hbar}{\omega}_{pl} = 1.29 \text{ eV}$ for the thin-film AG simulations in Fig.~\ref{fig:Tc_vs_rho0_AGcomp_films}.  The effective SCOP anisotropy $\chi = 1$ was held at the same fixed value as in the case of bulk samples, although as we stressed earlier, within AG-like pair-breaking theories there are redundant dependencies of the calculated $T_{\text{c}}(\rho_{0})$ behavior on $\{\chi,\text{ }\omega_{pl}\}$, so other solutions would be possible that would fit the existing data equally well.  Finally, we assumed that any epitaxial-strain-induced increase (or decrease) in $T_{c0}$ for \SROnospace/\NGOnospace(110) relative to the $T_{c0}$ of bulk \SRO results from a $\mathbf{k}$-independent multiplicative enhancement (or suppression) of $|\Delta(\textbf{k})|$, along with negligible strain-dependent changes in the Fermi surface (cf.~Fig.~\ref{fig:SdH_film_one_dose}). Under this assumption, we uniformly scaled up (or down) the phenomenological AG-like $T_{\text{c}}(\rho_{0})$ curve based on various hypothetical $T_{c0}$ values, keeping $\hbar\omega_{pl} = 1.29 \text{ eV}$ constant, to produce simulated $T_{\text{c}}(\rho_{0})$ curves against which to compare the \SRO thin-film $(\rho_{0}, T_{\text{c}})$ data points.  

Figure~\ref{fig:Tc_vs_rho0_AGcomp_films} illustrates the results of this type of analysis. For hypothetical $T_{c0}$ curves towards the high end of our allowed range of $T_{c0}$, e.g., for the blue simulated curve with $T_{c0} = 2.6\text{ K}$ ($\rho_{0}^{\text{crit.}} = 1.77 \text{ }\mu\Omega\text{-cm}$), the two irradiated thin-film samples with $T_{\text{c}} = 0.63\text{ K}$ and $0.68\text{ K}$ (which both received a 2.5 MeV $e^{-}$ dose of $0.6\text{ C}/\text{cm}^2$) would be small enough fractions of $T_{c0}$ (i.e., $T_{\text{c}}/T_{c0} < 0.3$) that they would be well within the range of a given AG-like pair-breaking-theory curve where non-linearities in the $T_{\text{c}}(\rho_{0})$ behavior would become obvious.  The absence of any such non-linear features in the experimental data suggests that $T_{c0}$ for \SROnospace/\NGOnospace(110) is likely less than $2.2\text{ K}$, provided that our assumption of $\chi \approx 1$ is, in fact, realized. 

On the other hand, the red simulated $T_{\text{c}}(\rho_{0})$ curve having $T_{c0} = 2.0\text{ K}$ (which combined with $\{\chi = 1,\text{ }\hbar\omega_{pl} = 1.29 \text{ eV}\}$ results in $\rho_{0}^{\text{crit.}} = 1.36 \text{ }\mu\Omega\text{-cm}$) matches the red experimental data squares (plotted with an \emph{ad hoc} $\rho_{0, \text{ offset}} = 0.57$\uOc) quite well.  The purple simulated $T_{\text{c}}(\rho_{0})$ curve having $T_{c0} = 1.4\text{ K}$ ($\rho_{0}^{\text{crit.}} = 0.95 \text{ }\mu\Omega\text{-cm}$) matches the purple experimental data squares (plotted with a much reduced \emph{ad hoc} $\rho_{0, \text{ offset}} = 0.02$\uOc) almost equally well. Down to the low end of the range of $T_{c0}$ that we believe are plausible ($T_{c0} \approx 1.4\text{ K}$), the three most heavily irradiated thin-film samples, with $T_{\text{c}} < 0.4\text{ K}$, never quite provide strong enough upper bounds on $T_{\text{c}}$ to exclude any hypothetical $T_{c0}$ curves. Apparently all such reductions in $T_{c0}$ can be compensated for, at least in principle, by postulating smaller values of $\rho_{0, \text{ offset}}$.  
Overall, a more precise and conclusive realization of the basic ideas presented in this section would require collecting more experimental data points in the region near $\rho_{0}^{\text{crit.}}$ where $T_{\text{c}}$ is supposed to rapidly evolve with changes in $\rho_{0}$.  

\end{appendix}

% Bibliography ----------
%apsrev4-2.bst 2019-01-14 (MD) hand-edited version of apsrev4-1.bst
%Control: key (0)
%Control: author (8) initials jnrlst
%Control: editor formatted (1) identically to author
%Control: production of article title (0) allowed
%Control: page (0) single
%Control: year (1) truncated
%Control: production of eprint (0) enabled
%

% Supplemental materials
\onecolumngrid
\clearpage
\begin{center}
\textbf{\large Supplemental Material for \\``Controllable suppression of the unconventional superconductivity in \SRO via high-energy electron irradiation''}\\[2.5ex]

Jacob P. Ruf,\textsuperscript{1,2} Hilary M.\,L. Noad,\textsuperscript{1} Romain Grasset,\textsuperscript{3} Ludi Miao,\textsuperscript{2} Elina Zhakina,\textsuperscript{1} Philippa \\H. McGuinness,\textsuperscript{1} Hari P. Nair,\textsuperscript{4} Nathaniel J. Schreiber,\textsuperscript{4} Naoki Kikugawa,\textsuperscript{5} Dmitry Sokolov,\textsuperscript{1} \\Marcin Konczykowski,\textsuperscript{3} Darrell G. Schlom,\textsuperscript{4,6,7} Kyle M. Shen,\textsuperscript{2,6} and Andrew P. Mackenzie,\textsuperscript{1,8}\\[1ex]

{\small
\textsuperscript{1} \textit{Max Planck Institute for Chemical Physics of Solids, 01187 Dresden, Germany}\\
\textsuperscript{2} \textit{Department of Physics, Laboratory of Atomic and Solid State Physics, Cornell University, Ithaca, NY 14853, USA}\\
\textsuperscript{3} \textit{Laboratoire des Solides Irradi\'{e}s, CEA/DRF/IRAMIS, \'{E}cole Polytechnique,\\ CNRS, Institut Polytechnique de Paris, 91128 Palaiseau, France}\\
\textsuperscript{4} \textit{Department of Materials Science and Engineering, Cornell University, Ithaca, NY 14853, USA}\\
\textsuperscript{5} \textit{National Institute for Materials Science, Tsukuba 305-0003, Japan}\\
\textsuperscript{6} \textit{Kavli Institute at Cornell for Nanoscale Science, Ithaca, NY 14853, USA}\\
\textsuperscript{7} \textit{Leibniz-Institut f\"{u}r Kristallzüchtung, 12489 Berlin, Germany}\\
\textsuperscript{8} \textit{SUPA, School of Physics and Astronomy, University of St Andrews, St Andrews KY16 9SS, United Kingdom}\\
}

\end{center}

\setcounter{equation}{0}
\setcounter{figure}{0}
\setcounter{table}{0}
\setcounter{section}{0}
\setcounter{page}{1}
\makeatletter
\renewcommand{\theequation}{S\arabic{equation}}
\renewcommand{\thefigure}{S\arabic{figure}}
\renewcommand{\thetable}{S\arabic{table}}
\makeatother

\section{FIB-sculpted single crystals \label{SIsec:singlecrystals}}

\begin{figure*}[h]
\includegraphics{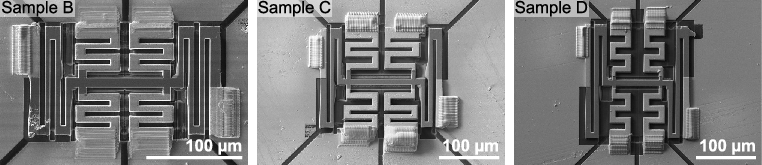}
\caption{\label{SIfig:other_three_microstructures} Scanning electron microscope images of three of the four microstructures of bulk, single-crystal \SRO that were studied in this work. Sample A is shown in Fig.~\ref{fig:rt_films_crystals_one_dose}(a) of the main text.}. 
\end{figure*}

\begin{table*}[h]
\renewcommand{\thetable}{S\arabic{table}}

\caption{\label{table:microstructure_dims} Dimensions of the FIB-sculpted single-crystal \SRO Hall bars, determined from scanning electron microscope images, and irradiation doses.}
\begin{ruledtabular}
\begin{tabular}{cccccc}
Sample & Length ($\mu$m) & Width ($\mu$m) & Thickness ($\mu$m) & $1^{\text{st}}$ dose (C/cm$^{2})$ & $2^{\text{nd}}$ dose (C/cm$^{2})$\\
\colrule
A & 109.8 & 9.7 & 6.5 & 0.45 & 0.60\\ % irr04
B & 80.0 & 9.7 & 5.5 & 1.177 & --\\ % irr03
C & 101.8 & 8.7 & 6.5 & 0.60 & 0.30\\ % irr05
D & 71.2 & 8.5 & 4.1 & 0.30 & 0.45\\ % irr06
\end{tabular}
\end{ruledtabular}
\end{table*}

\section{Growth and structural characterization of epitaxial thin films \label{SIsec:thinfilms_structure}}

The \SROnospace/\NGOnospace(110) film was grown by codepositing an elemental strontium flux of $2.2 \times 10^{13} \text{ atoms }\text{cm}^{-2} \text{ s}^{-1}$ and an elemental ruthenium flux of $1.6 \times 10^{13} \text{ atoms } \text{cm}^{-2} \text{ s}^{-1}$ in a background partial pressure of $3\times10^{-6}\text{ Torr}$ of distilled ozone ($\approx 80\%\text{ O}_{3} + 20\%\text{ O}_{2}$).  

\begin{figure}
\includegraphics{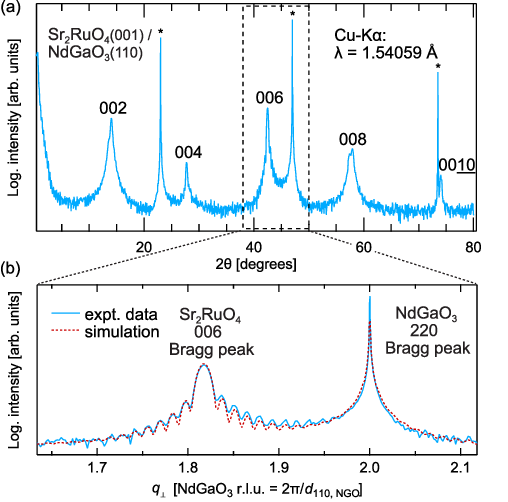}
\caption{\label{SIfig:xrd_film_thickness_determination} \SRO film thickness determination. (a) XRD data along the specular crystal truncation rod of the \SROnospace(001) thin film/\NGOnospace(110) substrate, acquired with Cu-K$\alpha$ radiation at room temperature.  Substrate Bragg peaks are marked by asterisks, and the family of $00L$ peaks ($L$ even) arising from scattering off the \SRO film are labeled accordingly.  (b) Zoomed-in view of the experimental data (blue) near the \SROnospace~$006$ and \NGOnospace~$220$ Bragg peaks, with the best-fit simulation overlaid (dashed red).}
\end{figure} 

To determine the thickness of the \SROnospace/\NGOnospace(110) film, we used a lab-based x-ray diffractometer (Rigaku) and Cu-K$\alpha$ radiation to measure x-ray diffraction (XRD) data at room temperature along the specular crystal truncation rod of the \SRO thin film/\NGO substrate, as shown in Fig.~\ref{SIfig:xrd_film_thickness_determination}(a).  Coherent scattering from film crystallites having finite sizes along the out-of-plane direction produces characteristic interference fringes\textemdash i.e., secondary maxima and minima\textemdash in the XRD intensity surrounding each primary film Bragg peak.  The spacing between these fringes scales inversely with the total number of layers in the crystallites, $N$; meanwhile, the average interplanar spacing between the layers in each crystallite, $c/2$, determines where each primary film Bragg peak is centered along the $q_{\perp}$ axis (here and elsewhere, we use $c$ to denote the out-of-plane lattice constant of the conventional unit cell of \SROnospace).  Together these two quantities determine the total film thickness according to $t = Nc/2$, which is later needed to convert measured film resistances to resistivities.

We utilized the GenX software package~\cite{bjorck_genx_2007} to directly fit the experimental XRD data over a region of momentum-transfer space surrounding the \SROnospace~$006$ and \NGOnospace~$220$ Bragg peaks, ($q_{||}\approx 0$, $q_{\perp} = 1.63 - 2.12 \text{ \NGO r.l.u.} = 2.66 - 3.45 \text{ 1/\AA}$).  From these simulations, displayed as a dashed red line in Fig.~\ref{SIfig:xrd_film_thickness_determination}(b), we determined that the \SRO film crystallites consisted of $22 \pm 1$ conventional unit cells\textemdash i.e., $N = 44 \, \pm \, 2$ RuO$_{2}$ planes/layers\textemdash stacked along the out-of-plane direction, having an average $c$-axis lattice constant of $12.76~\text{\AA}$, and thus a total film thickness of $t = (22 \pm 1) \times 12.76~\text{\AA} = 281 \pm 13 ~\text{\AA}$.  The confidence intervals listed here for $t$ refer to the estimated scale of film roughnesses, which were incorporated as a parameter in the XRD simulations; thickness variations are likely the largest systematic uncertainty in determining the electrical resistivity of the film in absolute units.  

We note that $c = 12.76~\text{\AA}$ for a thin film of \SRO corresponds to expansion of the room-temperature bulk single-crystal value ($c = 12.746~\text{\AA}$~\cite{walz_refinement_1993}) by $\epsilon_{zz} = +0.11\%$. Using the relevant Poisson ratio for \SRO at 300 K, $\nu_{zx} \equiv -\epsilon_{zz}/\epsilon_{xx} = 0.207$~\cite{paglione_elastic_2002}, this measured $c$-axis expansion is consistent with an elastically strained thin film subject to an average in-plane compressive strain of $(\epsilon_{xx} + \epsilon_{yy})/2 = -0.27\%$, which agrees well with the nominal in-plane lattice mismatch of \SRO and \NGOnospace(110) at room temperature, $-0.28\%$~\cite{walz_refinement_1993, schmidbauer_high-precision_2012}, as we discuss in more detail below.

The (110) surface of the \NGO crystal structure (described using the non-standard $Pbnm$ setting of space group $\#62$) is spanned by the mutually orthogonal $[001]$ and $[1\overline{1}0]$ lattice translation vectors, which impart modest amounts of biaxial ($A_{1g} \equiv (\epsilon_{xx} + \epsilon_{yy})/2$) in-plane compressive strain and uniaxial ($B_{1g} \equiv |\epsilon_{xx} - \epsilon_{yy}|$) in-plane strain on the \SRO thin film. Specifically, at $295 \text{ K}$, we expect the two in-plane Ru-O-Ru bonding directions along $x\, ||$~\SROnospace[100] and $y \, ||$~\SROnospace[010] to be compressed by $-0.39\%\text{ }||\text{ }$\NGOnospace$[001]$ and $-0.16\%\text{ }||\text{ }$\NGOnospace$[1\overline{1}0]$ relative to unstrained bulk single crystals of \SRO $(a = 3.8694~\text{\AA})$~\cite{walz_refinement_1993, schmidbauer_high-precision_2012}. These deformations correspond to in-plane $A_{1g}$ and $B_{1g}$ strains of $-0.28\%$ and $0.23\%$.  At $ 4 \text{ K}$, the in-plane $A_{1g}$ and $B_{1g}$ components of substrate-imposed misfit strains on \SRO become $-0.22\%$ and $0.30\%$, respectively, accounting for differences in the thermal expansion of \NGO and \SROnospace~\cite{senyshyn_anomalous_2009,vogt_low-temperature_1995}. 

\section{Electrical characterization of epitaxial thin films \label{SIsec:thinfilms_electrical}}

\subsection{As-grown films \label{SIssec:thinfilms_electrical_asgrown}}

\begin{figure*}
\includegraphics{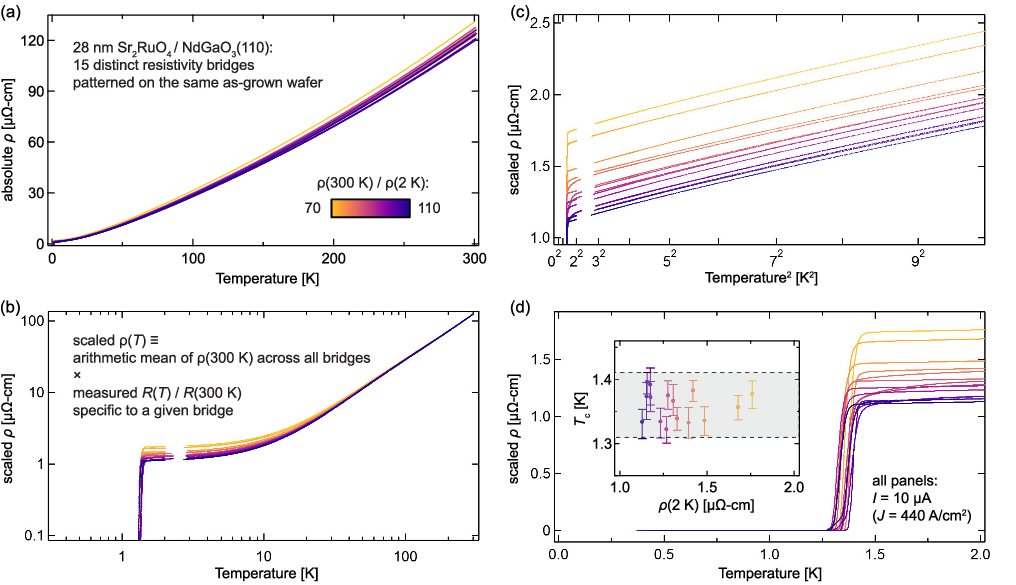}
\caption{\label{SIfig:as_grown_films_properties} Electrical transport properties across 15 distinct as-grown \SRO thin-film resistivity bridges. Data for a given sample are plotted using the same color in all panels; this color visually encodes the $RRR \equiv \rho(300\text{ K})/\rho(2\text{ K})$ measured for that sample, as indicated by the inset legend in (a). (a) Absolute resistivities versus temperature. (b) Log-log plot of the scaled resistivities versus temperature. Data are omitted for temperatures between about $2\text{ K} - 3 \text{ K}$ where helium-3 circulation starts/stops in the cryostat used for these measurements; the according rapid changes in temperature prohibit proper thermalization of the samples in this range of $T$. (c) Scaled resistivities versus temperature squared. All traces deviate in a subtle, but apparently systematic, fashion at low temperatures from the oft-prescribed $\rho_{0} + AT^2$ power-series expansion, which precludes extrapolating the resistivity below $T_{\text{c}}$ based on fits to such a power law above $T_{\text{c}}$. (d) Scaled resistivities versus temperature, zoomed in for $T \leq 2\text{ K}$ to visualize the superconducting transitions across distinct samples. Inset displays the resistively-measured $T_{\text{c}}$s [criterion: where $\rho(T)$ crosses 50\% of $\rho(2\text{ K})$], plotted against $\rho(2\text{ K})$, which is employed as an approximation of the residual resistivity. Error bars on $T_{\text{c}}$ are the temperatures where $\rho(T)$ crosses the 10\% and 80\% thresholds of $\rho(2\text{ K})$. Neither $T_{\text{c}}$ nor the superconducting transition widths exhibit any discernible correlation with $\rho(2\text{ K})$.}
\end{figure*}

In Fig.~\ref{SIfig:as_grown_films_properties}, we summarize the electrical resistivity versus temperature behavior of 15 distinct resistivity bridges patterned on the same \SROnospace/\NGOnospace(110) mother wafer, measured between $0.4 \text{ K}$ and $300 \text{ K}$. Hereafter we denote the resistivity by $\rho$, which is understood to be shorthand for the $\rho_{yy}$ component of a resistivity tensor that is defined with respect to a crystallographic coordinate system having $x \, ||$~\SROnospace[100] $\, || \,$ \NGOnospace[001], $y \, ||$~\SROnospace[010] $\, || \,$ \NGOnospace$[1\overline{1}0]$, and $z \, ||$~\SROnospace[001] $\, || \, $ \NGOnospace(110).  We acquired all data displayed in Fig.~\ref{SIfig:as_grown_films_properties} with the \SRO thin-film bridges in their as-grown states\textemdash i.e., before subjecting the samples to any doses of high-energy electron irradiation.  The graded color scale of the data traces visually represents and corresponds to the ordering of the samples when they are sorted by their residual resistivity ratios (RRRs), defined here as $RRR \equiv \rho(300\text{ K})/\rho(2\text{ K})$, which range from about 70 (yellow-orange) to 110 (dark purple).

Following previous work~\cite{mackenzie_extremely_1998}, we assume that the variation in absolute $\rho(300 \text{ K})$ observed across distinct as-grown resistivity bridges in Fig.~\ref{SIfig:as_grown_films_properties}(a), with values ranging from 120 to 131 \uOc, is primarily caused by systematic uncertainties in the sample preparation and in the measurements, rather than by any intrinsic differences in the scattering mechanisms relevant to \SRO that relax the charge-carrier momentum in this highly incoherent regime of electrical transport.  Likely sources of such errors include variations in the effective cross-sectional thickness that the electrical current is constricted to flow through, as well as small misalignments and discrepancies in the lithographically patterned dimensions of the electrical contacts to different bridges. More microscopically, such variations in the effective cross-sectional thicknesses of the resistance bridges may result from total film thickness variations, out-of-phase boundaries in the film that preferentially nucleate in the vicinity of substrate step edges~\cite{fang_quantum_2021, goodge_disentangling_2022, zurbuchen_suppression_2001, wu_electronic_2020}, and/or locally higher-$N$ Ruddlesden-Popper inclusions/stacking faults that form to accommodate non-stoichiometry in Ru-rich (nominally \SROnospace) epitaxial thin films~\cite{uchida_molecular_2017, nair_demystifying_2018, kim_superconducting_2021, goodge_disentangling_2022}.

Therefore, for all other curves plotted in Fig.~\ref{SIfig:as_grown_films_properties}(b-d), and throughout the main text and the remainder of the Supplemental Material, we plot so-called \textquoteleft scaled $\rho(T)$\textquoteright \space curves: we first normalize the measured resistance versus temperature $R(T)$ data for a given bridge by its resistance measured at $300\text{ K}$, and then we uniformly multiply this normalized data by the arithmetic mean of $\rho(300 \text{ K})$ measured across all as-grown bridges, $\overline{\rho(300\text{ K})} = 124$\uOc.  We note that the value of $\overline{\rho(300\text{ K})} = 124$\uOc~(standard deviation about the mean $= 3$\uOc) obtained for the \SRO thin-film resistivity bridges studied in this work agrees well with the established literature value appropriate to bulk \SRO single crystals, $\overline{\rho(300\text{ K})} = 121$\uOc~(standard deviation about the mean $= 17$\uOc)~\cite{mackenzie_extremely_1998}.  This quantitative level of agreement indicates that these \SRO films do not contain significant volume fractions of higher-$N$ Ruddlesden-Popper intergrowths (SrO)(SrRuO$_{3}$)$_{\text{N}}$, which for all $N\neq 1$ compounds have characteristic values of $\rho(300\text{ K}) \approx 200$\uOc~\cite{wang_separated_2021, capogna_sensitivity_2002}.  Additionally, we note that the above procedure of computing appropriately scaled $\rho(T)$ values is important only if one wishes to interpret $\rho$ in absolute units\textemdash for example, as a proxy for computing the transport lifetime (or transport scattering length) of charged quasiparticle excitations in \SROnospace.  By contrast, many of the results contained in this manuscript for irradiation-induced changes in quantities like $T_{\text{c}}$ are extracted from relative changes in the measured resistance as a function of temperature, which are independent of any remaining systematic errors in the deduced temperature-independent scale factors that enter into conversions of $R$ to $\rho$.

From the data in Fig.~\ref{SIfig:as_grown_films_properties}, we determined (arithmetic mean $\pm$ one standard deviation) values of $\overline{\rho(300\text{ K})} = 124\,\pm\,3$\uOc, $\overline{\rho(2\text{ K})} = 1.33\,\pm\,0.17$\uOc, and $\overline{T_{\text{c}}} = 1.359\,\pm\,0.024\text{ K}$.  Here $T_{\text{c}}$ is reported as the temperature at which the resistance of each \SRO thin-film bridge measured at a current of $I = 10 \text{ }\mu\text{A}$\textemdash or equivalently, at a nominal current density of $|\mathbf{J}| = 440 \text{ A}/\text{cm}^2$\textemdash\space crosses $50\%$ of its value measured at $2\text{ K}$ Sec.~\ref{SIsec:determine_rho0_Tc} of the Supplemental Material contains further discussion of the definitions and analysis methods used to extract the values of $T_{\text{c}}$ shown in this work.  The uniformity of the resistively measured $T_{\text{c}}$s across the wafer $(0.024/1.359 = 1.8\%)$ justifies our use of different bridges as nearly-identical \textquoteleft initial conditions\textquoteright \space for gauging the effects of different electron irradiation doses on $T_{\text{c}}$.  On the other hand, the comparatively larger spread in $\rho(2\text{ K})$ across the wafer $(0.17/1.33 = 13\%)$, the lack of any discernible correlation between $T_{\text{c}}$ and $\rho(2\text{ K})$ for this set of as-grown samples [see inset to Fig.~\ref{SIfig:as_grown_films_properties}(d)], and the large absolute values of $\rho(2\text{ K})$ at which superconductivity remains robustly observed (``large'' relative to the phenomenological dependence of $T_{\text{c}}$ versus $\rho_{0}$ established for bulk \SRO single crystals, such as in Fig.~3 of the main text), all suggest that scattering from most of the native defects existing in these \SRO epitaxial thin films does not appreciably contribute to $T_{\text{c}}$ suppression.  

\subsection{Films before and after irradiation \label{SIssec:thinfilms_electrical_irradiation}}

\begin{figure}
\includegraphics{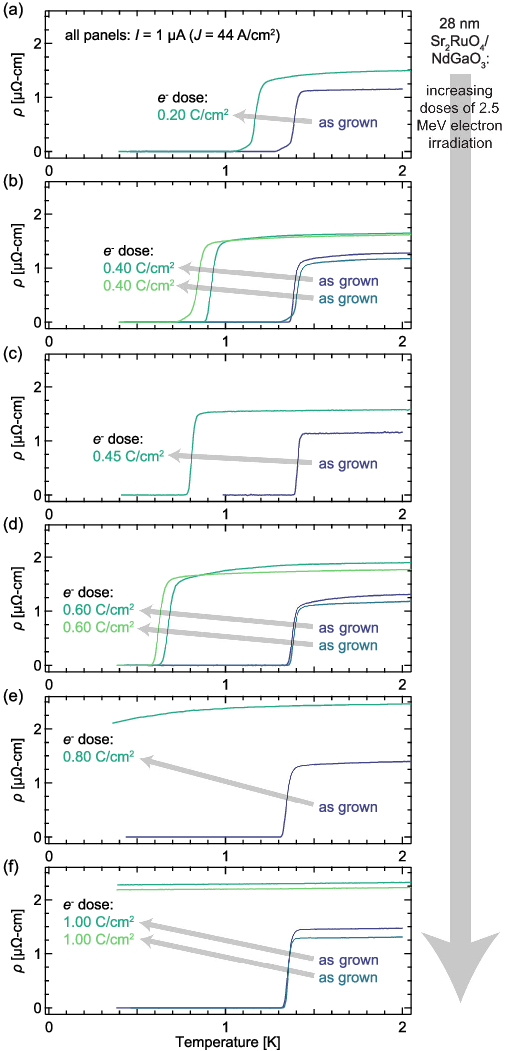}
\caption{\label{fig:before_after_irradiation_rhoVsT_films_all} Electrical resistivity versus temperature of \SRO thin films, before and after irradiation. Panels (a-f) are sorted from top to bottom in order of increasing dose of 2.5 MeV $e^{-}$ irradiation, ranging from (a)~$0.20 \text{ C}/\text{cm}^2$ to (f)~$1.00\text{ C}/\text{cm}^2$.  The vertical and horizontal scales of all graphs are identical, to facilitate visualization of $e^{-}$-dose-dependent changes in $\rho$ and $T_{\text{c}}$. Gray arrows within each panel are guides to the eye, schematically indicating the irradiation-induced changes in $\rho_{0}$ and $T_{\text{c}}$ for each sample.}
\end{figure}

In Fig.~\ref{fig:before_after_irradiation_rhoVsT_films_all}, we show all of the low-temperature, zero-field electrical transport data for \SRO thin films that underlie the data points included in Fig.~2(b,d) and Fig.~3 of the main text.  Each panel of Fig.~\ref{fig:before_after_irradiation_rhoVsT_films_all} focuses on one\textemdash or sometimes two, to check reproducibility\textemdash \space distinct \SRO thin-film resistivity bridge(s), and plots the $\rho(T)$ data trace(s) measured before and after incrementing the dose $D$ of 2.5 MeV electron irradiation received by the sample(s). We used excitation currents of $I = 1 \,\mu\text{A}$ ($J = 44 \text{ A}/\text{cm}^2$) to measure all data displayed in Fig.~\ref{fig:before_after_irradiation_rhoVsT_films_all}. No discernible shifts of the measured $T_{\text{c}}$s occur for $I \leq 1 \, \mu\text{A}$ in either the as-grown or irradiated samples; on the other hand, increasing the excitation current to $I = 10 \,\mu\text{A}$ causes all measured $T_{\text{c}}$s to decrease by about $0.02\text{ K}$ [cf.~Fig.~\ref{SIfig:as_grown_films_properties}(d)]. 

All $\rho(T)$ data displayed in Fig.~\ref{fig:before_after_irradiation_rhoVsT_films_all} are scaled resistivities, computed according to the procedure described in Fig.~\ref{SIfig:as_grown_films_properties} and Sec.~\ref{SIssec:thinfilms_electrical_asgrown} above. By normalizing all raw $R(300 \text{ K})$ values (experimentally measured before and after irradiation) to a common value of $\overline{\rho(300\text{ K})} = 124 $\uOc, we intended to disentangle and to isolate the irradiation-induced modifications in $RRR$s from potential irradiation-induced modifications to the effective cross-sectional area of the conducting channel specific to each thin-film resistivity bridge. We believe that the former effects are the physically relevant changes needed to understand and to interpret the response of superconductivity in \SRO to high-energy electron irradiation, which is the focus of this manuscript.  

We do (somewhat irregularly) observe hints suggestive of the latter type of effects, however, such as nearly-temperature-independent scaling of the pre-irradiation $R(T)$ curve for a given bridge by a factor of $1.02 - 1.03$ after irradiation.  Although other explanations of this are possible and should be investigated in more detail in future work, we note that one of the simplest scenarios\textemdash as suggested, e.g., for SrRuO$_{3}$ thin films in Ref.~\cite{haham_testing_2013}\textemdash would involve migration of irradiation-induced interstitial defects to the surfaces of the thin-film samples in a way that reduces the conducting film thickness by about one RuO$_{2}$ monolayer.  This would result in a fractional increase in measured resistances of about $1/44 = 2.3\%$ for a \SRO thin film that originally consisted of about $44$ RuO$_{2}$ monolayers, in addition to the irradiation-induced changes in the temperature-dependent resistivity of the remaining intact $43$ layers.

Independent of the above considerations about how seriously to interpret the plotted values of $\rho$ in absolute units for the \SRO epitaxial thin-film samples, the low and intermediate irradiation doses ($D = 0.20 - 0.60\text{ C}/\text{cm}^2$) displayed in Fig.~\ref{fig:before_after_irradiation_rhoVsT_films_all} clearly cause modest reductions of $T_{\text{c}}$ by several tenths of a Kelvin, and the highest irradiation doses ($D = 1.00 \text{ C}/\text{cm}^2$) are sufficient to reduce $T_{\text{c}}$ from $1.30 - 1.40 \text{ K}$ (as grown) to values after irradiation that are well below the base temperature of the helium-3 cryostat used for these measurements, $0.4\text{ K}$.

\section{Determining $\rho_{0}$ and $T_{\text{c}}$ from resistivity \label{SIsec:determine_rho0_Tc}}

\subsection{FIB-sculpted single crystals \label{SIssec:determine_rho0_Tc_singlecrystal}}

\begin{figure}[h]
\includegraphics{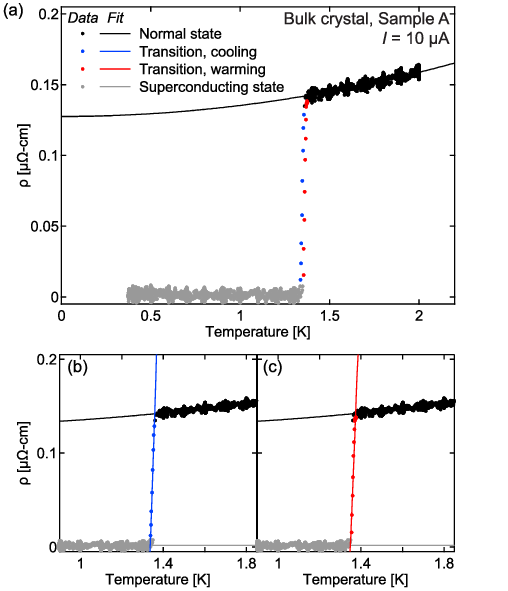}
\caption{\label{SIfig:appendix_rho0_Tc_singlecrystal} Determining ${\rho}_{0}$ and $T_{\text{c}}$ from low-temperature resistivity data acquired for \SRO single crystals. The data set shown here was also plotted in Fig.~1(c) of the main text. (a) Entire dataset and fit to normal-state data for finding ${\rho}_{0}$, the residual resistivity at $T = 0$. (b,c) Same dataset as in (a), now showing fits used to determine $T_{\text{c}}$, with (b) warming and (c) cooling transitions excluded for clarity.}
\end{figure}

For the FIB-sculpted single crystals of \SROnospace, we determined the extrapolated zero-temperature normal-state (ns) residual resistivities, ${\rho}_{0} \equiv \rho_{\text{ns}}(T \rightarrow 0 \text{ K})$, and superconducting $T_{\text{c}}$s from resistivity data collected upon cooling and warming between $2 \text{ K}$ and $0.4 \text{ K}$. We found ${\rho}_{0}$ by fitting the $\rho(T>T_{\text{c}})$ cooling and warming data together to the quadratic functional form ${\rho}_{\text{ns}}(T) = {\rho}_{0} + AT^{2}$ (Fig.~\ref{SIfig:appendix_rho0_Tc_singlecrystal}, black), following previous work in the literature~\cite{mackenzie_extremely_1998}. Across four distinct as-grown samples, we obtained fitted $A$ coefficients ranging from $6.9$ to $8.2 \text{ n}\Omega\text{-cm}/\text{K}^2$, in agreement with the values $A = 4$ to $12~\text{n}\Omega\text{-cm}/\text{K}^{2}$ reported in the literature~\cite{forsythe_evolution_2002, akima_intrinsic_1999, maeno_two-dimensional_1997, hussey_normal-state_1998, barber_resistivity_2018}. 

To extract $T_{\text{c}}$, we initially treated the superconducting transitions measured upon warming and cooling separately (red and blue data points in Fig.~\ref{SIfig:appendix_rho0_Tc_singlecrystal}, respectively), in order to check for hysteresis. We fitted $\rho(T)$ data in the transition region to a linear function ${\rho}_{\text{trans}}(T) = x_{0} + x_{1}T$, and data in the superconducting state (Fig.~\ref{SIfig:appendix_rho0_Tc_singlecrystal}, gray) to a constant function ${\rho}_{\text{sc}} = x_{2}$. From the fitted coefficients $\{x_{0}, x_{1}, x_{2}\}$, we then calculated the temperatures where ${\rho}_{ns}(T)$ intersects ${\rho}_{\text{trans}}(T)$ and where ${\rho}_{\text{trans}}(T)$ intersects ${\rho}_{\text{sc}}$, to find the onset and completion transition temperatures, respectively. Since the experimental ${\rho}_{\text{trans}}(T)$ behavior is described well by straight lines for all samples we have measured, we took $T_{\text{c}}$ to be the midpoint between the onset and completion transition temperatures. Because the thermal hysteresis between cooling and warming data is quite small\textemdash at most $0.017\text{ K}$ and typically $< 0.010\text{ K}$\textemdash throughout the manuscript we report $T_{\text{c}}$ as the arithmetic mean of the values obtained upon cooling and warming.

\subsection{Epitaxial thin films \label{SIssec:determine_rho0_Tc_thinfilm}}

In \SRO epitaxial thin films, the measured normal-state resistivities ${\rho}_{ns}(T)$ (Fig.~\ref{SIfig:as_grown_films_properties}) do not follow the oft-prescribed $\rho_{0} + AT^{2}$ power-series expansion over a region that is wide enough in temperature to justify fitting the data to this functional form, which precludes using the results of such fits to extrapolate ${\rho}_{ns}(T)$ below $T_{\text{c}}$. Accordingly, we adopted slightly more empirical approaches to extract values of $\rho_{0}$ and $T_{\text{c}}$ to use in plots and analyses throughout the main text and Supplemental Material: we set $\rho_{0} = \rho(2 \text{ K})$ (i.e., simply equal to the measured resistivity at $2 \text{ K}$); we define $T_{\text{c}}$ to be the temperature at which $\rho(T)$ crosses the $50\%$ threshold of $\rho(2 \text{ K})$; finally, we quote ``error bars'' on $T_{\text{c}}$ as the temperatures at which $\rho(T)$ crosses the $80\%$ and $10\%$ thresholds of $\rho(2 \text{ K})$, respectively.  Any effects of thermal hysteresis on the resistively measured superconducting transitions for \SRO thin films are negligibly small compared to other, likely extrinsic, factors that broaden these transitions, such as static material inhomogeneities; therefore, all plotted and analyzed $\rho(T)$ data for $T < 2.3\text{ K}$ were acquired upon gradually warming up the thin-film samples, starting from near the base temperature of the helium-3 cryostat. 

We note in passing that if one insists on fitting the \SRO thin-film transport data to $\rho_{ns}(T) = \rho_{0} + AT^{2}$, e.g., over the interval $T$:~$[3 \text{ K} - 5\text{ K}]$, the resulting fit coefficients range from $A = 7.7$ to $8.4~\text{n}\Omega\text{-cm}/\text{K}^{2}$ and $\rho_{0} = 1.10$ to $1.73~\mu\Omega\text{-cm}$ across the 15 as-grown samples shown in Fig.~\ref{SIfig:as_grown_films_properties}(c).  Extending the fitting range to higher temperatures yields systematically smaller fitted $A$ coefficients and larger fitted values of $\rho_{0}$.  Additionally, we note that the relative contribution of temperature-dependent scattering mechanisms to $\rho(2\text{ K})$ in \SRO can be estimated using published results of fits to ${\rho}_{\text{ns}}(T) = {\rho}_{0} + AT^{2}$, which find that $A = 4$ to $12~\text{n}\Omega\text{-cm}/\text{K}^{2}$~\cite{forsythe_evolution_2002, akima_intrinsic_1999, maeno_two-dimensional_1997, hussey_normal-state_1998, barber_resistivity_2018}, and hence $\rho(2\text{ K}) - \rho_{0} = 16$ to $48~\text{n}\Omega\text{-cm}$.  On the other hand, the as-grown and irradiated \SRO thin films studied in this work have measured $\rho(2 \text{ K}) = 1$ to $2~\mu\Omega\text{-cm}$ (Figs.~\ref{SIfig:as_grown_films_properties} and ~\ref{fig:before_after_irradiation_rhoVsT_films_all}).  Thus, our \textit{ad hoc} assignment of $\rho_{0} = \rho(2 \text{ K})$ will generally overestimate the hypothetical ``true'' values of $\rho_{0}$ by $1$ to $5\%$.


\begin{thebibliography}{79}%
\makeatletter
\providecommand \@ifxundefined [1]{%
 \@ifx{#1\undefined}
}%
\providecommand \@ifnum [1]{%
 \ifnum #1\expandafter \@firstoftwo
 \else \expandafter \@secondoftwo
 \fi
}%
\providecommand \@ifx [1]{%
 \ifx #1\expandafter \@firstoftwo
 \else \expandafter \@secondoftwo
 \fi
}%
\providecommand \natexlab [1]{#1}%
\providecommand \enquote  [1]{``#1''}%
\providecommand \bibnamefont  [1]{#1}%
\providecommand \bibfnamefont [1]{#1}%
\providecommand \citenamefont [1]{#1}%
\providecommand \href@noop [0]{\@secondoftwo}%
\providecommand \href [0]{\begingroup \@sanitize@url \@href}%
\providecommand \@href[1]{\@@startlink{#1}\@@href}%
\providecommand \@@href[1]{\endgroup#1\@@endlink}%
\providecommand \@sanitize@url [0]{\catcode `\\12\catcode `\$12\catcode
  `\&12\catcode `\#12\catcode `\^12\catcode `\_12\catcode `\%12\relax}%
\providecommand \@@startlink[1]{}%
\providecommand \@@endlink[0]{}%
\providecommand \url  [0]{\begingroup\@sanitize@url \@url }%
\providecommand \@url [1]{\endgroup\@href {#1}{\urlprefix }}%
\providecommand \urlprefix  [0]{URL }%
\providecommand \Eprint [0]{\href }%
\providecommand \doibase [0]{https://doi.org/}%
\providecommand \selectlanguage [0]{\@gobble}%
\providecommand \bibinfo  [0]{\@secondoftwo}%
\providecommand \bibfield  [0]{\@secondoftwo}%
\providecommand \translation [1]{[#1]}%
\providecommand \BibitemOpen [0]{}%
\providecommand \bibitemStop [0]{}%
\providecommand \bibitemNoStop [0]{.\EOS\space}%
\providecommand \EOS [0]{\spacefactor3000\relax}%
\providecommand \BibitemShut  [1]{\csname bibitem#1\endcsname}%
\let\auto@bib@innerbib\@empty
%</preamble>
\bibitem [{\citenamefont {Mackenzie}\ and\ \citenamefont
  {Maeno}(2003)}]{mackenzie_superconductivity_2003}%
  \BibitemOpen
  \bibfield  {author} {\bibinfo {author} {\bibfnamefont {A.~P.}\ \bibnamefont
  {Mackenzie}}\ and\ \bibinfo {author} {\bibfnamefont {Y.}~\bibnamefont
  {Maeno}},\ }\bibfield  {title} {\bibinfo {title} {The superconductivity of
  {Sr$_{2}$RuO$_{4}$} and the physics of spin-triplet pairing},\ }\href
  {https://doi.org/10.1103/RevModPhys.75.657} {\bibfield  {journal} {\bibinfo
  {journal} {Rev. Mod. Phys.}\ }\textbf {\bibinfo {volume} {75}},\ \bibinfo
  {pages} {657} (\bibinfo {year} {2003})}\BibitemShut {NoStop}%
\bibitem [{\citenamefont {Mackenzie}\ \emph {et~al.}(2017)\citenamefont
  {Mackenzie}, \citenamefont {Scaffidi}, \citenamefont {Hicks},\ and\
  \citenamefont {Maeno}}]{mackenzie_even_2017}%
  \BibitemOpen
  \bibfield  {author} {\bibinfo {author} {\bibfnamefont {A.~P.}\ \bibnamefont
  {Mackenzie}}, \bibinfo {author} {\bibfnamefont {T.}~\bibnamefont {Scaffidi}},
  \bibinfo {author} {\bibfnamefont {C.~W.}\ \bibnamefont {Hicks}},\ and\
  \bibinfo {author} {\bibfnamefont {Y.}~\bibnamefont {Maeno}},\ }\bibfield
  {title} {\bibinfo {title} {Even odder after twenty-three years: the
  superconducting order parameter puzzle of {Sr$_{2}$RuO$_{4}$}},\ }\href
  {https://doi.org/10.1038/s41535-017-0045-4} {\bibfield  {journal} {\bibinfo
  {journal} {npj Quantum Mater.}\ }\textbf {\bibinfo {volume} {2}},\ \bibinfo
  {pages} {1} (\bibinfo {year} {2017})}\BibitemShut {NoStop}%
\bibitem [{\citenamefont {Mackenzie}\ \emph {et~al.}(1998)\citenamefont
  {Mackenzie}, \citenamefont {Haselwimmer}, \citenamefont {Tyler},
  \citenamefont {Lonzarich}, \citenamefont {Mori}, \citenamefont {Nishizaki},\
  and\ \citenamefont {Maeno}}]{mackenzie_extremely_1998}%
  \BibitemOpen
  \bibfield  {author} {\bibinfo {author} {\bibfnamefont {A.~P.}\ \bibnamefont
  {Mackenzie}}, \bibinfo {author} {\bibfnamefont {R.~K.~W.}\ \bibnamefont
  {Haselwimmer}}, \bibinfo {author} {\bibfnamefont {A.~W.}\ \bibnamefont
  {Tyler}}, \bibinfo {author} {\bibfnamefont {G.~G.}\ \bibnamefont
  {Lonzarich}}, \bibinfo {author} {\bibfnamefont {Y.}~\bibnamefont {Mori}},
  \bibinfo {author} {\bibfnamefont {S.}~\bibnamefont {Nishizaki}},\ and\
  \bibinfo {author} {\bibfnamefont {Y.}~\bibnamefont {Maeno}},\ }\bibfield
  {title} {\bibinfo {title} {Extremely {Strong} {Dependence} of
  {Superconductivity} on {Disorder} in {Sr$_{2}$RuO$_{4}$}},\ }\href
  {https://doi.org/10.1103/PhysRevLett.80.161} {\bibfield  {journal} {\bibinfo
  {journal} {Phys. Rev. Lett.}\ }\textbf {\bibinfo {volume} {80}},\ \bibinfo
  {pages} {161} (\bibinfo {year} {1998})}\BibitemShut {NoStop}%
\bibitem [{\citenamefont {Kikugawa}\ \emph {et~al.}(2002)\citenamefont
  {Kikugawa}, \citenamefont {Saita}, \citenamefont {Minakata},\ and\
  \citenamefont {Maeno}}]{kikugawa_effect_2002}%
  \BibitemOpen
  \bibfield  {author} {\bibinfo {author} {\bibfnamefont {N.}~\bibnamefont
  {Kikugawa}}, \bibinfo {author} {\bibfnamefont {S.}~\bibnamefont {Saita}},
  \bibinfo {author} {\bibfnamefont {M.}~\bibnamefont {Minakata}},\ and\
  \bibinfo {author} {\bibfnamefont {Y.}~\bibnamefont {Maeno}},\ }\bibfield
  {title} {\bibinfo {title} {Effect of {Ti} substitution on the residual
  resistivity in the spin-triplet superconductor {Sr$_{2}$RuO$_{4}$}},\ }\href
  {https://doi.org/10.1016/S0921-45260101208-X} {\bibfield  {journal} {\bibinfo
   {journal} {Physica B}\ }\textbf {\bibinfo {volume} {312-313}},\ \bibinfo
  {pages} {803} (\bibinfo {year} {2002})}\BibitemShut {NoStop}%
\bibitem [{\citenamefont {Kikugawa}\ \emph {et~al.}(2003)\citenamefont
  {Kikugawa}, \citenamefont {Peter~Mackenzie},\ and\ \citenamefont
  {Maeno}}]{kikugawa_effects_2003}%
  \BibitemOpen
  \bibfield  {author} {\bibinfo {author} {\bibfnamefont {N.}~\bibnamefont
  {Kikugawa}}, \bibinfo {author} {\bibfnamefont {A.}~\bibnamefont
  {Peter~Mackenzie}},\ and\ \bibinfo {author} {\bibfnamefont {Y.}~\bibnamefont
  {Maeno}},\ }\bibfield  {title} {\bibinfo {title} {Effects of {In}-{Plane}
  {Impurity} {Substitution} in {Sr$_{2}$RuO$_{4}$}},\ }\href
  {https://doi.org/10.1143/JPSJ.72.237} {\bibfield  {journal} {\bibinfo
  {journal} {J. Phys. Soc. Jpn.}\ }\textbf {\bibinfo {volume} {72}},\ \bibinfo
  {pages} {237} (\bibinfo {year} {2003})}\BibitemShut {NoStop}%
\bibitem [{\citenamefont {Kikugawa}\ \emph {et~al.}(2004)\citenamefont
  {Kikugawa}, \citenamefont {Mackenzie}, \citenamefont {Bergemann},
  \citenamefont {Borzi}, \citenamefont {Grigera},\ and\ \citenamefont
  {Maeno}}]{kikugawa_rigid-band_2004}%
  \BibitemOpen
  \bibfield  {author} {\bibinfo {author} {\bibfnamefont {N.}~\bibnamefont
  {Kikugawa}}, \bibinfo {author} {\bibfnamefont {A.~P.}\ \bibnamefont
  {Mackenzie}}, \bibinfo {author} {\bibfnamefont {C.}~\bibnamefont
  {Bergemann}}, \bibinfo {author} {\bibfnamefont {R.~A.}\ \bibnamefont
  {Borzi}}, \bibinfo {author} {\bibfnamefont {S.~A.}\ \bibnamefont {Grigera}},\
  and\ \bibinfo {author} {\bibfnamefont {Y.}~\bibnamefont {Maeno}},\ }\bibfield
   {title} {\bibinfo {title} {Rigid-band shift of the {Fermi} level in the
  strongly correlated metal: {Sr$_{2-y}$La$_{y}$RuO$_{4}$}},\ }\href
  {https://doi.org/10.1103/PhysRevB.70.060508} {\bibfield  {journal} {\bibinfo
  {journal} {Phys. Rev. B}\ }\textbf {\bibinfo {volume} {70}},\ \bibinfo
  {pages} {060508} (\bibinfo {year} {2004})}\BibitemShut {NoStop}%
\bibitem [{\citenamefont {Suderow}\ \emph {et~al.}(1998)\citenamefont
  {Suderow}, \citenamefont {Brison}, \citenamefont {Floquet}, \citenamefont
  {Tyler},\ and\ \citenamefont {Maeno}}]{suderow_thermalcond_1998}%
  \BibitemOpen
  \bibfield  {author} {\bibinfo {author} {\bibfnamefont {H.}~\bibnamefont
  {Suderow}}, \bibinfo {author} {\bibfnamefont {J.~P.}\ \bibnamefont {Brison}},
  \bibinfo {author} {\bibfnamefont {J.}~\bibnamefont {Floquet}}, \bibinfo
  {author} {\bibfnamefont {A.~W.}\ \bibnamefont {Tyler}},\ and\ \bibinfo
  {author} {\bibfnamefont {Y.}~\bibnamefont {Maeno}},\ }\bibfield  {title}
  {\bibinfo {title} {{Very low temperature thermal conductivity in the layered
  perovskite superconductor Sr$_{2}$RuO$_{4}$}},\ }\href
  {https://doi.org/10.1088/0953-8984/10/34/004} {\bibfield  {journal} {\bibinfo
   {journal} {J. Phys.: Condens. Matter}\ }\textbf {\bibinfo {volume} {10}},\
  \bibinfo {pages} {L597} (\bibinfo {year} {1998})}\BibitemShut {NoStop}%
\bibitem [{\citenamefont {Mao}\ \emph {et~al.}(1999)\citenamefont {Mao},
  \citenamefont {Mori},\ and\ \citenamefont {Maeno}}]{mao_suppression_1999}%
  \BibitemOpen
  \bibfield  {author} {\bibinfo {author} {\bibfnamefont {Z.~Q.}\ \bibnamefont
  {Mao}}, \bibinfo {author} {\bibfnamefont {Y.}~\bibnamefont {Mori}},\ and\
  \bibinfo {author} {\bibfnamefont {Y.}~\bibnamefont {Maeno}},\ }\bibfield
  {title} {\bibinfo {title} {Suppression of superconductivity in
  {Sr$_{2}$RuO$_{4}$} caused by defects},\ }\href
  {https://doi.org/10.1103/PhysRevB.60.610} {\bibfield  {journal} {\bibinfo
  {journal} {Phys. Rev. B}\ }\textbf {\bibinfo {volume} {60}},\ \bibinfo
  {pages} {610} (\bibinfo {year} {1999})}\BibitemShut {NoStop}%
\bibitem [{\citenamefont {Uchida}\ \emph {et~al.}(2020)\citenamefont {Uchida},
  \citenamefont {Sakuraba}, \citenamefont {Kawamura}, \citenamefont {Ide},
  \citenamefont {Takahashi}, \citenamefont {Tokura},\ and\ \citenamefont
  {Kawasaki}}]{uchida_characterization_2020}%
  \BibitemOpen
  \bibfield  {author} {\bibinfo {author} {\bibfnamefont {M.}~\bibnamefont
  {Uchida}}, \bibinfo {author} {\bibfnamefont {I.}~\bibnamefont {Sakuraba}},
  \bibinfo {author} {\bibfnamefont {M.}~\bibnamefont {Kawamura}}, \bibinfo
  {author} {\bibfnamefont {M.}~\bibnamefont {Ide}}, \bibinfo {author}
  {\bibfnamefont {K.~S.}\ \bibnamefont {Takahashi}}, \bibinfo {author}
  {\bibfnamefont {Y.}~\bibnamefont {Tokura}},\ and\ \bibinfo {author}
  {\bibfnamefont {M.}~\bibnamefont {Kawasaki}},\ }\bibfield  {title} {\bibinfo
  {title} {Characterization of {Sr$_{2}$RuO$_{4}$} {Josephson} junctions made
  of epitaxial films},\ }\href {https://doi.org/10.1103/PhysRevB.101.035107}
  {\bibfield  {journal} {\bibinfo  {journal} {Phys. Rev. B}\ }\textbf {\bibinfo
  {volume} {101}},\ \bibinfo {pages} {035107} (\bibinfo {year}
  {2020})}\BibitemShut {NoStop}%
\bibitem [{\citenamefont {Fang}\ \emph {et~al.}(2021)\citenamefont {Fang},
  \citenamefont {Nair}, \citenamefont {Miao}, \citenamefont {Goodge},
  \citenamefont {Schreiber}, \citenamefont {Ruf}, \citenamefont {Kourkoutis},
  \citenamefont {Shen}, \citenamefont {Schlom},\ and\ \citenamefont
  {Ramshaw}}]{fang_quantum_2021}%
  \BibitemOpen
  \bibfield  {author} {\bibinfo {author} {\bibfnamefont {Y.}~\bibnamefont
  {Fang}}, \bibinfo {author} {\bibfnamefont {H.~P.}\ \bibnamefont {Nair}},
  \bibinfo {author} {\bibfnamefont {L.}~\bibnamefont {Miao}}, \bibinfo {author}
  {\bibfnamefont {B.}~\bibnamefont {Goodge}}, \bibinfo {author} {\bibfnamefont
  {N.~J.}\ \bibnamefont {Schreiber}}, \bibinfo {author} {\bibfnamefont {J.~P.}\
  \bibnamefont {Ruf}}, \bibinfo {author} {\bibfnamefont {L.~F.}\ \bibnamefont
  {Kourkoutis}}, \bibinfo {author} {\bibfnamefont {K.~M.}\ \bibnamefont
  {Shen}}, \bibinfo {author} {\bibfnamefont {D.~G.}\ \bibnamefont {Schlom}},\
  and\ \bibinfo {author} {\bibfnamefont {B.~J.}\ \bibnamefont {Ramshaw}},\
  }\bibfield  {title} {\bibinfo {title} {Quantum oscillations and quasiparticle
  properties of thin film {Sr$_{2}$RuO$_{4}$}},\ }\href
  {https://doi.org/10.1103/PhysRevB.104.045152} {\bibfield  {journal} {\bibinfo
   {journal} {Phys. Rev. B}\ }\textbf {\bibinfo {volume} {104}},\ \bibinfo
  {pages} {045152} (\bibinfo {year} {2021})}\BibitemShut {NoStop}%
\bibitem [{\citenamefont {Noad}\ \emph {et~al.}(2016)\citenamefont {Noad},
  \citenamefont {Spanton}, \citenamefont {Nowack}, \citenamefont {Inoue},
  \citenamefont {Kim}, \citenamefont {Merz}, \citenamefont {Bell},
  \citenamefont {Hikita}, \citenamefont {Xu}, \citenamefont {Liu},
  \citenamefont {Vailionis}, \citenamefont {Hwang},\ and\ \citenamefont
  {Moler}}]{noad_variation_2016}%
  \BibitemOpen
  \bibfield  {author} {\bibinfo {author} {\bibfnamefont {H.}~\bibnamefont
  {Noad}}, \bibinfo {author} {\bibfnamefont {E.~M.}\ \bibnamefont {Spanton}},
  \bibinfo {author} {\bibfnamefont {K.~C.}\ \bibnamefont {Nowack}}, \bibinfo
  {author} {\bibfnamefont {H.}~\bibnamefont {Inoue}}, \bibinfo {author}
  {\bibfnamefont {M.}~\bibnamefont {Kim}}, \bibinfo {author} {\bibfnamefont
  {T.~A.}\ \bibnamefont {Merz}}, \bibinfo {author} {\bibfnamefont
  {C.}~\bibnamefont {Bell}}, \bibinfo {author} {\bibfnamefont {Y.}~\bibnamefont
  {Hikita}}, \bibinfo {author} {\bibfnamefont {R.}~\bibnamefont {Xu}}, \bibinfo
  {author} {\bibfnamefont {W.}~\bibnamefont {Liu}}, \bibinfo {author}
  {\bibfnamefont {A.}~\bibnamefont {Vailionis}}, \bibinfo {author}
  {\bibfnamefont {H.~Y.}\ \bibnamefont {Hwang}},\ and\ \bibinfo {author}
  {\bibfnamefont {K.~A.}\ \bibnamefont {Moler}},\ }\bibfield  {title} {\bibinfo
  {title} {Variation in superconducting transition temperature due to
  tetragonal domains in two-dimensionally doped {SrTiO$_{3}$}},\ }\href
  {https://doi.org/10.1103/PhysRevB.94.174516} {\bibfield  {journal} {\bibinfo
  {journal} {Phys. Rev. B}\ }\textbf {\bibinfo {volume} {94}},\ \bibinfo
  {pages} {174516} (\bibinfo {year} {2016})}\BibitemShut {NoStop}%
\bibitem [{\citenamefont {Watson}\ \emph {et~al.}(2018)\citenamefont {Watson},
  \citenamefont {Gibbs}, \citenamefont {Mackenzie}, \citenamefont {Hicks},\
  and\ \citenamefont {Moler}}]{watson_micron-scale_2018}%
  \BibitemOpen
  \bibfield  {author} {\bibinfo {author} {\bibfnamefont {C.~A.}\ \bibnamefont
  {Watson}}, \bibinfo {author} {\bibfnamefont {A.~S.}\ \bibnamefont {Gibbs}},
  \bibinfo {author} {\bibfnamefont {A.~P.}\ \bibnamefont {Mackenzie}}, \bibinfo
  {author} {\bibfnamefont {C.~W.}\ \bibnamefont {Hicks}},\ and\ \bibinfo
  {author} {\bibfnamefont {K.~A.}\ \bibnamefont {Moler}},\ }\bibfield  {title}
  {\bibinfo {title} {Micron-scale measurements of low anisotropic strain
  response of local {$T_{c}$} in {Sr$_{2}$RuO$_{4}$}},\ }\href
  {https://doi.org/10.1103/PhysRevB.98.094521} {\bibfield  {journal} {\bibinfo
  {journal} {Phys. Rev. B}\ }\textbf {\bibinfo {volume} {98}},\ \bibinfo
  {pages} {094521} (\bibinfo {year} {2018})}\BibitemShut {NoStop}%
\bibitem [{\citenamefont {Burganov}\ \emph {et~al.}(2016)\citenamefont
  {Burganov}, \citenamefont {Adamo}, \citenamefont {Mulder}, \citenamefont
  {Uchida}, \citenamefont {King}, \citenamefont {Harter}, \citenamefont {Shai},
  \citenamefont {Gibbs}, \citenamefont {Mackenzie}, \citenamefont {Uecker},
  \citenamefont {Bruetzam}, \citenamefont {Beasley}, \citenamefont {Fennie},
  \citenamefont {Schlom},\ and\ \citenamefont {Shen}}]{burganov_strain_2016}%
  \BibitemOpen
  \bibfield  {author} {\bibinfo {author} {\bibfnamefont {B.}~\bibnamefont
  {Burganov}}, \bibinfo {author} {\bibfnamefont {C.}~\bibnamefont {Adamo}},
  \bibinfo {author} {\bibfnamefont {A.}~\bibnamefont {Mulder}}, \bibinfo
  {author} {\bibfnamefont {M.}~\bibnamefont {Uchida}}, \bibinfo {author}
  {\bibfnamefont {P.~D.~C.}\ \bibnamefont {King}}, \bibinfo {author}
  {\bibfnamefont {J.~W.}\ \bibnamefont {Harter}}, \bibinfo {author}
  {\bibfnamefont {D.~E.}\ \bibnamefont {Shai}}, \bibinfo {author}
  {\bibfnamefont {A.~S.}\ \bibnamefont {Gibbs}}, \bibinfo {author}
  {\bibfnamefont {A.~P.}\ \bibnamefont {Mackenzie}}, \bibinfo {author}
  {\bibfnamefont {R.}~\bibnamefont {Uecker}}, \bibinfo {author} {\bibfnamefont
  {M.}~\bibnamefont {Bruetzam}}, \bibinfo {author} {\bibfnamefont {M.~R.}\
  \bibnamefont {Beasley}}, \bibinfo {author} {\bibfnamefont {C.~J.}\
  \bibnamefont {Fennie}}, \bibinfo {author} {\bibfnamefont {D.~G.}\
  \bibnamefont {Schlom}},\ and\ \bibinfo {author} {\bibfnamefont {K.~M.}\
  \bibnamefont {Shen}},\ }\bibfield  {title} {\bibinfo {title} {Strain
  {Control} of {Fermiology} and {Many}-{Body} {Interactions} in
  {Two}-{Dimensional} {Ruthenates}},\ }\href
  {https://doi.org/10.1103/PhysRevLett.116.197003} {\bibfield  {journal}
  {\bibinfo  {journal} {Phys. Rev. Lett.}\ }\textbf {\bibinfo {volume} {116}},\
  \bibinfo {pages} {197003} (\bibinfo {year} {2016})}\BibitemShut {NoStop}%
\bibitem [{\citenamefont {Wang}\ \emph {et~al.}(2021)\citenamefont {Wang},
  \citenamefont {Nair}, \citenamefont {Schreiber}, \citenamefont {Ruf},
  \citenamefont {Cheng}, \citenamefont {Schlom}, \citenamefont {Shen},\ and\
  \citenamefont {Armitage}}]{wang_separated_2021}%
  \BibitemOpen
  \bibfield  {author} {\bibinfo {author} {\bibfnamefont {Y.}~\bibnamefont
  {Wang}}, \bibinfo {author} {\bibfnamefont {H.~P.}\ \bibnamefont {Nair}},
  \bibinfo {author} {\bibfnamefont {N.~J.}\ \bibnamefont {Schreiber}}, \bibinfo
  {author} {\bibfnamefont {J.~P.}\ \bibnamefont {Ruf}}, \bibinfo {author}
  {\bibfnamefont {B.}~\bibnamefont {Cheng}}, \bibinfo {author} {\bibfnamefont
  {D.~G.}\ \bibnamefont {Schlom}}, \bibinfo {author} {\bibfnamefont {K.~M.}\
  \bibnamefont {Shen}},\ and\ \bibinfo {author} {\bibfnamefont {N.~P.}\
  \bibnamefont {Armitage}},\ }\bibfield  {title} {\bibinfo {title} {Separated
  transport relaxation scales and interband scattering in thin films of
  {SrRuO$_{3}$}, {CaRuO$_{3}$}, and {Sr$_{2}$RuO$_{4}$}},\ }\href
  {https://doi.org/10.1103/PhysRevB.103.205109} {\bibfield  {journal} {\bibinfo
   {journal} {Phys. Rev. B}\ }\textbf {\bibinfo {volume} {103}},\ \bibinfo
  {pages} {205109} (\bibinfo {year} {2021})}\BibitemShut {NoStop}%
\bibitem [{\citenamefont {Tsuei}\ and\ \citenamefont
  {Kirtley}(2000)}]{tsuei_symmetry_2000}%
  \BibitemOpen
  \bibfield  {author} {\bibinfo {author} {\bibfnamefont {C.~C.}\ \bibnamefont
  {Tsuei}}\ and\ \bibinfo {author} {\bibfnamefont {J.~R.}\ \bibnamefont
  {Kirtley}},\ }\bibfield  {title} {\bibinfo {title} {Pairing symmetry in
  cuprate superconductors},\ }\href {https://doi.org/10.1103/RevModPhys.72.969}
  {\bibfield  {journal} {\bibinfo  {journal} {Rev. Mod. Phys.}\ }\textbf
  {\bibinfo {volume} {72}},\ \bibinfo {pages} {969} (\bibinfo {year}
  {2000})}\BibitemShut {NoStop}%
\bibitem [{\citenamefont {Maeno}\ \emph {et~al.}(1994)\citenamefont {Maeno},
  \citenamefont {Hashimoto}, \citenamefont {Yoshida}, \citenamefont
  {Nishizaki}, \citenamefont {Fujita}, \citenamefont {Bednorz},\ and\
  \citenamefont {Lichtenberg}}]{maeno_superconductivity_1994}%
  \BibitemOpen
  \bibfield  {author} {\bibinfo {author} {\bibfnamefont {Y.}~\bibnamefont
  {Maeno}}, \bibinfo {author} {\bibfnamefont {H.}~\bibnamefont {Hashimoto}},
  \bibinfo {author} {\bibfnamefont {K.}~\bibnamefont {Yoshida}}, \bibinfo
  {author} {\bibfnamefont {S.}~\bibnamefont {Nishizaki}}, \bibinfo {author}
  {\bibfnamefont {T.}~\bibnamefont {Fujita}}, \bibinfo {author} {\bibfnamefont
  {J.~G.}\ \bibnamefont {Bednorz}},\ and\ \bibinfo {author} {\bibfnamefont
  {F.}~\bibnamefont {Lichtenberg}},\ }\bibfield  {title} {\bibinfo {title}
  {Superconductivity in a layered perovskite without copper},\ }\href
  {https://doi.org/10.1038/372532a0} {\bibfield  {journal} {\bibinfo  {journal}
  {Nature}\ }\textbf {\bibinfo {volume} {372}},\ \bibinfo {pages} {532}
  (\bibinfo {year} {1994})}\BibitemShut {NoStop}%
\bibitem [{\citenamefont {Krockenberger}\ \emph {et~al.}(2010)\citenamefont
  {Krockenberger}, \citenamefont {Uchida}, \citenamefont {Takahashi},
  \citenamefont {Nakamura}, \citenamefont {Kawasaki},\ and\ \citenamefont
  {Tokura}}]{krockenberger_growth_2010}%
  \BibitemOpen
  \bibfield  {author} {\bibinfo {author} {\bibfnamefont {Y.}~\bibnamefont
  {Krockenberger}}, \bibinfo {author} {\bibfnamefont {M.}~\bibnamefont
  {Uchida}}, \bibinfo {author} {\bibfnamefont {K.~S.}\ \bibnamefont
  {Takahashi}}, \bibinfo {author} {\bibfnamefont {M.}~\bibnamefont {Nakamura}},
  \bibinfo {author} {\bibfnamefont {M.}~\bibnamefont {Kawasaki}},\ and\
  \bibinfo {author} {\bibfnamefont {Y.}~\bibnamefont {Tokura}},\ }\bibfield
  {title} {\bibinfo {title} {Growth of superconducting {Sr$_{2}$RuO$_{4}$} thin
  films},\ }\href {https://doi.org/10.1063/1.3481363} {\bibfield  {journal}
  {\bibinfo  {journal} {Appl. Phys. Lett.}\ }\textbf {\bibinfo {volume} {97}},\
  \bibinfo {pages} {082502} (\bibinfo {year} {2010})}\BibitemShut {NoStop}%
\bibitem [{\citenamefont {Uchida}\ \emph {et~al.}(2017)\citenamefont {Uchida},
  \citenamefont {Ide}, \citenamefont {Watanabe}, \citenamefont {Takahashi},
  \citenamefont {Tokura},\ and\ \citenamefont
  {Kawasaki}}]{uchida_molecular_2017}%
  \BibitemOpen
  \bibfield  {author} {\bibinfo {author} {\bibfnamefont {M.}~\bibnamefont
  {Uchida}}, \bibinfo {author} {\bibfnamefont {M.}~\bibnamefont {Ide}},
  \bibinfo {author} {\bibfnamefont {H.}~\bibnamefont {Watanabe}}, \bibinfo
  {author} {\bibfnamefont {K.~S.}\ \bibnamefont {Takahashi}}, \bibinfo {author}
  {\bibfnamefont {Y.}~\bibnamefont {Tokura}},\ and\ \bibinfo {author}
  {\bibfnamefont {M.}~\bibnamefont {Kawasaki}},\ }\bibfield  {title} {\bibinfo
  {title} {Molecular beam epitaxy growth of superconducting {Sr$_{2}$RuO$_{4}$}
  films},\ }\href {https://doi.org/10.1063/1.5007342} {\bibfield  {journal}
  {\bibinfo  {journal} {APL Mater.}\ }\textbf {\bibinfo {volume} {5}},\
  \bibinfo {pages} {106108} (\bibinfo {year} {2017})}\BibitemShut {NoStop}%
\bibitem [{\citenamefont {Nair}\ \emph
  {et~al.}(2018{\natexlab{a}})\citenamefont {Nair}, \citenamefont {Ruf},
  \citenamefont {Schreiber}, \citenamefont {Miao}, \citenamefont {Grandon},
  \citenamefont {Baek}, \citenamefont {Goodge}, \citenamefont {Ruff},
  \citenamefont {Kourkoutis}, \citenamefont {Shen},\ and\ \citenamefont
  {Schlom}}]{nair_demystifying_2018}%
  \BibitemOpen
  \bibfield  {author} {\bibinfo {author} {\bibfnamefont {H.~P.}\ \bibnamefont
  {Nair}}, \bibinfo {author} {\bibfnamefont {J.~P.}\ \bibnamefont {Ruf}},
  \bibinfo {author} {\bibfnamefont {N.~J.}\ \bibnamefont {Schreiber}}, \bibinfo
  {author} {\bibfnamefont {L.}~\bibnamefont {Miao}}, \bibinfo {author}
  {\bibfnamefont {M.~L.}\ \bibnamefont {Grandon}}, \bibinfo {author}
  {\bibfnamefont {D.~J.}\ \bibnamefont {Baek}}, \bibinfo {author}
  {\bibfnamefont {B.~H.}\ \bibnamefont {Goodge}}, \bibinfo {author}
  {\bibfnamefont {J.~P.~C.}\ \bibnamefont {Ruff}}, \bibinfo {author}
  {\bibfnamefont {L.~F.}\ \bibnamefont {Kourkoutis}}, \bibinfo {author}
  {\bibfnamefont {K.~M.}\ \bibnamefont {Shen}},\ and\ \bibinfo {author}
  {\bibfnamefont {D.~G.}\ \bibnamefont {Schlom}},\ }\bibfield  {title}
  {\bibinfo {title} {Demystifying the growth of superconducting
  {Sr$_{2}$RuO$_{4}$} thin films},\ }\href {https://doi.org/10.1063/1.5053084}
  {\bibfield  {journal} {\bibinfo  {journal} {APL Mater.}\ }\textbf {\bibinfo
  {volume} {6}},\ \bibinfo {pages} {101108} (\bibinfo {year}
  {2018}{\natexlab{a}})}\BibitemShut {NoStop}%
\bibitem [{\citenamefont {Garcia}\ \emph {et~al.}(2020)\citenamefont {Garcia},
  \citenamefont {Di~Bernardo}, \citenamefont {Kimbell}, \citenamefont
  {Vickers}, \citenamefont {Massabuau}, \citenamefont {Komori}, \citenamefont
  {Divitini}, \citenamefont {Yasui}, \citenamefont {Lee}, \citenamefont {Kim},
  \citenamefont {Kim}, \citenamefont {Blamire}, \citenamefont {Vecchione},
  \citenamefont {Fittipaldi}, \citenamefont {Maeno}, \citenamefont {Noh},\ and\
  \citenamefont {Robinson}}]{garcia_pair_2020}%
  \BibitemOpen
  \bibfield  {author} {\bibinfo {author} {\bibfnamefont {C.~M.~P.}\
  \bibnamefont {Garcia}}, \bibinfo {author} {\bibfnamefont {A.}~\bibnamefont
  {Di~Bernardo}}, \bibinfo {author} {\bibfnamefont {G.}~\bibnamefont
  {Kimbell}}, \bibinfo {author} {\bibfnamefont {M.~E.}\ \bibnamefont
  {Vickers}}, \bibinfo {author} {\bibfnamefont {F.~C.-P.}\ \bibnamefont
  {Massabuau}}, \bibinfo {author} {\bibfnamefont {S.}~\bibnamefont {Komori}},
  \bibinfo {author} {\bibfnamefont {G.}~\bibnamefont {Divitini}}, \bibinfo
  {author} {\bibfnamefont {Y.}~\bibnamefont {Yasui}}, \bibinfo {author}
  {\bibfnamefont {H.~G.}\ \bibnamefont {Lee}}, \bibinfo {author} {\bibfnamefont
  {J.}~\bibnamefont {Kim}}, \bibinfo {author} {\bibfnamefont {B.}~\bibnamefont
  {Kim}}, \bibinfo {author} {\bibfnamefont {M.~G.}\ \bibnamefont {Blamire}},
  \bibinfo {author} {\bibfnamefont {A.}~\bibnamefont {Vecchione}}, \bibinfo
  {author} {\bibfnamefont {R.}~\bibnamefont {Fittipaldi}}, \bibinfo {author}
  {\bibfnamefont {Y.}~\bibnamefont {Maeno}}, \bibinfo {author} {\bibfnamefont
  {T.~W.}\ \bibnamefont {Noh}},\ and\ \bibinfo {author} {\bibfnamefont
  {J.~W.~A.}\ \bibnamefont {Robinson}},\ }\bibfield  {title} {\bibinfo {title}
  {Pair suppression caused by mosaic-twist defects in superconducting
  {Sr$_{2}$RuO$_{4}$} thin-films prepared using pulsed laser deposition},\
  }\href {https://doi.org/10.1038/s43246-020-0026-1} {\bibfield  {journal}
  {\bibinfo  {journal} {Commun. Mater.}\ }\textbf {\bibinfo {volume} {1}},\
  \bibinfo {pages} {23} (\bibinfo {year} {2020})}\BibitemShut {NoStop}%
\bibitem [{\citenamefont {Kim}\ \emph {et~al.}(2021)\citenamefont {Kim},
  \citenamefont {Mun}, \citenamefont {Palomares~Garc\'{i}a}, \citenamefont
  {Kim}, \citenamefont {Perry}, \citenamefont {Jo}, \citenamefont {Im},
  \citenamefont {Lee}, \citenamefont {Ko}, \citenamefont {Chang}, \citenamefont
  {Chung}, \citenamefont {Kim}, \citenamefont {Robinson}, \citenamefont
  {Yonezawa}, \citenamefont {Maeno}, \citenamefont {Wang},\ and\ \citenamefont
  {Noh}}]{kim_superconducting_2021}%
  \BibitemOpen
  \bibfield  {author} {\bibinfo {author} {\bibfnamefont {J.}~\bibnamefont
  {Kim}}, \bibinfo {author} {\bibfnamefont {J.}~\bibnamefont {Mun}}, \bibinfo
  {author} {\bibfnamefont {C.~M.}\ \bibnamefont {Palomares~Garc\'{i}a}},
  \bibinfo {author} {\bibfnamefont {B.}~\bibnamefont {Kim}}, \bibinfo {author}
  {\bibfnamefont {R.~S.}\ \bibnamefont {Perry}}, \bibinfo {author}
  {\bibfnamefont {Y.}~\bibnamefont {Jo}}, \bibinfo {author} {\bibfnamefont
  {H.}~\bibnamefont {Im}}, \bibinfo {author} {\bibfnamefont {H.~G.}\
  \bibnamefont {Lee}}, \bibinfo {author} {\bibfnamefont {E.~K.}\ \bibnamefont
  {Ko}}, \bibinfo {author} {\bibfnamefont {S.~H.}\ \bibnamefont {Chang}},
  \bibinfo {author} {\bibfnamefont {S.~B.}\ \bibnamefont {Chung}}, \bibinfo
  {author} {\bibfnamefont {M.}~\bibnamefont {Kim}}, \bibinfo {author}
  {\bibfnamefont {J.~W.~A.}\ \bibnamefont {Robinson}}, \bibinfo {author}
  {\bibfnamefont {S.}~\bibnamefont {Yonezawa}}, \bibinfo {author}
  {\bibfnamefont {Y.}~\bibnamefont {Maeno}}, \bibinfo {author} {\bibfnamefont
  {L.}~\bibnamefont {Wang}},\ and\ \bibinfo {author} {\bibfnamefont {T.~W.}\
  \bibnamefont {Noh}},\ }\bibfield  {title} {\bibinfo {title} {Superconducting
  {Sr$_{2}$RuO$_{4}$} {Thin} {Films} without {Out}-of-{Phase} {Boundaries} by
  {Higher}-{Order} {Ruddlesden}–{Popper} {Intergrowth}},\ }\href
  {https://doi.org/10.1021/acs.nanolett.0c04963} {\bibfield  {journal}
  {\bibinfo  {journal} {Nano Lett.}\ }\textbf {\bibinfo {volume} {21}},\
  \bibinfo {pages} {4185} (\bibinfo {year} {2021})}\BibitemShut {NoStop}%
\bibitem [{\citenamefont {Sunko}\ \emph {et~al.}(2020)\citenamefont {Sunko},
  \citenamefont {McGuinness}, \citenamefont {Chang}, \citenamefont {Zhakina},
  \citenamefont {Khim}, \citenamefont {Dreyer}, \citenamefont {Konczykowski},
  \citenamefont {Borrmann}, \citenamefont {Moll}, \citenamefont {K\"onig},
  \citenamefont {Muller},\ and\ \citenamefont
  {Mackenzie}}]{sunko_controlled_2020}%
  \BibitemOpen
  \bibfield  {author} {\bibinfo {author} {\bibfnamefont {V.}~\bibnamefont
  {Sunko}}, \bibinfo {author} {\bibfnamefont {P.~H.}\ \bibnamefont
  {McGuinness}}, \bibinfo {author} {\bibfnamefont {C.~S.}\ \bibnamefont
  {Chang}}, \bibinfo {author} {\bibfnamefont {E.}~\bibnamefont {Zhakina}},
  \bibinfo {author} {\bibfnamefont {S.}~\bibnamefont {Khim}}, \bibinfo {author}
  {\bibfnamefont {C.~E.}\ \bibnamefont {Dreyer}}, \bibinfo {author}
  {\bibfnamefont {M.}~\bibnamefont {Konczykowski}}, \bibinfo {author}
  {\bibfnamefont {H.}~\bibnamefont {Borrmann}}, \bibinfo {author}
  {\bibfnamefont {P.~J.~W.}\ \bibnamefont {Moll}}, \bibinfo {author}
  {\bibfnamefont {M.}~\bibnamefont {K\"onig}}, \bibinfo {author} {\bibfnamefont
  {D.~A.}\ \bibnamefont {Muller}},\ and\ \bibinfo {author} {\bibfnamefont
  {A.~P.}\ \bibnamefont {Mackenzie}},\ }\bibfield  {title} {\bibinfo {title}
  {Controlled {Introduction} of {Defects} to {Delafossite} {Metals} by
  {Electron} {Irradiation}},\ }\href
  {https://doi.org/10.1103/PhysRevX.10.021018} {\bibfield  {journal} {\bibinfo
  {journal} {Phys. Rev. X}\ }\textbf {\bibinfo {volume} {10}},\ \bibinfo
  {pages} {021018} (\bibinfo {year} {2020})}\BibitemShut {NoStop}%
\bibitem [{\citenamefont {Bobowski}\ \emph {et~al.}(2019)\citenamefont
  {Bobowski}, \citenamefont {Kikugawa}, \citenamefont {Miyoshi}, \citenamefont
  {Suwa}, \citenamefont {Xu}, \citenamefont {Yonezawa}, \citenamefont
  {Sokolov}, \citenamefont {Mackenzie},\ and\ \citenamefont
  {Maeno}}]{bobowski_improved_2019}%
  \BibitemOpen
  \bibfield  {author} {\bibinfo {author} {\bibfnamefont {J.~S.}\ \bibnamefont
  {Bobowski}}, \bibinfo {author} {\bibfnamefont {N.}~\bibnamefont {Kikugawa}},
  \bibinfo {author} {\bibfnamefont {T.}~\bibnamefont {Miyoshi}}, \bibinfo
  {author} {\bibfnamefont {H.}~\bibnamefont {Suwa}}, \bibinfo {author}
  {\bibfnamefont {H.-s.}\ \bibnamefont {Xu}}, \bibinfo {author} {\bibfnamefont
  {S.}~\bibnamefont {Yonezawa}}, \bibinfo {author} {\bibfnamefont {D.~A.}\
  \bibnamefont {Sokolov}}, \bibinfo {author} {\bibfnamefont {A.~P.}\
  \bibnamefont {Mackenzie}},\ and\ \bibinfo {author} {\bibfnamefont
  {Y.}~\bibnamefont {Maeno}},\ }\bibfield  {title} {\bibinfo {title} {Improved
  {Single}-{Crystal} {Growth} of {Sr$_{2}$RuO$_{4}$}},\ }\href
  {https://doi.org/10.3390/condmat4010006} {\bibfield  {journal} {\bibinfo
  {journal} {Condens. Matter}\ }\textbf {\bibinfo {volume} {4}},\ \bibinfo
  {pages} {6} (\bibinfo {year} {2019})}\BibitemShut {NoStop}%
\bibitem [{Sup()}]{SupplementalMaterial}%
  \BibitemOpen
  \href@noop {} {}\bibinfo {note} {See Supplemental Material at [URL will be
  inserted by publisher] for further details on samples, electrical
  characterization of films, and a discussion of the determination of
  $\rho_{0}$ and $T_{\text{c}}$. Includes references \cite{bjorck_genx_2007,
  paglione_elastic_2002, senyshyn_anomalous_2009, vogt_low-temperature_1995,
  wu_electronic_2020, capogna_sensitivity_2002, forsythe_evolution_2002,
  akima_intrinsic_1999, maeno_two-dimensional_1997, hussey_normal-state_1998,
  barber_resistivity_2018}.}\BibitemShut {Stop}%
\bibitem [{\citenamefont {Nair}\ \emph
  {et~al.}(2018{\natexlab{b}})\citenamefont {Nair}, \citenamefont {Liu},
  \citenamefont {Ruf}, \citenamefont {Schreiber}, \citenamefont {Shang},
  \citenamefont {Baek}, \citenamefont {Goodge}, \citenamefont {Kourkoutis},
  \citenamefont {Liu}, \citenamefont {Shen},\ and\ \citenamefont
  {Schlom}}]{nair_synthesis_2018}%
  \BibitemOpen
  \bibfield  {author} {\bibinfo {author} {\bibfnamefont {H.~P.}\ \bibnamefont
  {Nair}}, \bibinfo {author} {\bibfnamefont {Y.}~\bibnamefont {Liu}}, \bibinfo
  {author} {\bibfnamefont {J.~P.}\ \bibnamefont {Ruf}}, \bibinfo {author}
  {\bibfnamefont {N.~J.}\ \bibnamefont {Schreiber}}, \bibinfo {author}
  {\bibfnamefont {S.-L.}\ \bibnamefont {Shang}}, \bibinfo {author}
  {\bibfnamefont {D.~J.}\ \bibnamefont {Baek}}, \bibinfo {author}
  {\bibfnamefont {B.~H.}\ \bibnamefont {Goodge}}, \bibinfo {author}
  {\bibfnamefont {L.~F.}\ \bibnamefont {Kourkoutis}}, \bibinfo {author}
  {\bibfnamefont {Z.-K.}\ \bibnamefont {Liu}}, \bibinfo {author} {\bibfnamefont
  {K.~M.}\ \bibnamefont {Shen}},\ and\ \bibinfo {author} {\bibfnamefont
  {D.~G.}\ \bibnamefont {Schlom}},\ }\bibfield  {title} {\bibinfo {title}
  {Synthesis science of {SrRuO$_{3}$} and {CaRuO$_{3}$} epitaxial films with
  high residual resistivity ratios},\ }\href
  {https://doi.org/10.1063/1.5023477} {\bibfield  {journal} {\bibinfo
  {journal} {APL Mater.}\ }\textbf {\bibinfo {volume} {6}},\ \bibinfo {pages}
  {046101} (\bibinfo {year} {2018}{\natexlab{b}})}\BibitemShut {NoStop}%
\bibitem [{\citenamefont {Walz}\ and\ \citenamefont
  {Lichtenberg}(1993)}]{walz_refinement_1993}%
  \BibitemOpen
  \bibfield  {author} {\bibinfo {author} {\bibfnamefont {L.}~\bibnamefont
  {Walz}}\ and\ \bibinfo {author} {\bibfnamefont {F.}~\bibnamefont
  {Lichtenberg}},\ }\bibfield  {title} {\bibinfo {title} {Refinement of the
  structure of {Sr$_{2}$RuO$_{4}$} with 100 and 295 {K} {X}-ray data},\ }\href
  {https://doi.org/10.1107/S0108270192013143} {\bibfield  {journal} {\bibinfo
  {journal} {Acta Crystallogr. Sect. C}\ }\textbf {\bibinfo {volume} {49}},\
  \bibinfo {pages} {1268} (\bibinfo {year} {1993})}\BibitemShut {NoStop}%
\bibitem [{\citenamefont {Schmidbauer}\ \emph {et~al.}(2012)\citenamefont
  {Schmidbauer}, \citenamefont {Kwasniewski},\ and\ \citenamefont
  {Schwarzkopf}}]{schmidbauer_high-precision_2012}%
  \BibitemOpen
  \bibfield  {author} {\bibinfo {author} {\bibfnamefont {M.}~\bibnamefont
  {Schmidbauer}}, \bibinfo {author} {\bibfnamefont {A.}~\bibnamefont
  {Kwasniewski}},\ and\ \bibinfo {author} {\bibfnamefont {J.}~\bibnamefont
  {Schwarzkopf}},\ }\bibfield  {title} {\bibinfo {title} {High-precision
  absolute lattice parameter determination of {SrTiO$_{3}$}, {DyScO$_{3}$} and
  {NdGaO$_{3}$} single crystals},\ }\href
  {https://doi.org/10.1107/S0108768111046738} {\bibfield  {journal} {\bibinfo
  {journal} {Acta Crystallogr. Sect. B}\ }\textbf {\bibinfo {volume} {68}},\
  \bibinfo {pages} {8} (\bibinfo {year} {2012})}\BibitemShut {NoStop}%
\bibitem [{\citenamefont {Barber}(2017)}]{BarberPhD17}%
  \BibitemOpen
  \bibfield  {author} {\bibinfo {author} {\bibfnamefont {M.~E.}\ \bibnamefont
  {Barber}},\ }\emph {\bibinfo {title} {Uniaxial stress technique and
  investigations into correlated electron systems}},\ \href@noop {} {Ph.D.
  thesis},\ \bibinfo  {school} {University of St. Andrews} (\bibinfo {year}
  {2017})\BibitemShut {NoStop}%
\bibitem [{\citenamefont {Iseler}\ \emph {et~al.}(1966)\citenamefont {Iseler},
  \citenamefont {Dawson}, \citenamefont {Mehner},\ and\ \citenamefont
  {Kauffman}}]{iseler_production_1966}%
  \BibitemOpen
  \bibfield  {author} {\bibinfo {author} {\bibfnamefont {G.~W.}\ \bibnamefont
  {Iseler}}, \bibinfo {author} {\bibfnamefont {H.~I.}\ \bibnamefont {Dawson}},
  \bibinfo {author} {\bibfnamefont {A.~S.}\ \bibnamefont {Mehner}},\ and\
  \bibinfo {author} {\bibfnamefont {J.~W.}\ \bibnamefont {Kauffman}},\
  }\bibfield  {title} {\bibinfo {title} {Production {Rates} of {Electrical}
  {Resistivity} in {Copper} and {Aluminum} {Induced} by {Electron}
  {Irradiation}},\ }\href {https://doi.org/10.1103/PhysRev.146.468} {\bibfield
  {journal} {\bibinfo  {journal} {Phys. Rev.}\ }\textbf {\bibinfo {volume}
  {146}},\ \bibinfo {pages} {468} (\bibinfo {year} {1966})}\BibitemShut
  {NoStop}%
\bibitem [{\citenamefont {Konczykowski}\ and\ \citenamefont
  {Gilchrist}(1991)}]{konczykowski_electron_1991}%
  \BibitemOpen
  \bibfield  {author} {\bibinfo {author} {\bibfnamefont {M.}~\bibnamefont
  {Konczykowski}}\ and\ \bibinfo {author} {\bibfnamefont {J.}~\bibnamefont
  {Gilchrist}},\ }\bibfield  {title} {\bibinfo {title} {Electron irradiation of
  {YBa$_{2}$Cu$_{3}$O$_{7}$} ceramics},\ }\href
  {https://doi.org/10.1051/jp3:1991230} {\bibfield  {journal} {\bibinfo
  {journal} {J. Phys. (Paris) III}\ }\textbf {\bibinfo {volume} {1}},\ \bibinfo
  {pages} {1765} (\bibinfo {year} {1991})}\BibitemShut {NoStop}%
\bibitem [{\citenamefont {Berger}\ \emph {et~al.}(2017)\citenamefont {Berger},
  \citenamefont {Couresy}, \citenamefont {Zucker},\ and\ \citenamefont
  {Chang}}]{nist_tables}%
  \BibitemOpen
  \bibfield  {author} {\bibinfo {author} {\bibfnamefont {M.~J.}\ \bibnamefont
  {Berger}}, \bibinfo {author} {\bibfnamefont {J.~S.}\ \bibnamefont {Couresy}},
  \bibinfo {author} {\bibfnamefont {M.~A.}\ \bibnamefont {Zucker}},\ and\
  \bibinfo {author} {\bibfnamefont {J.}~\bibnamefont {Chang}},\ }\href
  {https://dx.doi.org/10.18434/T4NC7P} {\emph {\bibinfo {title} {NIST Standard
  Reference Database 124}}}\ (\bibinfo  {publisher} {National Institute of
  Standards and Technology},\ \bibinfo {year} {2017})\BibitemShut {NoStop}%
\bibitem [{\citenamefont {Kim}\ \emph {et~al.}(2019)\citenamefont {Kim},
  \citenamefont {Suyolcu}, \citenamefont {Herrero-Martin}, \citenamefont
  {Putzky}, \citenamefont {Nair}, \citenamefont {Ruf}, \citenamefont
  {Schreiber}, \citenamefont {Dietl}, \citenamefont {Christiani}, \citenamefont
  {Logvenov}, \citenamefont {Minola}, \citenamefont {van Aken}, \citenamefont
  {Shen}, \citenamefont {Schlom},\ and\ \citenamefont
  {Keimer}}]{kim_electronic_2019}%
  \BibitemOpen
  \bibfield  {author} {\bibinfo {author} {\bibfnamefont {G.}~\bibnamefont
  {Kim}}, \bibinfo {author} {\bibfnamefont {Y.~E.}\ \bibnamefont {Suyolcu}},
  \bibinfo {author} {\bibfnamefont {J.}~\bibnamefont {Herrero-Martin}},
  \bibinfo {author} {\bibfnamefont {D.}~\bibnamefont {Putzky}}, \bibinfo
  {author} {\bibfnamefont {H.~P.}\ \bibnamefont {Nair}}, \bibinfo {author}
  {\bibfnamefont {J.~P.}\ \bibnamefont {Ruf}}, \bibinfo {author} {\bibfnamefont
  {N.~J.}\ \bibnamefont {Schreiber}}, \bibinfo {author} {\bibfnamefont
  {C.}~\bibnamefont {Dietl}}, \bibinfo {author} {\bibfnamefont
  {G.}~\bibnamefont {Christiani}}, \bibinfo {author} {\bibfnamefont
  {G.}~\bibnamefont {Logvenov}}, \bibinfo {author} {\bibfnamefont
  {M.}~\bibnamefont {Minola}}, \bibinfo {author} {\bibfnamefont {P.~A.}\
  \bibnamefont {van Aken}}, \bibinfo {author} {\bibfnamefont {K.~M.}\
  \bibnamefont {Shen}}, \bibinfo {author} {\bibfnamefont {D.~G.}\ \bibnamefont
  {Schlom}},\ and\ \bibinfo {author} {\bibfnamefont {B.}~\bibnamefont
  {Keimer}},\ }\bibfield  {title} {\bibinfo {title} {Electronic and vibrational
  signatures of ruthenium vacancies in {Sr$_{2}$RuO$_{4}$} thin films},\ }\href
  {https://doi.org/10.1103/PhysRevMaterials.3.094802} {\bibfield  {journal}
  {\bibinfo  {journal} {Phys. Rev. Materials}\ }\textbf {\bibinfo {volume}
  {3}},\ \bibinfo {pages} {094802} (\bibinfo {year} {2019})}\BibitemShut
  {NoStop}%
\bibitem [{\citenamefont {Goodge}\ \emph {et~al.}(2022)\citenamefont {Goodge},
  \citenamefont {Nair}, \citenamefont {Baek}, \citenamefont {Schreiber},
  \citenamefont {Miao}, \citenamefont {Ruf}, \citenamefont {Waite},
  \citenamefont {Carubia}, \citenamefont {Shen}, \citenamefont {Schlom},\ and\
  \citenamefont {Kourkoutis}}]{goodge_disentangling_2022}%
  \BibitemOpen
  \bibfield  {author} {\bibinfo {author} {\bibfnamefont {B.~H.}\ \bibnamefont
  {Goodge}}, \bibinfo {author} {\bibfnamefont {H.~P.}\ \bibnamefont {Nair}},
  \bibinfo {author} {\bibfnamefont {D.~J.}\ \bibnamefont {Baek}}, \bibinfo
  {author} {\bibfnamefont {N.~J.}\ \bibnamefont {Schreiber}}, \bibinfo {author}
  {\bibfnamefont {L.}~\bibnamefont {Miao}}, \bibinfo {author} {\bibfnamefont
  {J.~P.}\ \bibnamefont {Ruf}}, \bibinfo {author} {\bibfnamefont {E.~N.}\
  \bibnamefont {Waite}}, \bibinfo {author} {\bibfnamefont {P.~M.}\ \bibnamefont
  {Carubia}}, \bibinfo {author} {\bibfnamefont {K.~M.}\ \bibnamefont {Shen}},
  \bibinfo {author} {\bibfnamefont {D.~G.}\ \bibnamefont {Schlom}},\ and\
  \bibinfo {author} {\bibfnamefont {L.~F.}\ \bibnamefont {Kourkoutis}},\
  }\bibfield  {title} {\bibinfo {title} {Disentangling types of lattice
  disorder impacting superconductivity in {Sr}$_{2}${Ru}{O}$_{4}$ by
  quantitative local probes},\ }\href {https://doi.org/10.1063/5.0085279}
  {\bibfield  {journal} {\bibinfo  {journal} {APL Mater.}\ }\textbf {\bibinfo
  {volume} {10}},\ \bibinfo {pages} {041114} (\bibinfo {year}
  {2022})}\BibitemShut {NoStop}%
\bibitem [{\citenamefont {Zurbuchen}\ \emph {et~al.}(2001)\citenamefont
  {Zurbuchen}, \citenamefont {Jia}, \citenamefont {Knapp}, \citenamefont
  {Carim}, \citenamefont {Schlom}, \citenamefont {Zou},\ and\ \citenamefont
  {Liu}}]{zurbuchen_suppression_2001}%
  \BibitemOpen
  \bibfield  {author} {\bibinfo {author} {\bibfnamefont {M.~A.}\ \bibnamefont
  {Zurbuchen}}, \bibinfo {author} {\bibfnamefont {Y.}~\bibnamefont {Jia}},
  \bibinfo {author} {\bibfnamefont {S.}~\bibnamefont {Knapp}}, \bibinfo
  {author} {\bibfnamefont {A.~H.}\ \bibnamefont {Carim}}, \bibinfo {author}
  {\bibfnamefont {D.~G.}\ \bibnamefont {Schlom}}, \bibinfo {author}
  {\bibfnamefont {L.-N.}\ \bibnamefont {Zou}},\ and\ \bibinfo {author}
  {\bibfnamefont {Y.}~\bibnamefont {Liu}},\ }\bibfield  {title} {\bibinfo
  {title} {Suppression of superconductivity by crystallographic defects in
  epitaxial {Sr$_{2}$RuO$_{4}$} films},\ }\href
  {https://doi.org/10.1063/1.1364659} {\bibfield  {journal} {\bibinfo
  {journal} {Appl. Phys. Lett.}\ }\textbf {\bibinfo {volume} {78}},\ \bibinfo
  {pages} {2351} (\bibinfo {year} {2001})}\BibitemShut {NoStop}%
\bibitem [{\citenamefont {Zurbuchen}\ \emph {et~al.}(2003)\citenamefont
  {Zurbuchen}, \citenamefont {Jia}, \citenamefont {Knapp}, \citenamefont
  {Carim}, \citenamefont {Schlom},\ and\ \citenamefont
  {Pan}}]{zurbuchen_defect_2003}%
  \BibitemOpen
  \bibfield  {author} {\bibinfo {author} {\bibfnamefont {M.~A.}\ \bibnamefont
  {Zurbuchen}}, \bibinfo {author} {\bibfnamefont {Y.}~\bibnamefont {Jia}},
  \bibinfo {author} {\bibfnamefont {S.}~\bibnamefont {Knapp}}, \bibinfo
  {author} {\bibfnamefont {A.~H.}\ \bibnamefont {Carim}}, \bibinfo {author}
  {\bibfnamefont {D.~G.}\ \bibnamefont {Schlom}},\ and\ \bibinfo {author}
  {\bibfnamefont {X.~Q.}\ \bibnamefont {Pan}},\ }\bibfield  {title} {\bibinfo
  {title} {Defect generation by preferred nucleation in epitaxial
  {Sr$_{2}$RuO$_{4}$}/{LaAlO$_{3}$}},\ }\href
  {https://doi.org/10.1063/1.1624631} {\bibfield  {journal} {\bibinfo
  {journal} {Appl. Phys. Lett.}\ }\textbf {\bibinfo {volume} {83}},\ \bibinfo
  {pages} {3891} (\bibinfo {year} {2003})}\BibitemShut {NoStop}%
\bibitem [{\citenamefont {Zurbuchen}\ \emph {et~al.}(2007)\citenamefont
  {Zurbuchen}, \citenamefont {Tian}, \citenamefont {Pan}, \citenamefont {Fong},
  \citenamefont {Streiffer}, \citenamefont {Hawley}, \citenamefont {Lettieri},
  \citenamefont {Jia}, \citenamefont {Asayama}, \citenamefont {Fulk},
  \citenamefont {Comstock}, \citenamefont {Knapp}, \citenamefont {Carim},\ and\
  \citenamefont {Schlom}}]{zurbuchen_morphology_2007}%
  \BibitemOpen
  \bibfield  {author} {\bibinfo {author} {\bibfnamefont {M.~A.}\ \bibnamefont
  {Zurbuchen}}, \bibinfo {author} {\bibfnamefont {W.}~\bibnamefont {Tian}},
  \bibinfo {author} {\bibfnamefont {X.~Q.}\ \bibnamefont {Pan}}, \bibinfo
  {author} {\bibfnamefont {D.}~\bibnamefont {Fong}}, \bibinfo {author}
  {\bibfnamefont {S.~K.}\ \bibnamefont {Streiffer}}, \bibinfo {author}
  {\bibfnamefont {M.~E.}\ \bibnamefont {Hawley}}, \bibinfo {author}
  {\bibfnamefont {J.}~\bibnamefont {Lettieri}}, \bibinfo {author}
  {\bibfnamefont {Y.}~\bibnamefont {Jia}}, \bibinfo {author} {\bibfnamefont
  {G.}~\bibnamefont {Asayama}}, \bibinfo {author} {\bibfnamefont {S.~J.}\
  \bibnamefont {Fulk}}, \bibinfo {author} {\bibfnamefont {D.~J.}\ \bibnamefont
  {Comstock}}, \bibinfo {author} {\bibfnamefont {S.}~\bibnamefont {Knapp}},
  \bibinfo {author} {\bibfnamefont {A.~H.}\ \bibnamefont {Carim}},\ and\
  \bibinfo {author} {\bibfnamefont {D.~G.}\ \bibnamefont {Schlom}},\ }\bibfield
   {title} {\bibinfo {title} {Morphology, structure, and nucleation of
  out-of-phase boundaries ({OPBs}) in epitaxial films of layered oxides},\
  }\href {https://doi.org/10.1557/JMR.2007.0198} {\bibfield  {journal}
  {\bibinfo  {journal} {J. Mater. Res.}\ }\textbf {\bibinfo {volume} {22}},\
  \bibinfo {pages} {1439} (\bibinfo {year} {2007})}\BibitemShut {NoStop}%
\bibitem [{\citenamefont {Steppke}\ \emph {et~al.}(2017)\citenamefont
  {Steppke}, \citenamefont {Zhao}, \citenamefont {Barber}, \citenamefont
  {Scaffidi}, \citenamefont {Jerzembeck}, \citenamefont {Rosner}, \citenamefont
  {Gibbs}, \citenamefont {Maeno}, \citenamefont {Simon}, \citenamefont
  {Mackenzie},\ and\ \citenamefont {Hicks}}]{steppke_strong_2017}%
  \BibitemOpen
  \bibfield  {author} {\bibinfo {author} {\bibfnamefont {A.}~\bibnamefont
  {Steppke}}, \bibinfo {author} {\bibfnamefont {L.}~\bibnamefont {Zhao}},
  \bibinfo {author} {\bibfnamefont {M.~E.}\ \bibnamefont {Barber}}, \bibinfo
  {author} {\bibfnamefont {T.}~\bibnamefont {Scaffidi}}, \bibinfo {author}
  {\bibfnamefont {F.}~\bibnamefont {Jerzembeck}}, \bibinfo {author}
  {\bibfnamefont {H.}~\bibnamefont {Rosner}}, \bibinfo {author} {\bibfnamefont
  {A.~S.}\ \bibnamefont {Gibbs}}, \bibinfo {author} {\bibfnamefont
  {Y.}~\bibnamefont {Maeno}}, \bibinfo {author} {\bibfnamefont {S.~H.}\
  \bibnamefont {Simon}}, \bibinfo {author} {\bibfnamefont {A.~P.}\ \bibnamefont
  {Mackenzie}},\ and\ \bibinfo {author} {\bibfnamefont {C.~W.}\ \bibnamefont
  {Hicks}},\ }\bibfield  {title} {\bibinfo {title} {Strong peak in {$T_{c}$} of
  {Sr$_{2}$RuO$_{4}$} under uniaxial pressure},\ }\href
  {https://doi.org/10.1126/science.aaf9398} {\bibfield  {journal} {\bibinfo
  {journal} {Science}\ }\textbf {\bibinfo {volume} {355}},\ \bibinfo {pages}
  {eaaf9398} (\bibinfo {year} {2017})}\BibitemShut {NoStop}%
\bibitem [{\citenamefont {Mackenzie}(1993)}]{mackenzie_recent_1993}%
  \BibitemOpen
  \bibfield  {author} {\bibinfo {author} {\bibfnamefont {A.~P.}\ \bibnamefont
  {Mackenzie}},\ }\bibfield  {title} {\bibinfo {title} {Recent progress in
  electron probe microanalysis},\ }\href
  {https://doi.org/10.1088/0034-4885/56/4/002} {\bibfield  {journal} {\bibinfo
  {journal} {Rep. Prog. Phys.}\ }\textbf {\bibinfo {volume} {56}},\ \bibinfo
  {pages} {557} (\bibinfo {year} {1993})}\BibitemShut {NoStop}%
\bibitem [{\citenamefont {NishiZaki}\ \emph {et~al.}(1999)\citenamefont
  {NishiZaki}, \citenamefont {Maeno},\ and\ \citenamefont
  {Mao}}]{nishizaki_effect_1999}%
  \BibitemOpen
  \bibfield  {author} {\bibinfo {author} {\bibfnamefont {S.}~\bibnamefont
  {NishiZaki}}, \bibinfo {author} {\bibfnamefont {Y.}~\bibnamefont {Maeno}},\
  and\ \bibinfo {author} {\bibfnamefont {Z.}~\bibnamefont {Mao}},\ }\bibfield
  {title} {\bibinfo {title} {Effect of {Impurities} on the {Specific} {Heat} of
  the {Spin}-{Triplet} {Superconductor} {Sr$_{2}$RuO$_{4}$}},\ }\href
  {https://doi.org/10.1023/A:1022551313401} {\bibfield  {journal} {\bibinfo
  {journal} {J. Low Temp. Phys.}\ }\textbf {\bibinfo {volume} {117}},\ \bibinfo
  {pages} {1581} (\bibinfo {year} {1999})}\BibitemShut {NoStop}%
\bibitem [{\citenamefont {Li}\ \emph {et~al.}(2021)\citenamefont {Li},
  \citenamefont {Kikugawa}, \citenamefont {Sokolov}, \citenamefont
  {Jerzembeck}, \citenamefont {Gibbs}, \citenamefont {Maeno}, \citenamefont
  {Hicks}, \citenamefont {Schmalian}, \citenamefont {Nicklas},\ and\
  \citenamefont {Mackenzie}}]{li_high-sensitivity_2021}%
  \BibitemOpen
  \bibfield  {author} {\bibinfo {author} {\bibfnamefont {Y.-S.}\ \bibnamefont
  {Li}}, \bibinfo {author} {\bibfnamefont {N.}~\bibnamefont {Kikugawa}},
  \bibinfo {author} {\bibfnamefont {D.~A.}\ \bibnamefont {Sokolov}}, \bibinfo
  {author} {\bibfnamefont {F.}~\bibnamefont {Jerzembeck}}, \bibinfo {author}
  {\bibfnamefont {A.~S.}\ \bibnamefont {Gibbs}}, \bibinfo {author}
  {\bibfnamefont {Y.}~\bibnamefont {Maeno}}, \bibinfo {author} {\bibfnamefont
  {C.~W.}\ \bibnamefont {Hicks}}, \bibinfo {author} {\bibfnamefont
  {J.}~\bibnamefont {Schmalian}}, \bibinfo {author} {\bibfnamefont
  {M.}~\bibnamefont {Nicklas}},\ and\ \bibinfo {author} {\bibfnamefont {A.~P.}\
  \bibnamefont {Mackenzie}},\ }\bibfield  {title} {\bibinfo {title}
  {High-sensitivity heat-capacity measurements on {Sr$_{2}$RuO$_{4}$} under
  uniaxial pressure},\ }\href {https://doi.org/10.1073/pnas.2020492118}
  {\bibfield  {journal} {\bibinfo  {journal} {Proc. Natl. Acad. Sci. USA}\
  }\textbf {\bibinfo {volume} {118}},\ \bibinfo {pages} {e2020492118} (\bibinfo
  {year} {2021})}\BibitemShut {NoStop}%
\bibitem [{\citenamefont {Li}\ \emph {et~al.}(2022)\citenamefont {Li},
  \citenamefont {Garst}, \citenamefont {Schmalian}, \citenamefont {Ghosh},
  \citenamefont {Kikugawa}, \citenamefont {Sokolov}, \citenamefont {Hicks},
  \citenamefont {Jerzembeck}, \citenamefont {Ikeda}, \citenamefont {Hu},
  \citenamefont {Ramshaw}, \citenamefont {Rost}, \citenamefont {Nicklas},\ and\
  \citenamefont {Mackenzie}}]{li_elastocaloric_2022}%
  \BibitemOpen
  \bibfield  {author} {\bibinfo {author} {\bibfnamefont {Y.-S.}\ \bibnamefont
  {Li}}, \bibinfo {author} {\bibfnamefont {M.}~\bibnamefont {Garst}}, \bibinfo
  {author} {\bibfnamefont {J.}~\bibnamefont {Schmalian}}, \bibinfo {author}
  {\bibfnamefont {S.}~\bibnamefont {Ghosh}}, \bibinfo {author} {\bibfnamefont
  {N.}~\bibnamefont {Kikugawa}}, \bibinfo {author} {\bibfnamefont {D.~A.}\
  \bibnamefont {Sokolov}}, \bibinfo {author} {\bibfnamefont {C.~W.}\
  \bibnamefont {Hicks}}, \bibinfo {author} {\bibfnamefont {F.}~\bibnamefont
  {Jerzembeck}}, \bibinfo {author} {\bibfnamefont {M.~S.}\ \bibnamefont
  {Ikeda}}, \bibinfo {author} {\bibfnamefont {Z.}~\bibnamefont {Hu}}, \bibinfo
  {author} {\bibfnamefont {B.~J.}\ \bibnamefont {Ramshaw}}, \bibinfo {author}
  {\bibfnamefont {A.~W.}\ \bibnamefont {Rost}}, \bibinfo {author}
  {\bibfnamefont {M.}~\bibnamefont {Nicklas}},\ and\ \bibinfo {author}
  {\bibfnamefont {A.~P.}\ \bibnamefont {Mackenzie}},\ }\bibfield  {title}
  {\bibinfo {title} {Elastocaloric determination of the phase diagram of
  {Sr$_{2}$RuO$_{4}$}},\ }\href {https://doi.org/10.1038/s41586-022-04820-z}
  {\bibfield  {journal} {\bibinfo  {journal} {Nature}\ }\textbf {\bibinfo
  {volume} {607}},\ \bibinfo {pages} {276} (\bibinfo {year}
  {2022})}\BibitemShut {NoStop}%
\bibitem [{\citenamefont {Ghosh}\ \emph {et~al.}(2021)\citenamefont {Ghosh},
  \citenamefont {Shekhter}, \citenamefont {Jerzembeck}, \citenamefont
  {Kikugawa}, \citenamefont {Sokolov}, \citenamefont {Brando}, \citenamefont
  {Mackenzie}, \citenamefont {Hicks},\ and\ \citenamefont
  {Ramshaw}}]{ghosh_thermodynamic_2021}%
  \BibitemOpen
  \bibfield  {author} {\bibinfo {author} {\bibfnamefont {S.}~\bibnamefont
  {Ghosh}}, \bibinfo {author} {\bibfnamefont {A.}~\bibnamefont {Shekhter}},
  \bibinfo {author} {\bibfnamefont {F.}~\bibnamefont {Jerzembeck}}, \bibinfo
  {author} {\bibfnamefont {N.}~\bibnamefont {Kikugawa}}, \bibinfo {author}
  {\bibfnamefont {D.~A.}\ \bibnamefont {Sokolov}}, \bibinfo {author}
  {\bibfnamefont {M.}~\bibnamefont {Brando}}, \bibinfo {author} {\bibfnamefont
  {A.~P.}\ \bibnamefont {Mackenzie}}, \bibinfo {author} {\bibfnamefont {C.~W.}\
  \bibnamefont {Hicks}},\ and\ \bibinfo {author} {\bibfnamefont {B.~J.}\
  \bibnamefont {Ramshaw}},\ }\bibfield  {title} {\bibinfo {title}
  {Thermodynamic evidence for a two-component superconducting order parameter
  in {Sr$_{2}$RuO$_{4}$}},\ }\href {https://doi.org/10.1038/s41567-020-1032-4}
  {\bibfield  {journal} {\bibinfo  {journal} {Nat. Phys.}\ }\textbf {\bibinfo
  {volume} {17}},\ \bibinfo {pages} {199} (\bibinfo {year} {2021})}\BibitemShut
  {NoStop}%
\bibitem [{\citenamefont {Benhabib}\ \emph {et~al.}(2021)\citenamefont
  {Benhabib}, \citenamefont {Lupien}, \citenamefont {Paul}, \citenamefont
  {Berges}, \citenamefont {Dion}, \citenamefont {Nardone}, \citenamefont
  {Zitouni}, \citenamefont {Mao}, \citenamefont {Maeno}, \citenamefont
  {Georges}, \citenamefont {Taillefer},\ and\ \citenamefont
  {Proust}}]{benhabib_ultrasound_2021}%
  \BibitemOpen
  \bibfield  {author} {\bibinfo {author} {\bibfnamefont {S.}~\bibnamefont
  {Benhabib}}, \bibinfo {author} {\bibfnamefont {C.}~\bibnamefont {Lupien}},
  \bibinfo {author} {\bibfnamefont {I.}~\bibnamefont {Paul}}, \bibinfo {author}
  {\bibfnamefont {L.}~\bibnamefont {Berges}}, \bibinfo {author} {\bibfnamefont
  {M.}~\bibnamefont {Dion}}, \bibinfo {author} {\bibfnamefont {M.}~\bibnamefont
  {Nardone}}, \bibinfo {author} {\bibfnamefont {A.}~\bibnamefont {Zitouni}},
  \bibinfo {author} {\bibfnamefont {Z.~Q.}\ \bibnamefont {Mao}}, \bibinfo
  {author} {\bibfnamefont {Y.}~\bibnamefont {Maeno}}, \bibinfo {author}
  {\bibfnamefont {A.}~\bibnamefont {Georges}}, \bibinfo {author} {\bibfnamefont
  {L.}~\bibnamefont {Taillefer}},\ and\ \bibinfo {author} {\bibfnamefont
  {C.}~\bibnamefont {Proust}},\ }\bibfield  {title} {\bibinfo {title}
  {Ultrasound evidence for a two-component superconducting order parameter in
  {Sr$_{2}$RuO$_{4}$}},\ }\href {https://doi.org/10.1038/s41567-020-1033-3}
  {\bibfield  {journal} {\bibinfo  {journal} {Nat. Phys.}\ }\textbf {\bibinfo
  {volume} {17}},\ \bibinfo {pages} {194} (\bibinfo {year} {2021})}\BibitemShut
  {NoStop}%
\bibitem [{\citenamefont {Kivelson}\ \emph {et~al.}(2020)\citenamefont
  {Kivelson}, \citenamefont {Yuan}, \citenamefont {Ramshaw},\ and\
  \citenamefont {Thomale}}]{kivelson_proposal_2020}%
  \BibitemOpen
  \bibfield  {author} {\bibinfo {author} {\bibfnamefont {S.~A.}\ \bibnamefont
  {Kivelson}}, \bibinfo {author} {\bibfnamefont {A.~C.}\ \bibnamefont {Yuan}},
  \bibinfo {author} {\bibfnamefont {B.}~\bibnamefont {Ramshaw}},\ and\ \bibinfo
  {author} {\bibfnamefont {R.}~\bibnamefont {Thomale}},\ }\bibfield  {title}
  {\bibinfo {title} {A proposal for reconciling diverse experiments on the
  superconducting state in {Sr$_{2}$RuO$_{4}$}},\ }\href
  {https://doi.org/10.1038/s41535-020-0245-1} {\bibfield  {journal} {\bibinfo
  {journal} {npj Quantum Mater.}\ }\textbf {\bibinfo {volume} {5}},\ \bibinfo
  {pages} {1} (\bibinfo {year} {2020})}\BibitemShut {NoStop}%
\bibitem [{\citenamefont {Luke}\ \emph {et~al.}(1998)\citenamefont {Luke},
  \citenamefont {Fudamoto}, \citenamefont {Kojima}, \citenamefont {Larkin},
  \citenamefont {Merrin}, \citenamefont {Nachumi}, \citenamefont {Uemura},
  \citenamefont {Maeno}, \citenamefont {Mao}, \citenamefont {Mori},
  \citenamefont {Nakamura},\ and\ \citenamefont
  {Sigrist}}]{luke_time-reversal_1998}%
  \BibitemOpen
  \bibfield  {author} {\bibinfo {author} {\bibfnamefont {G.~M.}\ \bibnamefont
  {Luke}}, \bibinfo {author} {\bibfnamefont {Y.}~\bibnamefont {Fudamoto}},
  \bibinfo {author} {\bibfnamefont {K.~M.}\ \bibnamefont {Kojima}}, \bibinfo
  {author} {\bibfnamefont {M.~I.}\ \bibnamefont {Larkin}}, \bibinfo {author}
  {\bibfnamefont {J.}~\bibnamefont {Merrin}}, \bibinfo {author} {\bibfnamefont
  {B.}~\bibnamefont {Nachumi}}, \bibinfo {author} {\bibfnamefont {Y.~J.}\
  \bibnamefont {Uemura}}, \bibinfo {author} {\bibfnamefont {Y.}~\bibnamefont
  {Maeno}}, \bibinfo {author} {\bibfnamefont {Z.~Q.}\ \bibnamefont {Mao}},
  \bibinfo {author} {\bibfnamefont {Y.}~\bibnamefont {Mori}}, \bibinfo {author}
  {\bibfnamefont {H.}~\bibnamefont {Nakamura}},\ and\ \bibinfo {author}
  {\bibfnamefont {M.}~\bibnamefont {Sigrist}},\ }\bibfield  {title} {\bibinfo
  {title} {Time-reversal symmetry-breaking superconductivity in
  {Sr$_{2}$RuO$_{4}$}},\ }\href {https://doi.org/10.1038/29038} {\bibfield
  {journal} {\bibinfo  {journal} {Nature}\ }\textbf {\bibinfo {volume} {394}},\
  \bibinfo {pages} {558} (\bibinfo {year} {1998})}\BibitemShut {NoStop}%
\bibitem [{\citenamefont {Grinenko}\ \emph
  {et~al.}(2021{\natexlab{a}})\citenamefont {Grinenko}, \citenamefont {Ghosh},
  \citenamefont {Sarkar}, \citenamefont {Orain}, \citenamefont {Nikitin},
  \citenamefont {Elender}, \citenamefont {Das}, \citenamefont {Guguchia},
  \citenamefont {Br\"{u}ckner}, \citenamefont {Barber}, \citenamefont {Park},
  \citenamefont {Kikugawa}, \citenamefont {Sokolov}, \citenamefont {Bobowski},
  \citenamefont {Miyoshi}, \citenamefont {Maeno}, \citenamefont {Mackenzie},
  \citenamefont {Luetkens}, \citenamefont {Hicks},\ and\ \citenamefont
  {Klauss}}]{grinenko_split_2021}%
  \BibitemOpen
  \bibfield  {author} {\bibinfo {author} {\bibfnamefont {V.}~\bibnamefont
  {Grinenko}}, \bibinfo {author} {\bibfnamefont {S.}~\bibnamefont {Ghosh}},
  \bibinfo {author} {\bibfnamefont {R.}~\bibnamefont {Sarkar}}, \bibinfo
  {author} {\bibfnamefont {J.-C.}\ \bibnamefont {Orain}}, \bibinfo {author}
  {\bibfnamefont {A.}~\bibnamefont {Nikitin}}, \bibinfo {author} {\bibfnamefont
  {M.}~\bibnamefont {Elender}}, \bibinfo {author} {\bibfnamefont
  {D.}~\bibnamefont {Das}}, \bibinfo {author} {\bibfnamefont {Z.}~\bibnamefont
  {Guguchia}}, \bibinfo {author} {\bibfnamefont {F.}~\bibnamefont
  {Br\"{u}ckner}}, \bibinfo {author} {\bibfnamefont {M.~E.}\ \bibnamefont
  {Barber}}, \bibinfo {author} {\bibfnamefont {J.}~\bibnamefont {Park}},
  \bibinfo {author} {\bibfnamefont {N.}~\bibnamefont {Kikugawa}}, \bibinfo
  {author} {\bibfnamefont {D.~A.}\ \bibnamefont {Sokolov}}, \bibinfo {author}
  {\bibfnamefont {J.~S.}\ \bibnamefont {Bobowski}}, \bibinfo {author}
  {\bibfnamefont {T.}~\bibnamefont {Miyoshi}}, \bibinfo {author} {\bibfnamefont
  {Y.}~\bibnamefont {Maeno}}, \bibinfo {author} {\bibfnamefont {A.~P.}\
  \bibnamefont {Mackenzie}}, \bibinfo {author} {\bibfnamefont {H.}~\bibnamefont
  {Luetkens}}, \bibinfo {author} {\bibfnamefont {C.~W.}\ \bibnamefont
  {Hicks}},\ and\ \bibinfo {author} {\bibfnamefont {H.-H.}\ \bibnamefont
  {Klauss}},\ }\bibfield  {title} {\bibinfo {title} {Split superconducting and
  time-reversal symmetry-breaking transitions in {Sr$_{2}$RuO$_{4}$} under
  stress},\ }\href {https://doi.org/10.1038/s41567-021-01182-7} {\bibfield
  {journal} {\bibinfo  {journal} {Nat. Phys.}\ }\textbf {\bibinfo {volume}
  {17}},\ \bibinfo {pages} {748} (\bibinfo {year}
  {2021}{\natexlab{a}})}\BibitemShut {NoStop}%
\bibitem [{\citenamefont {Grinenko}\ \emph
  {et~al.}(2021{\natexlab{b}})\citenamefont {Grinenko}, \citenamefont {Das},
  \citenamefont {Gupta}, \citenamefont {Zinkl}, \citenamefont {Kikugawa},
  \citenamefont {Maeno}, \citenamefont {Hicks}, \citenamefont {Klauss},
  \citenamefont {Sigrist},\ and\ \citenamefont
  {Khasanov}}]{grinenko_unsplit_2021}%
  \BibitemOpen
  \bibfield  {author} {\bibinfo {author} {\bibfnamefont {V.}~\bibnamefont
  {Grinenko}}, \bibinfo {author} {\bibfnamefont {D.}~\bibnamefont {Das}},
  \bibinfo {author} {\bibfnamefont {R.}~\bibnamefont {Gupta}}, \bibinfo
  {author} {\bibfnamefont {B.}~\bibnamefont {Zinkl}}, \bibinfo {author}
  {\bibfnamefont {N.}~\bibnamefont {Kikugawa}}, \bibinfo {author}
  {\bibfnamefont {Y.}~\bibnamefont {Maeno}}, \bibinfo {author} {\bibfnamefont
  {C.~W.}\ \bibnamefont {Hicks}}, \bibinfo {author} {\bibfnamefont {H.-H.}\
  \bibnamefont {Klauss}}, \bibinfo {author} {\bibfnamefont {M.}~\bibnamefont
  {Sigrist}},\ and\ \bibinfo {author} {\bibfnamefont {R.}~\bibnamefont
  {Khasanov}},\ }\bibfield  {title} {\bibinfo {title} {Unsplit superconducting
  and time reversal symmetry breaking transitions in {Sr$_{2}$RuO$_{4}$} under
  hydrostatic pressure and disorder},\ }\href
  {https://doi.org/10.1038/s41467-021-24176-8} {\bibfield  {journal} {\bibinfo
  {journal} {Nat. Commun.}\ }\textbf {\bibinfo {volume} {12}},\ \bibinfo
  {pages} {3920} (\bibinfo {year} {2021}{\natexlab{b}})}\BibitemShut {NoStop}%
\bibitem [{\citenamefont {Xia}\ \emph {et~al.}(2006)\citenamefont {Xia},
  \citenamefont {Maeno}, \citenamefont {Beyersdorf}, \citenamefont {Fejer},\
  and\ \citenamefont {Kapitulnik}}]{xia_high_2006}%
  \BibitemOpen
  \bibfield  {author} {\bibinfo {author} {\bibfnamefont {J.}~\bibnamefont
  {Xia}}, \bibinfo {author} {\bibfnamefont {Y.}~\bibnamefont {Maeno}}, \bibinfo
  {author} {\bibfnamefont {P.~T.}\ \bibnamefont {Beyersdorf}}, \bibinfo
  {author} {\bibfnamefont {M.~M.}\ \bibnamefont {Fejer}},\ and\ \bibinfo
  {author} {\bibfnamefont {A.}~\bibnamefont {Kapitulnik}},\ }\bibfield  {title}
  {\bibinfo {title} {High {Resolution} {Polar} {Kerr} {Effect} {Measurements}
  of {Sr$_{2}$RuO$_{4}$}: {Evidence} for {Broken} {Time}-{Reversal} {Symmetry}
  in the {Superconducting} {State}},\ }\href
  {https://doi.org/10.1103/PhysRevLett.97.167002} {\bibfield  {journal}
  {\bibinfo  {journal} {Phys. Rev. Lett.}\ }\textbf {\bibinfo {volume} {97}},\
  \bibinfo {pages} {167002} (\bibinfo {year} {2006})}\BibitemShut {NoStop}%
\bibitem [{\citenamefont {Tolpygo}\ \emph
  {et~al.}(1996{\natexlab{a}})\citenamefont {Tolpygo}, \citenamefont {Lin},
  \citenamefont {Gurvitch}, \citenamefont {Hou},\ and\ \citenamefont
  {Phillips}}]{tolpygo_effect_1996}%
  \BibitemOpen
  \bibfield  {author} {\bibinfo {author} {\bibfnamefont {S.~K.}\ \bibnamefont
  {Tolpygo}}, \bibinfo {author} {\bibfnamefont {J.-Y.}\ \bibnamefont {Lin}},
  \bibinfo {author} {\bibfnamefont {M.}~\bibnamefont {Gurvitch}}, \bibinfo
  {author} {\bibfnamefont {S.~Y.}\ \bibnamefont {Hou}},\ and\ \bibinfo {author}
  {\bibfnamefont {J.~M.}\ \bibnamefont {Phillips}},\ }\bibfield  {title}
  {\bibinfo {title} {Effect of oxygen defects on transport properties and
  {$T_{c}$} of {YBa$_{2}$Cu$_{3}$O$_{6+x}$}: {Displacement} energy for plane
  and chain oxygen and implications for irradiation-induced resistivity and
  {$T_{c}$} suppression},\ }\href {https://doi.org/10.1103/PhysRevB.53.12462}
  {\bibfield  {journal} {\bibinfo  {journal} {Phys. Rev. B}\ }\textbf {\bibinfo
  {volume} {53}},\ \bibinfo {pages} {12462} (\bibinfo {year}
  {1996}{\natexlab{a}})}\BibitemShut {NoStop}%
\bibitem [{\citenamefont {Tolpygo}\ \emph
  {et~al.}(1996{\natexlab{b}})\citenamefont {Tolpygo}, \citenamefont {Lin},
  \citenamefont {Gurvitch}, \citenamefont {Hou},\ and\ \citenamefont
  {Phillips}}]{tolpygo_universal_1996}%
  \BibitemOpen
  \bibfield  {author} {\bibinfo {author} {\bibfnamefont {S.~K.}\ \bibnamefont
  {Tolpygo}}, \bibinfo {author} {\bibfnamefont {J.-Y.}\ \bibnamefont {Lin}},
  \bibinfo {author} {\bibfnamefont {M.}~\bibnamefont {Gurvitch}}, \bibinfo
  {author} {\bibfnamefont {S.~Y.}\ \bibnamefont {Hou}},\ and\ \bibinfo {author}
  {\bibfnamefont {J.~M.}\ \bibnamefont {Phillips}},\ }\bibfield  {title}
  {\bibinfo {title} {Universal {$T_{c}$} suppression by in-plane defects in
  high-temperature superconductors: {Implications} for pairing symmetry},\
  }\href {https://doi.org/10.1103/PhysRevB.53.12454} {\bibfield  {journal}
  {\bibinfo  {journal} {Phys. Rev. B}\ }\textbf {\bibinfo {volume} {53}},\
  \bibinfo {pages} {12454} (\bibinfo {year} {1996}{\natexlab{b}})}\BibitemShut
  {NoStop}%
\bibitem [{Note1()}]{Note1}%
  \BibitemOpen
  \bibinfo {note} {Dataset for Ruf \protect \textit {et al.} ``Controllable
  suppression of the unconventional superconductivity in Sr$_{2}$RuO$_{4}$ via
  high-energy electron irradiation'' (2024); DOI to be inserted
  here}\BibitemShut {NoStop}%
\bibitem [{\citenamefont {Klein}\ \emph {et~al.}(2001)\citenamefont {Klein},
  \citenamefont {Kats}, \citenamefont {Wiser}, \citenamefont {Konczykowski},
  \citenamefont {Reiner}, \citenamefont {Geballe}, \citenamefont {Beasley},\
  and\ \citenamefont {Kapitulnik}}]{klein_negative_2001}%
  \BibitemOpen
  \bibfield  {author} {\bibinfo {author} {\bibfnamefont {L.}~\bibnamefont
  {Klein}}, \bibinfo {author} {\bibfnamefont {Y.}~\bibnamefont {Kats}},
  \bibinfo {author} {\bibfnamefont {N.}~\bibnamefont {Wiser}}, \bibinfo
  {author} {\bibfnamefont {M.}~\bibnamefont {Konczykowski}}, \bibinfo {author}
  {\bibfnamefont {J.~W.}\ \bibnamefont {Reiner}}, \bibinfo {author}
  {\bibfnamefont {T.~H.}\ \bibnamefont {Geballe}}, \bibinfo {author}
  {\bibfnamefont {M.~R.}\ \bibnamefont {Beasley}},\ and\ \bibinfo {author}
  {\bibfnamefont {A.}~\bibnamefont {Kapitulnik}},\ }\bibfield  {title}
  {\bibinfo {title} {Negative deviations from {Matthiessen}'s rule for
  {SrRuO$_{3}$} and {CaRuO$_{3}$}},\ }\href
  {https://doi.org/10.1209/epl/i2001-00448-8} {\bibfield  {journal} {\bibinfo
  {journal} {Europhys. Lett.}\ }\textbf {\bibinfo {volume} {55}},\ \bibinfo
  {pages} {532} (\bibinfo {year} {2001})}\BibitemShut {NoStop}%
\bibitem [{\citenamefont {Haham}\ \emph {et~al.}(2013)\citenamefont {Haham},
  \citenamefont {Konczykowski}, \citenamefont {Kuiper}, \citenamefont
  {Koster},\ and\ \citenamefont {Klein}}]{haham_testing_2013}%
  \BibitemOpen
  \bibfield  {author} {\bibinfo {author} {\bibfnamefont {N.}~\bibnamefont
  {Haham}}, \bibinfo {author} {\bibfnamefont {M.}~\bibnamefont {Konczykowski}},
  \bibinfo {author} {\bibfnamefont {B.}~\bibnamefont {Kuiper}}, \bibinfo
  {author} {\bibfnamefont {G.}~\bibnamefont {Koster}},\ and\ \bibinfo {author}
  {\bibfnamefont {L.}~\bibnamefont {Klein}},\ }\bibfield  {title} {\bibinfo
  {title} {Testing dependence of anomalous {Hall} effect on resistivity in
  {SrRuO$_{3}$} by its increase with electron irradiation},\ }\href
  {https://doi.org/10.1103/PhysRevB.88.214431} {\bibfield  {journal} {\bibinfo
  {journal} {Phys. Rev. B}\ }\textbf {\bibinfo {volume} {88}},\ \bibinfo
  {pages} {214431} (\bibinfo {year} {2013})}\BibitemShut {NoStop}%
\bibitem [{\citenamefont {Legris}\ \emph {et~al.}(1993)\citenamefont {Legris},
  \citenamefont {Rullier-Albenque}, \citenamefont {Radeva},\ and\ \citenamefont
  {Lejay}}]{legris_effects_1993}%
  \BibitemOpen
  \bibfield  {author} {\bibinfo {author} {\bibfnamefont {A.}~\bibnamefont
  {Legris}}, \bibinfo {author} {\bibfnamefont {F.}~\bibnamefont
  {Rullier-Albenque}}, \bibinfo {author} {\bibfnamefont {E.}~\bibnamefont
  {Radeva}},\ and\ \bibinfo {author} {\bibfnamefont {P.}~\bibnamefont
  {Lejay}},\ }\bibfield  {title} {\bibinfo {title} {Effects of electron
  irradiation on {YBa$_{2}$Cu$_{3}$O$_{7-\delta}$} superconductor},\ }\href
  {https://doi.org/10.1051/jp1:1993203} {\bibfield  {journal} {\bibinfo
  {journal} {J. Phys. (Paris) I}\ }\textbf {\bibinfo {volume} {3}},\ \bibinfo
  {pages} {1605} (\bibinfo {year} {1993})}\BibitemShut {NoStop}%
\bibitem [{\citenamefont {Lijian}\ \emph {et~al.}(1995)\citenamefont {Lijian},
  \citenamefont {Qing},\ and\ \citenamefont
  {Xhengming}}]{lijian_analytic_1995}%
  \BibitemOpen
  \bibfield  {author} {\bibinfo {author} {\bibfnamefont {T.}~\bibnamefont
  {Lijian}}, \bibinfo {author} {\bibfnamefont {H.}~\bibnamefont {Qing}},\ and\
  \bibinfo {author} {\bibfnamefont {L.}~\bibnamefont {Xhengming}},\ }\bibfield
  {title} {\bibinfo {title} {Analytic {F}itting to the {M}ott {C}ross {S}ection
  of {E}lectrons},\ }\href {https://doi.org/10.1016/0969-806X(94)00063-8}
  {\bibfield  {journal} {\bibinfo  {journal} {Radiat. Phys. Chem.}\ }\textbf
  {\bibinfo {volume} {45}},\ \bibinfo {pages} {235} (\bibinfo {year}
  {1995})}\BibitemShut {NoStop}%
\bibitem [{\citenamefont {Boschini}\ \emph {et~al.}(2013)\citenamefont
  {Boschini}, \citenamefont {Consolandi}, \citenamefont {Gervasi},
  \citenamefont {Giani}, \citenamefont {Grandi}, \citenamefont {Ivanchenko},
  \citenamefont {Nieminem}, \citenamefont {Pensotti}, \citenamefont
  {Rancoita},\ and\ \citenamefont {Tacconi}}]{boschini_expression_2013}%
  \BibitemOpen
  \bibfield  {author} {\bibinfo {author} {\bibfnamefont {M.~J.}\ \bibnamefont
  {Boschini}}, \bibinfo {author} {\bibfnamefont {C.}~\bibnamefont
  {Consolandi}}, \bibinfo {author} {\bibfnamefont {M.}~\bibnamefont {Gervasi}},
  \bibinfo {author} {\bibfnamefont {S.}~\bibnamefont {Giani}}, \bibinfo
  {author} {\bibfnamefont {D.}~\bibnamefont {Grandi}}, \bibinfo {author}
  {\bibfnamefont {V.}~\bibnamefont {Ivanchenko}}, \bibinfo {author}
  {\bibfnamefont {P.}~\bibnamefont {Nieminem}}, \bibinfo {author}
  {\bibfnamefont {S.}~\bibnamefont {Pensotti}}, \bibinfo {author}
  {\bibfnamefont {P.~G.}\ \bibnamefont {Rancoita}},\ and\ \bibinfo {author}
  {\bibfnamefont {M.}~\bibnamefont {Tacconi}},\ }\bibfield  {title} {\bibinfo
  {title} {An expression for the {M}ott cross section for electrons and
  positrons on nuclei with {$Z$} up to 118},\ }\href
  {https://doi.org/10.1016/j.radphyschem.2013.04.020} {\bibfield  {journal}
  {\bibinfo  {journal} {Radiat. Phys. Chem.}\ }\textbf {\bibinfo {volume}
  {90}},\ \bibinfo {pages} {39} (\bibinfo {year} {2013})}\BibitemShut {NoStop}%
\bibitem [{\citenamefont {Tinkham}(1963)}]{tinkham_effect_1963}%
  \BibitemOpen
  \bibfield  {author} {\bibinfo {author} {\bibfnamefont {M.}~\bibnamefont
  {Tinkham}},\ }\bibfield  {title} {\bibinfo {title} {Effect of {Fluxoid}
  {Quantization} on {Transitions} of {Superconducting} {Films}},\ }\href
  {https://doi.org/10.1103/PhysRev.129.2413} {\bibfield  {journal} {\bibinfo
  {journal} {Phys. Rev.}\ }\textbf {\bibinfo {volume} {129}},\ \bibinfo {pages}
  {2413} (\bibinfo {year} {1963})}\BibitemShut {NoStop}%
\bibitem [{\citenamefont {Harper}\ and\ \citenamefont
  {Tinkham}(1968)}]{harper_mixed_1968}%
  \BibitemOpen
  \bibfield  {author} {\bibinfo {author} {\bibfnamefont {F.~E.}\ \bibnamefont
  {Harper}}\ and\ \bibinfo {author} {\bibfnamefont {M.}~\bibnamefont
  {Tinkham}},\ }\bibfield  {title} {\bibinfo {title} {The {Mixed} {State} in
  {Superconducting} {Thin} {Films}},\ }\href
  {https://doi.org/10.1103/PhysRev.172.441} {\bibfield  {journal} {\bibinfo
  {journal} {Phys. Rev.}\ }\textbf {\bibinfo {volume} {172}},\ \bibinfo {pages}
  {441} (\bibinfo {year} {1968})}\BibitemShut {NoStop}%
\bibitem [{\citenamefont {Uchida}\ \emph {et~al.}(2019)\citenamefont {Uchida},
  \citenamefont {Ide}, \citenamefont {Kawamura}, \citenamefont {Takahashi},
  \citenamefont {Kozuka}, \citenamefont {Tokura},\ and\ \citenamefont
  {Kawasaki}}]{uchida_anomalous_2019}%
  \BibitemOpen
  \bibfield  {author} {\bibinfo {author} {\bibfnamefont {M.}~\bibnamefont
  {Uchida}}, \bibinfo {author} {\bibfnamefont {M.}~\bibnamefont {Ide}},
  \bibinfo {author} {\bibfnamefont {M.}~\bibnamefont {Kawamura}}, \bibinfo
  {author} {\bibfnamefont {K.~S.}\ \bibnamefont {Takahashi}}, \bibinfo {author}
  {\bibfnamefont {Y.}~\bibnamefont {Kozuka}}, \bibinfo {author} {\bibfnamefont
  {Y.}~\bibnamefont {Tokura}},\ and\ \bibinfo {author} {\bibfnamefont
  {M.}~\bibnamefont {Kawasaki}},\ }\bibfield  {title} {\bibinfo {title}
  {Anomalous enhancement of upper critical field in {Sr$_{2}$RuO$_{4}$} thin
  films},\ }\href {https://doi.org/10.1103/PhysRevB.99.161111} {\bibfield
  {journal} {\bibinfo  {journal} {Phys. Rev. B}\ }\textbf {\bibinfo {volume}
  {99}},\ \bibinfo {pages} {161111} (\bibinfo {year} {2019})}\BibitemShut
  {NoStop}%
\bibitem [{\citenamefont {Hsu}\ and\ \citenamefont
  {Kapitulnik}(1992)}]{hsu_superconducting_1992}%
  \BibitemOpen
  \bibfield  {author} {\bibinfo {author} {\bibfnamefont {J.~W.~P.}\
  \bibnamefont {Hsu}}\ and\ \bibinfo {author} {\bibfnamefont {A.}~\bibnamefont
  {Kapitulnik}},\ }\bibfield  {title} {\bibinfo {title} {Superconducting
  transition, fluctuation, and vortex motion in a two-dimensional
  single-crystal {Nb} film},\ }\href {https://doi.org/10.1103/PhysRevB.45.4819}
  {\bibfield  {journal} {\bibinfo  {journal} {Phys. Rev. B}\ }\textbf {\bibinfo
  {volume} {45}},\ \bibinfo {pages} {4819} (\bibinfo {year}
  {1992})}\BibitemShut {NoStop}%
\bibitem [{\citenamefont {Abrikosov}\ and\ \citenamefont
  {Gor'kov}(1961)}]{abrikosov_contribution_1961}%
  \BibitemOpen
  \bibfield  {author} {\bibinfo {author} {\bibfnamefont {A.~A.}\ \bibnamefont
  {Abrikosov}}\ and\ \bibinfo {author} {\bibfnamefont {L.~P.}\ \bibnamefont
  {Gor'kov}},\ }\bibfield  {title} {\bibinfo {title} {{Contribution} to the
  theory of superconducting alloys with paramagnetic impurities},\ }\href@noop
  {} {\bibfield  {journal} {\bibinfo  {journal} {Sov. Phys. JETP}\ }\textbf
  {\bibinfo {volume} {12}},\ \bibinfo {pages} {1243} (\bibinfo {year}
  {1961})}\BibitemShut {NoStop}%
\bibitem [{\citenamefont {Hohenberg}(1964)}]{hohenberg_anisotropic_1964}%
  \BibitemOpen
  \bibfield  {author} {\bibinfo {author} {\bibfnamefont {P.}~\bibnamefont
  {Hohenberg}},\ }\bibfield  {title} {\bibinfo {title} {{Anisotropic}
  superconductors with nonmagnetic impurities},\ }\href@noop {} {\bibfield
  {journal} {\bibinfo  {journal} {Sov. Phys. JETP}\ }\textbf {\bibinfo {volume}
  {18}},\ \bibinfo {pages} {834} (\bibinfo {year} {1964})}\BibitemShut
  {NoStop}%
\bibitem [{\citenamefont {Abrikosov}(1993)}]{abrikosov_influence_1993}%
  \BibitemOpen
  \bibfield  {author} {\bibinfo {author} {\bibfnamefont {A.~A.}\ \bibnamefont
  {Abrikosov}},\ }\bibfield  {title} {\bibinfo {title} {Influence of the gap
  anisotropy on superconducting properties},\ }\href
  {https://doi.org/10.1016/0921-4534(93)90114-6} {\bibfield  {journal}
  {\bibinfo  {journal} {Physica C}\ }\textbf {\bibinfo {volume} {214}},\
  \bibinfo {pages} {107} (\bibinfo {year} {1993})}\BibitemShut {NoStop}%
\bibitem [{\citenamefont {Radtke}\ \emph {et~al.}(1993)\citenamefont {Radtke},
  \citenamefont {Levin}, \citenamefont {Sch\"{u}ttler},\ and\ \citenamefont
  {Norman}}]{radtke_predictions_1993}%
  \BibitemOpen
  \bibfield  {author} {\bibinfo {author} {\bibfnamefont {R.~J.}\ \bibnamefont
  {Radtke}}, \bibinfo {author} {\bibfnamefont {K.}~\bibnamefont {Levin}},
  \bibinfo {author} {\bibfnamefont {H.-B.}\ \bibnamefont {Sch\"{u}ttler}},\
  and\ \bibinfo {author} {\bibfnamefont {M.~R.}\ \bibnamefont {Norman}},\
  }\bibfield  {title} {\bibinfo {title} {Predictions for impurity-induced
  {$T_{c}$} suppression in the high-temperature superconductors},\ }\href
  {https://doi.org/10.1103/PhysRevB.48.653} {\bibfield  {journal} {\bibinfo
  {journal} {Phys. Rev. B}\ }\textbf {\bibinfo {volume} {48}},\ \bibinfo
  {pages} {653} (\bibinfo {year} {1993})}\BibitemShut {NoStop}%
\bibitem [{\citenamefont {Borkowski}\ and\ \citenamefont
  {Hirschfeld}(1994)}]{borkowski_distinguishing_1994}%
  \BibitemOpen
  \bibfield  {author} {\bibinfo {author} {\bibfnamefont {L.~S.}\ \bibnamefont
  {Borkowski}}\ and\ \bibinfo {author} {\bibfnamefont {P.~J.}\ \bibnamefont
  {Hirschfeld}},\ }\bibfield  {title} {\bibinfo {title} {Distinguishing
  $d$-wave superconductors from highly anisotropic $s$-wave superconductors},\
  }\href {https://doi.org/10.1103/PhysRevB.49.15404} {\bibfield  {journal}
  {\bibinfo  {journal} {Phys. Rev. B}\ }\textbf {\bibinfo {volume} {49}},\
  \bibinfo {pages} {15404} (\bibinfo {year} {1994})}\BibitemShut {NoStop}%
\bibitem [{\citenamefont {Openov}(1998)}]{openov_critical_1998}%
  \BibitemOpen
  \bibfield  {author} {\bibinfo {author} {\bibfnamefont {L.~A.}\ \bibnamefont
  {Openov}},\ }\bibfield  {title} {\bibinfo {title} {Critical temperature of an
  anisotropic superconductor containing both nonmagnetic and magnetic
  impurities},\ }\href {https://doi.org/10.1103/PhysRevB.58.9468} {\bibfield
  {journal} {\bibinfo  {journal} {Phys. Rev. B}\ }\textbf {\bibinfo {volume}
  {58}},\ \bibinfo {pages} {9468} (\bibinfo {year} {1998})}\BibitemShut
  {NoStop}%
\bibitem [{\citenamefont {Allen}(1978)}]{allen_new_1978}%
  \BibitemOpen
  \bibfield  {author} {\bibinfo {author} {\bibfnamefont {P.~B.}\ \bibnamefont
  {Allen}},\ }\bibfield  {title} {\bibinfo {title} {New method for solving
  {Boltzmann}'s equation for electrons in metals},\ }\href
  {https://doi.org/10.1103/PhysRevB.17.3725} {\bibfield  {journal} {\bibinfo
  {journal} {Phys. Rev. B}\ }\textbf {\bibinfo {volume} {17}},\ \bibinfo
  {pages} {3725} (\bibinfo {year} {1978})}\BibitemShut {NoStop}%
\bibitem [{\citenamefont {Golubov}\ and\ \citenamefont
  {Mazin}(1997)}]{golubov_effect_1997}%
  \BibitemOpen
  \bibfield  {author} {\bibinfo {author} {\bibfnamefont {A.~A.}\ \bibnamefont
  {Golubov}}\ and\ \bibinfo {author} {\bibfnamefont {I.~I.}\ \bibnamefont
  {Mazin}},\ }\bibfield  {title} {\bibinfo {title} {Effect of magnetic and
  nonmagnetic impurities on highly anisotropic superconductivity},\ }\href
  {https://doi.org/10.1103/PhysRevB.55.15146} {\bibfield  {journal} {\bibinfo
  {journal} {Phys. Rev. B}\ }\textbf {\bibinfo {volume} {55}},\ \bibinfo
  {pages} {15146} (\bibinfo {year} {1997})}\BibitemShut {NoStop}%
\bibitem [{\citenamefont {Bj\"{o}rck}\ and\ \citenamefont
  {Andersson}(2007)}]{bjorck_genx_2007}%
  \BibitemOpen
  \bibfield  {author} {\bibinfo {author} {\bibfnamefont {M.}~\bibnamefont
  {Bj\"{o}rck}}\ and\ \bibinfo {author} {\bibfnamefont {G.}~\bibnamefont
  {Andersson}},\ }\bibfield  {title} {\bibinfo {title} {{GenX}: an extensible
  {X}-ray reflectivity refinement program utilizing differential evolution},\
  }\href {https://doi.org/10.1107/S0021889807045086} {\bibfield  {journal}
  {\bibinfo  {journal} {J. Appl. Crystallogr.}\ }\textbf {\bibinfo {volume}
  {40}},\ \bibinfo {pages} {1174} (\bibinfo {year} {2007})}\BibitemShut
  {NoStop}%
\bibitem [{\citenamefont {Paglione}\ \emph {et~al.}(2002)\citenamefont
  {Paglione}, \citenamefont {Lupien}, \citenamefont {MacFarlane}, \citenamefont
  {Perz}, \citenamefont {Taillefer}, \citenamefont {Mao},\ and\ \citenamefont
  {Maeno}}]{paglione_elastic_2002}%
  \BibitemOpen
  \bibfield  {author} {\bibinfo {author} {\bibfnamefont {J.}~\bibnamefont
  {Paglione}}, \bibinfo {author} {\bibfnamefont {C.}~\bibnamefont {Lupien}},
  \bibinfo {author} {\bibfnamefont {W.~A.}\ \bibnamefont {MacFarlane}},
  \bibinfo {author} {\bibfnamefont {J.~M.}\ \bibnamefont {Perz}}, \bibinfo
  {author} {\bibfnamefont {L.}~\bibnamefont {Taillefer}}, \bibinfo {author}
  {\bibfnamefont {Z.~Q.}\ \bibnamefont {Mao}},\ and\ \bibinfo {author}
  {\bibfnamefont {Y.}~\bibnamefont {Maeno}},\ }\bibfield  {title} {\bibinfo
  {title} {Elastic tensor of {Sr$_{2}$RuO$_{4}$}},\ }\href
  {https://doi.org/10.1103/PhysRevB.65.220506} {\bibfield  {journal} {\bibinfo
  {journal} {Phys. Rev. B}\ }\textbf {\bibinfo {volume} {65}},\ \bibinfo
  {pages} {220506} (\bibinfo {year} {2002})}\BibitemShut {NoStop}%
\bibitem [{\citenamefont {Senyshyn}\ \emph {et~al.}(2009)\citenamefont
  {Senyshyn}, \citenamefont {Trots}, \citenamefont {Engel}, \citenamefont
  {Vasylechko}, \citenamefont {Ehrenberg}, \citenamefont {Hansen},
  \citenamefont {Berkowski},\ and\ \citenamefont
  {Fuess}}]{senyshyn_anomalous_2009}%
  \BibitemOpen
  \bibfield  {author} {\bibinfo {author} {\bibfnamefont {A.}~\bibnamefont
  {Senyshyn}}, \bibinfo {author} {\bibfnamefont {D.~M.}\ \bibnamefont {Trots}},
  \bibinfo {author} {\bibfnamefont {J.~M.}\ \bibnamefont {Engel}}, \bibinfo
  {author} {\bibfnamefont {L.}~\bibnamefont {Vasylechko}}, \bibinfo {author}
  {\bibfnamefont {H.}~\bibnamefont {Ehrenberg}}, \bibinfo {author}
  {\bibfnamefont {T.}~\bibnamefont {Hansen}}, \bibinfo {author} {\bibfnamefont
  {M.}~\bibnamefont {Berkowski}},\ and\ \bibinfo {author} {\bibfnamefont
  {H.}~\bibnamefont {Fuess}},\ }\bibfield  {title} {\bibinfo {title} {Anomalous
  thermal expansion in rare-earth gallium perovskites: a comprehensive powder
  diffraction study},\ }\href {https://doi.org/10.1088/0953-8984/21/14/145405}
  {\bibfield  {journal} {\bibinfo  {journal} {J. Phys. Condens. Matter}\
  }\textbf {\bibinfo {volume} {21}},\ \bibinfo {pages} {145405} (\bibinfo
  {year} {2009})}\BibitemShut {NoStop}%
\bibitem [{\citenamefont {Vogt}\ and\ \citenamefont
  {Buttrey}(1995)}]{vogt_low-temperature_1995}%
  \BibitemOpen
  \bibfield  {author} {\bibinfo {author} {\bibfnamefont {T.}~\bibnamefont
  {Vogt}}\ and\ \bibinfo {author} {\bibfnamefont {D.~J.}\ \bibnamefont
  {Buttrey}},\ }\bibfield  {title} {\bibinfo {title} {Low-temperature
  structural behavior of {Sr$_{2}$RuO$_{4}$}},\ }\href
  {https://doi.org/10.1103/PhysRevB.52.R9843} {\bibfield  {journal} {\bibinfo
  {journal} {Phys. Rev. B}\ }\textbf {\bibinfo {volume} {52}},\ \bibinfo
  {pages} {R9843} (\bibinfo {year} {1995})}\BibitemShut {NoStop}%
\bibitem [{\citenamefont {Wu}\ \emph {et~al.}(2020)\citenamefont {Wu},
  \citenamefont {Nair}, \citenamefont {Bollinger}, \citenamefont {He},
  \citenamefont {Robinson}, \citenamefont {Schreiber}, \citenamefont {Shen},
  \citenamefont {Schlom},\ and\ \citenamefont
  {Bo\v{z}ovi\'{c}}}]{wu_electronic_2020}%
  \BibitemOpen
  \bibfield  {author} {\bibinfo {author} {\bibfnamefont {J.}~\bibnamefont
  {Wu}}, \bibinfo {author} {\bibfnamefont {H.~P.}\ \bibnamefont {Nair}},
  \bibinfo {author} {\bibfnamefont {A.~T.}\ \bibnamefont {Bollinger}}, \bibinfo
  {author} {\bibfnamefont {X.}~\bibnamefont {He}}, \bibinfo {author}
  {\bibfnamefont {I.}~\bibnamefont {Robinson}}, \bibinfo {author}
  {\bibfnamefont {N.~J.}\ \bibnamefont {Schreiber}}, \bibinfo {author}
  {\bibfnamefont {K.~M.}\ \bibnamefont {Shen}}, \bibinfo {author}
  {\bibfnamefont {D.~G.}\ \bibnamefont {Schlom}},\ and\ \bibinfo {author}
  {\bibfnamefont {I.}~\bibnamefont {Bo\v{z}ovi\'{c}}},\ }\bibfield  {title}
  {\bibinfo {title} {Electronic nematicity in {Sr$_{2}$RuO$_{4}$}},\ }\href
  {https://doi.org/10.1073/pnas.1921713117} {\bibfield  {journal} {\bibinfo
  {journal} {Proc. Natl. Acad. Sci. USA}\ }\textbf {\bibinfo {volume} {117}},\
  \bibinfo {pages} {10654} (\bibinfo {year} {2020})}\BibitemShut {NoStop}%
\bibitem [{\citenamefont {Capogna}\ \emph {et~al.}(2002)\citenamefont
  {Capogna}, \citenamefont {Mackenzie}, \citenamefont {Perry}, \citenamefont
  {Grigera}, \citenamefont {Galvin}, \citenamefont {Raychaudhuri},
  \citenamefont {Schofield}, \citenamefont {Alexander}, \citenamefont {Cao},
  \citenamefont {Julian},\ and\ \citenamefont
  {Maeno}}]{capogna_sensitivity_2002}%
  \BibitemOpen
  \bibfield  {author} {\bibinfo {author} {\bibfnamefont {L.}~\bibnamefont
  {Capogna}}, \bibinfo {author} {\bibfnamefont {A.~P.}\ \bibnamefont
  {Mackenzie}}, \bibinfo {author} {\bibfnamefont {R.~S.}\ \bibnamefont
  {Perry}}, \bibinfo {author} {\bibfnamefont {S.~A.}\ \bibnamefont {Grigera}},
  \bibinfo {author} {\bibfnamefont {L.~M.}\ \bibnamefont {Galvin}}, \bibinfo
  {author} {\bibfnamefont {P.}~\bibnamefont {Raychaudhuri}}, \bibinfo {author}
  {\bibfnamefont {A.~J.}\ \bibnamefont {Schofield}}, \bibinfo {author}
  {\bibfnamefont {C.~S.}\ \bibnamefont {Alexander}}, \bibinfo {author}
  {\bibfnamefont {G.}~\bibnamefont {Cao}}, \bibinfo {author} {\bibfnamefont
  {S.~R.}\ \bibnamefont {Julian}},\ and\ \bibinfo {author} {\bibfnamefont
  {Y.}~\bibnamefont {Maeno}},\ }\bibfield  {title} {\bibinfo {title}
  {Sensitivity to {Disorder} of the {Metallic} {State} in the {Ruthenates}},\
  }\href {https://doi.org/10.1103/PhysRevLett.88.076602} {\bibfield  {journal}
  {\bibinfo  {journal} {Phys. Rev. Lett.}\ }\textbf {\bibinfo {volume} {88}},\
  \bibinfo {pages} {076602} (\bibinfo {year} {2002})}\BibitemShut {NoStop}%
\bibitem [{\citenamefont {Forsythe}\ \emph {et~al.}(2002)\citenamefont
  {Forsythe}, \citenamefont {Julian}, \citenamefont {Bergemann}, \citenamefont
  {Pugh}, \citenamefont {Steiner}, \citenamefont {Alireza}, \citenamefont
  {McMullan}, \citenamefont {Nakamura}, \citenamefont {Haselwimmer},
  \citenamefont {Walker}, \citenamefont {Saxena}, \citenamefont {Lonzarich},
  \citenamefont {Mackenzie}, \citenamefont {Mao},\ and\ \citenamefont
  {Maeno}}]{forsythe_evolution_2002}%
  \BibitemOpen
  \bibfield  {author} {\bibinfo {author} {\bibfnamefont {D.}~\bibnamefont
  {Forsythe}}, \bibinfo {author} {\bibfnamefont {S.~R.}\ \bibnamefont
  {Julian}}, \bibinfo {author} {\bibfnamefont {C.}~\bibnamefont {Bergemann}},
  \bibinfo {author} {\bibfnamefont {E.}~\bibnamefont {Pugh}}, \bibinfo {author}
  {\bibfnamefont {M.~J.}\ \bibnamefont {Steiner}}, \bibinfo {author}
  {\bibfnamefont {P.~L.}\ \bibnamefont {Alireza}}, \bibinfo {author}
  {\bibfnamefont {G.~J.}\ \bibnamefont {McMullan}}, \bibinfo {author}
  {\bibfnamefont {F.}~\bibnamefont {Nakamura}}, \bibinfo {author}
  {\bibfnamefont {R.~K.~W.}\ \bibnamefont {Haselwimmer}}, \bibinfo {author}
  {\bibfnamefont {I.~R.}\ \bibnamefont {Walker}}, \bibinfo {author}
  {\bibfnamefont {S.~S.}\ \bibnamefont {Saxena}}, \bibinfo {author}
  {\bibfnamefont {G.~G.}\ \bibnamefont {Lonzarich}}, \bibinfo {author}
  {\bibfnamefont {A.~P.}\ \bibnamefont {Mackenzie}}, \bibinfo {author}
  {\bibfnamefont {Z.~Q.}\ \bibnamefont {Mao}},\ and\ \bibinfo {author}
  {\bibfnamefont {Y.}~\bibnamefont {Maeno}},\ }\bibfield  {title} {\bibinfo
  {title} {Evolution of {Fermi}-{Liquid} {Interactions} in {Sr$_{2}$RuO$_{4}$}
  under {Pressure}},\ }\href {https://doi.org/10.1103/PhysRevLett.89.166402}
  {\bibfield  {journal} {\bibinfo  {journal} {Phys. Rev. Lett.}\ }\textbf
  {\bibinfo {volume} {89}},\ \bibinfo {pages} {166402} (\bibinfo {year}
  {2002})}\BibitemShut {NoStop}%
\bibitem [{\citenamefont {Akima}\ \emph {et~al.}(1999)\citenamefont {Akima},
  \citenamefont {NishiZaki},\ and\ \citenamefont
  {Maeno}}]{akima_intrinsic_1999}%
  \BibitemOpen
  \bibfield  {author} {\bibinfo {author} {\bibfnamefont {T.}~\bibnamefont
  {Akima}}, \bibinfo {author} {\bibfnamefont {S.}~\bibnamefont {NishiZaki}},\
  and\ \bibinfo {author} {\bibfnamefont {Y.}~\bibnamefont {Maeno}},\ }\bibfield
   {title} {\bibinfo {title} {Intrinsic {Superconducting} {Parameters} of
  {Sr$_{2}$RuO$_{4}$}},\ }\href {https://doi.org/10.1143/JPSJ.68.694}
  {\bibfield  {journal} {\bibinfo  {journal} {J. Phys. Soc. Jpn.}\ }\textbf
  {\bibinfo {volume} {68}},\ \bibinfo {pages} {694} (\bibinfo {year}
  {1999})}\BibitemShut {NoStop}%
\bibitem [{\citenamefont {Maeno}\ \emph {et~al.}(1997)\citenamefont {Maeno},
  \citenamefont {Yoshida}, \citenamefont {Hashimoto}, \citenamefont
  {Nishizaki}, \citenamefont {Ikeda}, \citenamefont {Nohara}, \citenamefont
  {Fujita}, \citenamefont {Mackenzie}, \citenamefont {Hussey}, \citenamefont
  {Bednorz},\ and\ \citenamefont {Lichtenberg}}]{maeno_two-dimensional_1997}%
  \BibitemOpen
  \bibfield  {author} {\bibinfo {author} {\bibfnamefont {Y.}~\bibnamefont
  {Maeno}}, \bibinfo {author} {\bibfnamefont {K.}~\bibnamefont {Yoshida}},
  \bibinfo {author} {\bibfnamefont {H.}~\bibnamefont {Hashimoto}}, \bibinfo
  {author} {\bibfnamefont {S.}~\bibnamefont {Nishizaki}}, \bibinfo {author}
  {\bibfnamefont {S.-i.}\ \bibnamefont {Ikeda}}, \bibinfo {author}
  {\bibfnamefont {M.}~\bibnamefont {Nohara}}, \bibinfo {author} {\bibfnamefont
  {T.}~\bibnamefont {Fujita}}, \bibinfo {author} {\bibfnamefont {A.~P.}\
  \bibnamefont {Mackenzie}}, \bibinfo {author} {\bibfnamefont {N.~E.}\
  \bibnamefont {Hussey}}, \bibinfo {author} {\bibfnamefont {J.~G.}\
  \bibnamefont {Bednorz}},\ and\ \bibinfo {author} {\bibfnamefont
  {F.}~\bibnamefont {Lichtenberg}},\ }\bibfield  {title} {\bibinfo {title}
  {Two-{Dimensional} {Fermi} {Liquid} {Behavior} of the {Superconductor}
  {Sr$_{2}$RuO$_{4}$}},\ }\href {https://doi.org/10.1143/JPSJ.66.1405}
  {\bibfield  {journal} {\bibinfo  {journal} {J. Phys. Soc. Jpn.}\ }\textbf
  {\bibinfo {volume} {66}},\ \bibinfo {pages} {1405} (\bibinfo {year}
  {1997})}\BibitemShut {NoStop}%
\bibitem [{\citenamefont {Hussey}\ \emph {et~al.}(1998)\citenamefont {Hussey},
  \citenamefont {Mackenzie}, \citenamefont {Cooper}, \citenamefont {Maeno},
  \citenamefont {Nishizaki},\ and\ \citenamefont
  {Fujita}}]{hussey_normal-state_1998}%
  \BibitemOpen
  \bibfield  {author} {\bibinfo {author} {\bibfnamefont {N.~E.}\ \bibnamefont
  {Hussey}}, \bibinfo {author} {\bibfnamefont {A.~P.}\ \bibnamefont
  {Mackenzie}}, \bibinfo {author} {\bibfnamefont {J.~R.}\ \bibnamefont
  {Cooper}}, \bibinfo {author} {\bibfnamefont {Y.}~\bibnamefont {Maeno}},
  \bibinfo {author} {\bibfnamefont {S.}~\bibnamefont {Nishizaki}},\ and\
  \bibinfo {author} {\bibfnamefont {T.}~\bibnamefont {Fujita}},\ }\bibfield
  {title} {\bibinfo {title} {Normal-state magnetoresistance of
  {Sr$_{2}$RuO$_{4}$}},\ }\href {https://doi.org/10.1103/PhysRevB.57.5505}
  {\bibfield  {journal} {\bibinfo  {journal} {Phys. Rev. B}\ }\textbf {\bibinfo
  {volume} {57}},\ \bibinfo {pages} {5505} (\bibinfo {year}
  {1998})}\BibitemShut {NoStop}%
\bibitem [{\citenamefont {Barber}\ \emph {et~al.}(2018)\citenamefont {Barber},
  \citenamefont {Gibbs}, \citenamefont {Maeno}, \citenamefont {Mackenzie},\
  and\ \citenamefont {Hicks}}]{barber_resistivity_2018}%
  \BibitemOpen
  \bibfield  {author} {\bibinfo {author} {\bibfnamefont {M.~E.}\ \bibnamefont
  {Barber}}, \bibinfo {author} {\bibfnamefont {A.~S.}\ \bibnamefont {Gibbs}},
  \bibinfo {author} {\bibfnamefont {Y.}~\bibnamefont {Maeno}}, \bibinfo
  {author} {\bibfnamefont {A.~P.}\ \bibnamefont {Mackenzie}},\ and\ \bibinfo
  {author} {\bibfnamefont {C.~W.}\ \bibnamefont {Hicks}},\ }\bibfield  {title}
  {\bibinfo {title} {Resistivity in the {Vicinity} of a van {Hove}
  {Singularity}: {Sr$_{2}$RuO$_{4}$} under {Uniaxial} {Pressure}},\ }\href
  {https://doi.org/10.1103/PhysRevLett.120.076602} {\bibfield  {journal}
  {\bibinfo  {journal} {Phys. Rev. Lett.}\ }\textbf {\bibinfo {volume} {120}},\
  \bibinfo {pages} {076602} (\bibinfo {year} {2018})}\BibitemShut {NoStop}%
\end{thebibliography}
\end{document}